\pdfoutput=1
\RequirePackage{ifpdf}
\ifpdf % We are running pdfTeX in pdf mode
\documentclass[pdftex]{sigma}
\else
\documentclass{sigma}
\fi

\usepackage{subfigure}

\numberwithin{theorem}{section}
\numberwithin{proposition}{section}
\numberwithin{lemma}{section}
\numberwithin{corollary}{section}
\numberwithin{definition}{section}
\numberwithin{example}{section}
\numberwithin{remark}{section}
\numberwithin{note}{section}

\newcommand{\N}{\mathbb{N}}
\newcommand{\Z}{\mathbb{Z}}

\newcommand{\cE}{\mathcal{E}}
\newcommand{\cI}{\mathcal{I}}
\newcommand{\cK}{\mathcal{K}}
\newcommand{\cL}{\mathcal{L}}
\newcommand{\cM}{\mathcal{M}}
\newcommand{\cH}{\mathcal{H}}
\newcommand{\cF}{\mathcal{F}}

\newcommand{\cJ}{\mathcal{J}}
\newcommand{\cP}{\mathcal{P}}
\newcommand{\cV}{\mathcal{V}}
\newcommand{\cR}{\mathcal{R}}
\newcommand{\cN}{\mathcal{N}}

\newcommand{\f}{\frac}

\newcommand{\cPJ}{\mathcal{P}\!\mathcal{J}}

\newcommand{ \cG }{{\cal G}}
\newcommand{ \cB }{{\cal B}}
\newcommand{ \cT }{{\cal T}}

\DeclareMathOperator{\tr}{Tr}
\DeclareMathOperator{\vol}{vol}
\DeclareMathOperator{\Vol}{Vol}

\newcommand{\vcup}{\,\vec{\cup}\,}

\newcommand{\dsty}{\displaystyle}

\numberwithin{equation}{section}

\begin{document}

\allowdisplaybreaks

\renewcommand{\thefootnote}{$\star$}

\renewcommand{\PaperNumber}{020}

\FirstPageHeading

\ShortArticleName{Colored Tensor Models -- a Review}

\ArticleName{Colored Tensor Models -- a Review\footnote{This
paper is a contribution to the Special Issue ``Loop Quantum Gravity and Cosmology''. The full collection is available at \href{http://www.emis.de/journals/SIGMA/LQGC.html}{http://www.emis.de/journals/SIGMA/LQGC.html}}}

\Author{Razvan GURAU~$^\dag$ and James P.~RYAN~$^\ddag$}

\AuthorNameForHeading{R.~Gurau and J.P.~Ryan}

\Address{$^\dag$~Perimeter Institute for Theoretical Physics,\\
\hphantom{$^\dag$}~31 Caroline St. N, ON N2L 2Y5, Waterloo, Canada}
\EmailD{\href{mailto:rgurau@perimeterinstitute.ca}{rgurau@perimeterinstitute.ca}}

\Address{$^\ddag$~MPI f\"ur Gravitationsphysik, Albert Einstein Institute,\\
\hphantom{$^\ddag$}~Am M\"uhlenberg 1, D-14476 Potsdam, Germany}
\EmailD{\href{mailto:james.ryan@aei.mpg.de}{james.ryan@aei.mpg.de}}

\ArticleDates{Received October 05, 2011, in f\/inal form March 13, 2012; Published online April 10, 2012}

\Abstract{Colored tensor models have recently burst onto the scene as a promising conceptual and computational
 tool in the investigation of problems of random geometry in dimension three and higher. We present a snapshot
 of the cutting edge in this rapidly expanding research f\/ield.  Colored tensor models have been shown to share
many of the properties of their direct ancestor, matrix models, which encode a theory of f\/luctuating two-dimensional
surfaces. These features include the possession of Feynman graphs encoding topological spaces, a $1/N$ expansion
of graph amplitudes,
embedded matrix models inside the tensor structure, a resumable leading order with critical behavior and a continuum
large volume limit,  Schwinger--Dyson equations satisfying a Lie algebra (akin to the Virasoro algebra in two dimensions),
non-trivial classical solutions and so on.  In this review, we give a~detailed introduction of colored tensor models and
 pointers to current and future research directions.}


\Classification{05C15;  05C75; 81Q30; 81T17; 81T18; 83C27; 83C45}

\tableofcontents

\renewcommand{\thefootnote}{\arabic{footnote}}
\setcounter{footnote}{0}

\section{Introduction}

Many problems one comes across in mathematics and theoretical physics, after one strips away the interpretational
dressing, reveal themselves to be enumerative in nature.  Simply speaking, one counts.  Meanwhile, the advent of quantum theory revealed the
evermore growing importance of probabilistic processes in physics. Much ink has been spilled in the mathematical community  pursuing
the development of probability theory in terms of measure-theoretic concepts \cite{kolmog}. Far from being divorced, these combinatorial and
probabilistic concepts appear to facilitate a harmonious co-existence. In fact, it is rather the case that they are becoming increasingly
intertwined.
Historically, the model processes of probability theory often involved an explicit combinatorial mechanism, for instance, counting the ways of
placing balls of varied colors into urns \cite{bern}. These seemingly antiquarian methods had direct physical implications relating to
 contemporary vo\-ting procedures
and models of gases.  One could perhaps view the theory of particles undergoing Brownian motion as the succinct culmination of these ef\/forts.
It also serves to illustrate these probabilistic approaches in a more modern setting: summing the (random) walks of the particles with
a probability
measure captures the dif\/fusive nature of this physical process.

\looseness=1
This marriage provides powerful tools to investigate a multitude of theories, nowhere more prevalently than in quantum f\/ield theory.
In this context, one has a priori three approaches within which to pose a given problem:  Schr\"odinger's analytic approach,
Heisenberg's algebraic
approach or Feynman's combinatorial approach.   While certain
conceptual problems may favor solution by analytic or algebraic methods, it is undoubtedly true that the combinatorial
formulation has proved to be the preeminent instrument for extracting data. One of its strengths as a computational tool stems from
 the fact that many applications just involve the ability to perform (generalized) Gaussian integrals. Furthermore, one need look no
 further than lattice QCD to f\/ind a scenario very well adapted to numerical methods in the non-perturbative regime.

Meanwhile, the combinatorial approach also has many attractive features from a conceptual viewpoint:  it implements the symmetries of
the theory directly; it allows one to incorporate constraints in a simple fashion and thus isolates relevant dynamical
variables. In fact, many thermodynamical quantities (free energies, critical exponents) and observables (Wilson loop observables)
have their most natural def\/inition in this language.

In this review, we shall discuss a class of quantum f\/ield theories known as colored tensor models. Since we develop the theory
in quite a linear fashion, we feel it is appropriate to  motivate each  section individually.

{\it Gaussian measures and Feynman graphs -- with what kind of theory are we dealing?}
The combinatorial formulation of quantum theory has exercised such an appealing allure that here, as in many other
texts, we shall def\/ine a quantum f\/ield theory as a probability measure for random functions.  In the spirit of the physics literature, we shall
call the random functions `f\/ields'. We start from rather general principles and introduce the theory
of probability measures adapted to our quantum f\/ield theoretic setting, that is Gaussian measures and their perturbation.  Having said that,
we do not stray into the more arid regions of measure theory involving $\sigma$-f\/ields, Borel sets and the like (for that, we invite the reader
to consult any thorough textbook on probability theory, for example \cite{grimmett}). Throughout this section, we supply ample simplif\/ied
examples, relevant for out subsequent study of colored tensor models.

The physical observables are the correlations of the probability measure.  Whenever the probability measure is a perturbed Gaussian measure,
the observables of a quantum f\/ield theory are computed via an expansion in Feynman graphs. The perturbation is encoded in an action, a
polynomial functional of the random f\/ields. The graphs, which are combinatorial objects, come equipped with combinatorial weights and amplitudes
f\/ixed by the details of the action: the measure and the vertex kernels of the perturbation. The graphs comprise of vertices and edges, hence
topologically, they are 1-complexes.

To conclude this section, we introduce the generic form of $(D+1)$-colored quantum f\/ield theory models.
They are probability measures for a collection of random f\/ields indexed by colors. In turn, this leads to graphs in which the edges
possess a color index ($(D+1)$-colored graphs).

 {\it A brief review of $($colored$)$ matrix models -- where can we look for hints?}
In this section, we present a quick review of matrix models, emphasizing the aspects that we shall later generalize to colored tensor models.
In the case of matrix models, the coloring does not play an important role and most of the notions we present for the colored models are
in fact relevant to non-colored ones. However, in order to prepare the reader for the higher-dimensional case, in which the colors play a
crucial role, we highlight the colored case also for matrices.

 {\it Topology of colored graphs -- what are the properties of our Feynman graphs?}
We examine in detail the characteristic properties of the $(D+1)$-colored graphs.  The nested structure of subgraphs indexed by colors
endows the graphs with a $D$-dimensional cellular complex structure. Most importantly, each cell complex associated to a graph is
 a simplicial pseudo-manifold,
consequently the colored models are statistical theories of random $D$-dimensional topological spaces.
We present the colored cellular homology and homotopy for our graphs and clarify the structure of
their boundaries, relevant for the analysis of observables of the colored models.

Subsequently, we proceed to the core of this section.  From the lessons one learns in the study of matrix models, it is of utmost
importance that one
identif\/ies combinatorial and topological quantities associated to graphs. For matrix models,
this is the genus of the (ribbon) Feynman graph and of its dual Riemann surface. Something very similar happens in higher dimensions
for the tensor models
we study.  For any colored graph there is a precisely def\/ined set of Riemann surfaces embedded in the graph, known as jackets.
Each such jacket
corresponds to a ribbon graph embedded in the $(D+1)$-colored graph. Each jacket has then a genus, and one def\/ines
the {\it degree} of the $(D+1)$-colored graph as the sum of all the genera of its jackets.

The picture, however, in higher dimensions is not as rosy as it might seem so far. The degree is {\it not} a topological invariant and the study
of the full perturbative expansion of tensor models requires more care. It transpires that there are a set of graph manipulations which preserve
certain combinatorial properties: these are the $k$-dipole creation/annihilation moves which were already known in the theory of manifold
crystallization \cite{FG, Lins}.  They allow one to def\/ine the notion of combinatorial equivalence and classify graphs into
equivalence classes. Moreover, a~subset of $k$-dipole moves encodes homeomorphisms and thus preserve the topology of the cell complex associated to the graph.
This subset leads to a classif\/ication of graphs taking into account not only the combinatorics but also the topology.

Our interest in the f\/ine structure of jackets does not end there. It appears that one can partition the set of jackets into subsets which individually capture
the degree of the graph.  This partition reappears when dealing with embedded matrix model regimes and classical solutions.

 {\it Tensor models.}
To a certain extent, the preceding three sections were mathematical preliminaries. In particular, no reference has so far been made to any specif\/ic higher-dimensional tensor model. We shall now deal with the f\/irst such specif\/ic example. A statistical theory of random
topological spaces requires no other ingredient than the colors. On the contrary the weights (amplitudes) of the graphs are strongly dependent on the details
of the model: the number of arguments of the f\/ields,  the connectivity of the arguments at the vertices, the Gaussian measure with respect to which we perturb,
an so on. We will concentrate our study on the simplest model one can consider, the independent identically distributed~(i.i.d.)~colored tensor model.

Before we go any further, let us spare a moment to place colored tensor models within the larger scheme of things. Often the random f\/ields in quantum f\/ield theory
are def\/ined on a group and one can formulate the same quantum f\/ield theory in terms of its Fourier modes. The random f\/ield thus becomes a random tensor. As
$\mathbb{R}^D$ is a group, most of the quantum f\/ield theories one encounters in f\/lat space time (quantum electrodynamics, non-Abelian gauge
theories, etc.) can be considered, in the broadest sense, tensor models. In opposition to this, we will call {\it tensor models} in this review those
quantum f\/ield theories whose graphs encode a topological space of dimension at least two. The simplest tensor models are thus the random matrix models.
We will use synonymously the name {\it group field theories} for these models\footnote{Less inclusive authors prefer to use the name group f\/ield
theory when referring to specif\/ic subclasses of such models.}.

Much of the analysis we conduct in this paper is the generalization in higher dimensions of the one undertaken in matrix
models (which are probability measures of random rank two tensors).  Random matrices f\/irst appeared in physics as an
attempt to understand the statistical behavior of slow neutron resonances.  Later, they were used in ef\/forts to
characterize chaotic systems, describe elastodynamic properties of structural materials, typify the conductivity attributes
of disordered metals, study the theory of strong interactions \cite{planar},
detail aspects of various putative physical systems such as two-dimensional quantum gravity (both pure or coupled to matter)
\cite{Boulatov:1986sb,Brezin:1989db,mm,Di_Francesco:1993nw,Kaz, Kazakov:1986hu,Knizhnik:1988ak}, conformal f\/iled theory
\cite{david2,DK,Dup}, string theory \cite{double,double1,double2}, illustrate the distribution of the values of the Riemann zeta function on
the critical line, count certain knots and links and the list keeps growing.
In all these applications, the power of the approach resides in the control one has on the statistical properties of large matrices, of size $N$.
This level of control stems from the so called `$1/N$ expansion', which is a (much better behaved) alternative to the usual small
coupling expansion in quantum f\/ield theory dominated by graphs corresponding to the simplest (spherical) topology \cite{Brez, mm, Kaz}.

Like matrix models, colored tensor models come equipped with a large parameter $N$, the size of the tensors. It is of utmost importance that we identify
in this case also an expansion in~$1/N$ and understand it as thoroughly as possible order by order. It is only then that we have any chance of comprehending
the statistical behavior of random topological spaces in higher dimensions.

The material discussed in this rest of this section  is dedicated to presenting
the $1/N$ expansion of colored tensor models. In fact in higher dimensions we must distinguish between two distinct expansions,
one taking into account only the amplitude of the graphs and one including also their topology.
To this end we show that the power counting in $N$ of the amplitudes of the i.i.d.~model is given by their degree.  We then make
use of the combinatorial and topological moves to def\/ine the notion of core graph.  Roughly speaking, these are the `simplest'
graphs (according to specif\/ic criteria) of a certain degree in a combinatorial/topological
equivalence class.  The power of these core graphs is that one can use them to order the~$1/N$ expansion. We give a~rundown of
several redundancies in these expansions and catalog the lowest order examples of core graphs. Critically, the leading order
graphs correspond to the simplest, spherical topology, in any dimension. However, not all graphs of spherical topology,
but only a very specif\/ic subclass, contribute at leading order.

 {\it Embedded matrix models.}
By the time we get to this point of the review, we shall have heard much about the similarities and dif\/ferences between matrix and tensor models.
There are certain algebraic and analytic tools we would like very much to generalize from the matrix to the tensor scenario vis \`a vis their exact
solution \cite{Di_Francesco:1993nw} (at least in a certain regime).  We take a f\/irst step in this direction here.  We utilize the concept of PJ-factorization to
partition the degrees of freedom of the tensor model.  Rather than viewing the tensor model as a simply generating weighted $D$-dimensional cellular
 complexes, we can identify already at the level of the action, the generators of the embedded jackets.  These generators correspond to matrix models
 embedded inside the tensor model.   This provides a launch pad to applying matrix model techniques directly to the certain tensor probability measures.

 {\it Critical behavior.}
In this section we investigate in detail the dominant contribution in the large $N$ limit of colored tensor models. We provide a purely combinatorial
characterization of the
`melon graphs' contributing to the leading order generalizing the planar graphs \cite{Brez} to arbitrary dimensions. By constructing an explicit
map between graphs and colored rooted $(D+1)$-ary trees, we present several
ways to resum the series. We show that this resumed leading order exhibits a critical behavior and undergoes an analytically controlled phase transition
from a~discrete to a continuum theory. We subsequently present an interpretation of the i.i.d.~colored tensor model in terms of dynamical
triangulations.  In this more geometric perspective the leading order continuum phase shares the critical behavior of the branched polymer phase
of dynamical triangulations. However, the average spectral and Hausdorf\/f dimensions of the melonic graphs ensemble have not been computed and one
can not yet conclude on the precise relationship between melonic graphs and branched polymers.

 {\it Bubble equations.}
A key objective of any quantum f\/ield theory is to identify the quantum equations of motion, that is, the equations
satisf\/ied by the correlation functions.  These are the Schwinger--Dyson equations and contain all the information
pertaining to the quantum dynamics.  Moreover, they can be re-interpreted as operators on the space of observables.
These operators form an algebra, just one representation of which is given by the quantum f\/ield theory correlation
functions.  This opens the door to f\/inding other faithful representations of the algebra, which will describe essentially
the same theory, although their aesthetic presentation might emphasize dif\/ferent aspects.  It is in this manner that
one maps between matrix models and Liouville gravity in two dimensions gravity since the Schwinger--Dyson (also called `loop')
equations form a representation of a sub-algebra of the Virasoro algebra \cite{Ambjorn:1990ji,Fukuma:1990jw,Makeenko:1991ry}.

 {\it Classical solutions.}
Any analysis of a  f\/ield theory would be incomplete if one did not attempt to uncover pertinent information about its
classical regime.  For some f\/ield theories, such as (quantum) electrodynamics, the classical action describes directly
rich physical phenomena.  For others, such as quantum chromodynamics, the classical action holds the key to unlocking
non-perturbative information. For example, instantonic solutions furnish genuinely non-perturbative gauge f\/ield conf\/igurations which
display a host of geometrical, topological and quantum ef\/fects with fundamental impact for the ground state and spectrum of non-Abelian gauge
 theories.  While we do not undertake an exhaustive study of the classical theory of colored tensor models, we shall analyze a class of ans\"atze
for solutions of the equations of motion.

Subsequently, we shall perturb the theory in various ways about this non-trivial solution and investigate the resulting quantum behavior.

 {\it Extended discussion and conclusion.}
We have not attempted here to present a complete literature review.  In particular, we have chosen not to include many interesting developments
that have taken place in specif\/ic tensor models, but which did not f\/it into the f\/low of the main review. For that reason, we have expanded the
discussion section so that we could at least brief\/ly coordinate these results with what we have presented here.

Much remains to be studied about colored tensor models, its relation with higher-dimensional conformal f\/ield theory, statistical physics and quantum
gravity, analysis of sub-dominant contributions, tighter control over the topological features of the graphs, just to name the f\/irst few that
 spring to mind.  We tender a forward-looking review to conclude.

\section{Gaussian measures and Feynman graphs}\label{sec:Gaussmeasure}

Ultimately, we shall be interested in analyzing  a class of quantum f\/ield theories known as  colored tensor models. To this end, we f\/irst def\/ine what we mean
by a quantum f\/ield theory and present the tools required to investigate its properties.

In the f\/irst part of this section, we review the relevant aspects of perturbed Gaussian probability measures, along with the Feynman graphs occurring in
the perturbative expansion of the free energy.  In the second part, we introduce the colored measures on which colored tensor models rely. Most of the
material we present is standard.

Let $X$ be a nonempty set and denote  its elements by $n\in X$.  We consider
$\phi:X \to \mathbb{R}$ (or $\mathbb{C}$,  or~$\mathbb{G})$, a real, complex or Grassmann-valued
 random function def\/ined on~$X$.\footnote{We consider
only Grassmann algebras endowed with an anti-involution $\phi \rightarrow \bar\phi $ with
$\overline{\phi\chi}=-\bar\chi \bar \phi$, $\bar{\bar \phi}=-\phi$.} If $\phi$ is a complex
(Grassmann) function, we denote $\bar \phi$ its complex conjugate (its involution). By~$\mathfrak{F}(X)$, we mean the
set of all functions on $X$ and by $\phi_n\equiv \phi(n)$ the value of $\phi$ at the point $n$.
\begin{definition}
 A {\it quantum field theory}
is a probability measure
$\dfrac{1}{Z}\, d\nu(\phi)$ for a random function (f\/ield) $\phi$, where $Z$ is the normalization factor.
\end{definition}

 In the sequel, we shall identify a probability measure through its {\it partition function $Z$} and {\it correlations}:
\begin{gather*}
 Z = \big{\langle} 1 \big{\rangle}  =  \int d\nu(\phi)   , \qquad
\big{\langle} \phi_{n_1} \cdots \phi_{n_p} \big{\rangle}  =  \int d\nu(\phi)  \, \phi_{n_1} \cdots \phi_{n_p}   .
\end{gather*}
The logarithm of the partition function is called the {\it free energy}, and is denoted $F=\ln Z$.
Importantly, generic correlations may be expressed in terms of a subset known as {\it connected correlations}
 (and signif\/ied by $\langle\cdot\rangle_{c}$):
\[
 \big{\langle} \phi_{n_1} \cdots \phi_{n_p} \big{\rangle} =
 \big{\langle} \phi_{n_1} \cdots \phi_{n_p} \big{\rangle}_c \big{\langle} 1 \big{\rangle} +
  \sum_{ P} \prod_{ S \in P } \big{\langle} \prod_{k \in S } \phi_{n_k}
\big{\rangle}_c \big{\langle} 1 \big{\rangle}   ,
\]
where $P$ denotes the partitions of the set $\{1,\dots, p\}$ into non-empty subsets $S$.

\begin{example}
 The set $X$ has a unique element $X = \{0\}$. The random function $\phi_0$ is then a~random variable.
A skewed coin with (real-numbered) outcomes $a$ and $b$, for instance, is the quantum f\/ield theory of the pure point
probability measure:
\begin{gather*}
 d\nu(\phi_0)  =   d\phi_0    \big( p   \delta(a -\phi_0) + q   \delta(b-\phi_0) \big) , \qquad
  Z  =  (p+q)   ,\qquad \left\langle \phi_0^n \right\rangle = p    a^n + q   b^n   .
\end{gather*}
\end{example}

\begin{example}
The set $X$ has a f\/inite number of elements $X=\{1,\dots, N\}$. A quantum f\/ield theory
is encoded in a probability density $P$:
\[
  d\nu(\phi) = \left( \prod_{n=1}^N d\phi_{n} \right)    P(\phi_{1},\dots,\phi_{N})  ,
\]
where $d\phi_{n}$ is the usual Lebesgue measure on $\mathbb{R}$.
\end{example}

Of course when the cardinality of $X$ becomes inf\/inite, the def\/inition of a probability measure is a thorny issue.
However, normalized Gaussian measures, can be def\/ined simply also in this case.  Suppose that $X$
 is a measurable space, and label the Lebesgue integral over $X$ by $\sum_n$.  A~linear operator $C:\mathfrak{F}(X) \rightarrow \mathfrak{F}(X)$
 is identif\/ied by its kernel $C_{n,n'}$:
\[
 (C\phi)_{n} = \sum_{n'} C_{n,n'}    \phi_{n'}   .
\]
For complex f\/ields, we will sometimes denote their variables by $\bar n$. Both $\bar n$ and $n$ belong to $X$, that is,
for the elements,  the bar does not denote complex conjugation; it is just a bookkeeping device used to track the indices
belonging to a complex-conjugated f\/ield.

\begin{example}
  Consider $X$ the set:
 \[
  X= \big\{ \vec n \;  \big{\vert} \; \vec n = (n_1,\dots, n_D) ,\;  n_k  = 1,\dots, N  \big \}  .
 \]
 The Lebesgue measure on $X$ is a pure point measure. A random function on $X$ is a random tensor with $D$ indices
 $\phi_{\vec n}$. The kernel of an operator $C$ is a $N^D \times N^D$ matrix $C_{\vec n,\vec n'}$.
\end{example}

\begin{definition}
A {\it normalized Gaussian probability measure of covariance $C$} is a measure whose only non-zero correlations are:
 \begin{itemize}\itemsep=0pt
  \item for a real f\/ield $\phi:X\rightarrow \mathbb{R}$,
    \[
 \left\langle \prod_{n=1}^{2p} \phi_{n} \right\rangle  =
\sum_{{\cal P}}\prod_{\langle n,n'\rangle \in {\cal P}} C_{n,n'}   ,
\]
where ${\cal P}$ denotes all the pairings (that is, distinct partitions of the set $\{1,\dots, 2p\}$ into 2-element
 subsets), and $\langle i,j\rangle$ denotes a pair.
  \item for a complex f\/ield $\phi, \bar\phi:X\rightarrow \mathbb{C}$,
  \[
 \left\langle \prod_{k=1}^p \bigl(   \phi_{ n_k }  \bar \phi_{\bar n_k } \bigr) \right\rangle
 = \sum_{\pi} \prod_k C_{   n_k , \bar n_{\pi(k)}  }  ,
\]
   where $\pi$ is a permutation of $p$ elements.
 \item for Grassmann f\/ields $\phi,\bar \phi  : X \rightarrow \mathbb{G}$,
  \[
 \big{\langle} \prod_{k=1}^p \bigl( \phi_{ n_k }  \bar \phi_{ \bar n_k }\bigr) \big{\rangle}
 = \sum_{\pi} \prod_k \epsilon(\pi)   C_{  n_k , \bar n_{\pi(k )} }   ,
\]
   where $\epsilon(\pi)$ is the signature of the permutation~$\pi$.
\end{itemize}
\end{definition}

The covariance $C$ needs neither be an invertible nor a hermitian operator.

\begin{example}
 If $X=\{1,\dots, N\}$, then the covariance is a matrix. Suppose the covariance is invertible and denote  its inverse by
 $ C_{i,j}^{-1}$. The normalized Gaussian probability measure of covariance $C$ can be written:
 \begin{itemize}\itemsep=0pt
  \item  for a real f\/ield $\phi:X\rightarrow \mathbb{R}$,
       \[
          d\mu_C(\phi) = \sqrt{\det(C)}  \;  \prod_{n=1}^N \frac{d\phi_{n}}{\sqrt{2\pi}}  \,
           e^{ -\frac{1}{2}\sum\limits_{n,n'=1}^N \phi_n C_{n,n'}^{-1} \phi_{n'}};
       \]
  \item  for a complex f\/ield $\phi:X\rightarrow \mathbb{C}$,
        \[
           d\mu_C(\phi, \bar \phi) = \det(C)     \prod_{n=1}^N \frac{d \bar \phi_{\bar n} d \phi_n}{ 2\pi } \,
           e^{- \sum\limits_{n,\bar n=1}^N \bar \phi_{\bar n} C_{ \bar n,n}^{-1} \phi_{n}  } ;
        \]
  \item  for a Grassmann f\/ield $\phi:X\rightarrow \mathbb{G}$,
        \[
          d\mu_C(\phi,\bar\phi)=\frac{1}{\det(C)}     \prod_{n=1}^N d \bar \phi_{\bar n} d \phi_n  \,
           e^{- \sum\limits_{n,\bar n=1}^N \bar \phi_{\bar n}  C_{\bar n,n}^{-1} \phi_{n}   } ,
        \]
         where, in this f\/inal case, the ordering matters\footnote{The integral over the Grassmann algebra is def\/ined
        as $\int d\phi = \partial_{\phi}$,
        and the derivative is def\/ined `to the left' $\partial_{\phi} (\phi A) = A$, if\/f $A\neq \phi B$.}.
 \end{itemize}
 \end{example}

A {\it polynomially perturbed} Gaussian complex or Grassmann measure is:
\begin{gather*}
  d\nu_C(\phi,\bar \phi)  =   d\mu_C(\phi,\bar \phi)  e^{ - S (\phi,\bar \phi) }   , \\
  S (\phi)  =  \sum_{s,\bar s=0}^{\infty} \lambda_{s,\bar s}   K_{n_1\dots n_s; \bar n_{\bar 1} \dots \bar n_{\bar s}}
     \phi_{n_1} \cdots \phi_{n_s}   \bar \phi_{ \bar n_{\bar 1}} \cdots \bar \phi_{ \bar n_{\bar s} }   ,
\end{gather*}
where the repeated indices $n$ and $\bar n$ are summed (note that $s$ and $\bar s$ are independent).

\begin{example}
 Let $X=\{0\}$ and $\phi_0$ a complex random variable. The {\it $D$-ary tree measure} is:
 \[
  d\nu^{J,g}(\phi_0,\bar\phi_0) = d\mu_{1} (\phi_0,\bar \phi_0)  e^{J \phi_0 + g  \phi_0  \bar\phi_0^{D}}   .
 \]
\end{example}

\begin{example}
 The {\it complex $($or Grassmann$)$ $\phi^{D+1}$ measure} is:
\[
  d\nu^{\lambda}( \phi, \bar \phi) = d\mu_C(\bar \phi, \phi)
  e^{-\lambda \sum_n    \phi_n^{D+1} - \bar \lambda \sum_{\bar n}    \bar \phi_{\bar n}^{D+1}}   .
\]
\end{example}

The partition function and correlations of perturbed Gaussian measures are evaluated through Feynman graphs.
 These graphs are obtained by Taylor expanding with respect to the perturbation parameters (the coupling constants)
 and computing the Gaussian integrals. The correlations of $t$ f\/ields $\phi$ and $\bar t$ f\/ields $\bar \phi$ can be written as:
\begin{gather*}
 \left \langle \phi_{n_1} \cdots \phi_{n_{t}}  \bar \phi_{\bar n_{\bar 1} } \cdots \bar \phi_{\bar n_{\bar t} } \right \rangle
 =
  \sum_{p_{s,\bar s} = 0}^{\infty} \left\{ \prod_{ \stackrel{s,\bar s}{p_{s,\bar s}\neq 0 } }
\left[  \frac{ (\lambda_{s,\bar s})^{p_{s, \bar s}} }{p_{s, \bar s }!}
   \prod_{l=1}^{p_{s,\bar s  }} K_{n_1^{(l)}\dots n_s^{(l)}; \bar n_{\bar 1}^{(l)} \dots \bar n_{ \bar s  }^{(l)}  }  \right]\right. \\
 \left.\qquad{} \times \int d\mu_{C}(\phi, \bar \phi)  \; \phi_{n_1} \cdots \phi_{n_{t}}
\bar \phi_{\bar n_{\bar 1} } \cdots \bar \phi_{\bar n_{\bar t}}
  \prod_{ \stackrel{s,\bar s}{p_{s,\bar s}\neq 0 } }
   \left[ \prod_{ l =1}^{p_{s,\bar s} }  \phi_{n_1^{(l)}} \cdots \phi_{n_s^{(l)}} \bar \phi_{\bar n_{\bar 1}^{ (l) }}
\cdots \bar \phi_{ \bar n_{\bar s }^{(l)} } \right] \right\}   ,
\end{gather*}
where again all the repeated $n, \bar n $ indices  are summed.
The Gaussian integrals are zero unless $t+ \sum_{p_{s,\bar s}\neq 0} s = \bar t +  \sum_{p_{ s, \bar s } \neq 0} \bar s$.
Denoting generically  the $n$ indices by $n_k$,  the $\bar n$ indices by $\bar n_k$,
and~$\pi$ a permutation of $k= t +\sum_{p_{s,\bar s}\neq 0} s $ elements:
\begin{gather*}
 \big \langle \phi_{n_1} \cdots \phi_{n_{t}}  \bar \phi_{\bar n_{\bar 1}} \cdots \bar \phi_{\bar n_{\bar t} } \big \rangle
    = \sum_{p_{s,\bar s} = 0}^{\infty} \sum_{\pi}   \prod_{ \stackrel{s,\bar s}{p_{s,\bar s}\neq 0 } }
\left[  \frac{ (\lambda_{s,\bar s})^{p_{s,\bar s}} }{p_{s,\bar s}!}
   \prod_{l=1}^{p_{s,\bar s}} K_{ n_1^{(l)}\dots n_s^{(l)}; \bar n_{\bar 1}^{(l)} \dots \bar n_{\bar s}^{(l)}  } \right]
     \prod_k C_{n_k,\pi(\bar n_k) }   .
\end{gather*}
Each term in the above sum can be represented as a graph.
The kernels $K$ are represented as $(s+\bar s)$-valent vertices, having $s$ half-lines $\phi$ and $\bar s$
half-lines $\bar \phi$. The insertions $\phi_{n}$ and $\bar \phi_{\bar n}$ are represented as external points.
A permutation $\pi$ corresponds to a contraction scheme in which one connects all indices $n$
with indices $\bar n$. Every connection is represented as a line. The sum may be rewritten as a sum over Feynman graphs with $p_{s,\bar s}$
vertices which are $(s+\bar s)$-valent:
\begin{gather*}
\big \langle \phi_{n_1} \cdots \phi_{n_{t}}  \bar \phi_{ \bar n_1} \cdots \bar \phi_{ \bar n_{\bar t} } \big \rangle
  =   \sum_{\cG} \prod_{ \stackrel{s,\bar s}{p_{s,\bar s}\neq 0 } }  \frac{ (\lambda_{s,\bar s})^{p_{s,\bar s}} }{p_{s,\bar s}!}
A_{n_1 \dots n_e; \bar n_1 \dots \bar n_{\bar \bar e} }(\cG)   , \\
 A_{n_1 \dots n_e; \bar n_1 \dots \bar n_{\bar e} }(\cG)   =
\prod_{ \stackrel{s,\bar s}{p_{s,\bar s}\neq 0 } } \left[
    \prod_{l=1}^{p_{s,\bar s}} K_{n_1^{(l)}\dots n_s^{(l)}; \bar n_1^{(l)} \dots \bar n_{\bar s }^{(l)}  } \right]
     \prod_k C_{n_k, \bar n_{\pi(k)} }  ,
\end{gather*}
where $A(\cG)$ is the amplitude of the graph $\cG$ (again recall that repeated indices are summed).
The class of graphs $\cG$ one sums over is
f\/ixed by the terms in the action. The free energy is the sum over {\it connected} graphs. Unless otherwise specif\/ied
we always deal with connected graphs.

\begin{example}\label{eq:darymeasure}
 Consider the $D$-ary tree measure. The partition function is:
\begin{gather*}
  Z  = \int d\mu_{\mathbb{I}} (\phi,\bar \phi)  e^{J \phi + g \phi \bar\phi^{D}}
  = \sum_{p,q=0}^{\infty} \frac{1}{p! q!} J^p g^q  \int d\mu_{\mathbb{I}}(\phi,\bar\phi)    \phi^{p+q} \bar\phi^{Dq} \\
\hphantom{Z}
  =  \sum_{p,q=0}^{\infty} \frac{(p+q)!}{p! q!} J^p g^q \delta_{p+q,Dq } =
\sum_{q=0}^{\infty} \binom{Dq}{q} J^{(D-1)q} g^q
  ,
\end{gather*}
and its 1-point correlation function is:
\begin{gather*}
\left\langle \bar \phi \right \rangle  =  \int d\mu_{\mathbb{I}} (\phi,\bar \phi)   \bar \phi
 e^{J \phi + g \phi \bar\phi^{D}}
= \sum_{p,q=0}^{\infty} \frac{1}{p! q!} J^p g^q  \int d\mu_{\mathbb{I}}(\phi,\bar\phi)   \phi^{p+q} \bar\phi^{Dq+1} \\
\hphantom{\left\langle \bar \phi \right \rangle}{}
=  \sum_{p,q=0}^{\infty} \frac{(p+q)!}{p! q!} J^p g^q \delta_{p+q,Dq+1 } =
\sum_{q=0}^{\infty} \binom{Dq+1}{q} J^{(D-1)q+1} g^q    .
\end{gather*}
The graphs contributing to the connected 1-point function
$ \left\langle \bar \phi \right \rangle_c =\frac{1}{Z} \left\langle \bar \phi \right\rangle $ are connected graphs
built as follows. They have a root vertex corresponding to~$\bar \phi$. The root either connects to
a~1-valent  $J$-vertex (in which case, the graph consist in exactly one line) or it connects
to the $\phi$ f\/ield in a $(D+1)$-valent $g$-vertex. All the other f\/ields of the $g$-vertex are $\bar \phi$. In turn, each of
 them either connects to a $J$-vertex  or to another $g$-vertex.  The graphs can not possess any loop line, since at every
 step in this iterative process,  all as yet uncontracted f\/ields are $\bar \phi$.  Thus, the graphs contributing to the
connected 1-point function of the $D$-ary tree measure are exactly the rooted $D$-ary trees. For further reference we note that:
\[
 \left\langle \bar \phi \right \rangle_c = \sum_{q=0}^{\infty} \frac{1}{Dq+1} \binom{Dq+1}{q} J^{(D-1)q+1} g^q \;,
\]
see \cite{Bonzom:2011zz}.
\end{example}

Although one obtains a summable series in the case of the $D$-ary tree measure, in general the expansion in Feynman graphs is
divergent. First of all, for numerous models, the amplitudes of individual graphs diverge. These are the well known ultra-violet
divergences. In order to render the amplitudes f\/inite, one introduces cutof\/fs and a renormalization procedure. A renormalization
procedure relies on the splitting of normalized Gaussian measures. We shall not enter here in the details of this procedure
 (see \cite{salmhofer} for an elementary introduction to the topic of renormalization). What we stress is that the ef\/fective theory
obtained
following a renormalization procedure is unavoidably an ef\/fective theory for the IR degrees of freedom corresponding to large
eigenvalues of the covariance.

After the UV divergences are dealt with, one needs to address the issue of summability of the series. In generic models, one
f\/inds loop lines in the Feynman graphs. Due to the presence of these loop lines, the number of graphs grows faster than
factorially (super exponentially) and the series has zero radius of convergence, being
at most just Borel summable.

The Feynman graphs of the simplest quantum f\/ield theories are made of vertices (0-cells) and lines (1-cells); hence, they
are 1-dimensional cellular complexes. Tensor models are quantum f\/ield theories such that their graphs have a richer
(combinatorial) topological structure, namely they also possess faces (2-cells).

\subsection{The $(D+1)$-colored models}

We come now to the def\/inition of a colored model. As we will see in a later section the graphs of that colored model with $D$ colors
automatically have a $D$-dimensional cellular structure (that is they have vertices, lines, faces and also 3-cells, 4-cells, and so on
up to $D$-cells), thus they are tensor models.

\begin{definition}\label{def:cmodel}
A {\it $(D+1)$-colored model} is a probability measure $d\nu$:
\begin{gather*}
  d\nu  =  \prod_{i} d\mu_{C^i}(\phi^i,\bar\phi^i)  e^{-S}   , \qquad
  S  =  \lambda \sum_{n_i \in X^i } K_{n_1 \dots n_D} \prod_{i=0}^D \phi^i_{n_i}
  + \bar \lambda \sum_{ \bar n_i \in X^i } \bar K_{ \bar n_1\dots \bar n_D}
\prod_{i=0}^D \phi^i_{ \bar n_i}   ,
\end{gather*}
where
\begin{itemize}\itemsep=0pt
\item $\phi^i:X^i\rightarrow \mathbb{C}$ are $D+1$ complex (Grassmann)
random f\/ields\footnote{The domain of def\/inition $X^i$ of the f\/ields need not be the same.};

\item $C^i:F(X^i)\rightarrow F(X^i)$ are $D+1$ covariances;

\item $K,\bar K: X^0\times \dots \times X^D \rightarrow \mathbb{C}$ are two vertex kernels.
\end{itemize}

\end{definition}

 The closed Feynman  graphs of these colored models are termed closed $(D+1)$-colored graphs. One may
give a constructive approach to their description but we prefer to state the end result.

\begin{definition}
A {\it closed $(D+1)$-colored graph} is a graph $\cG = (\cV,\cE)$ with vertex set $\cV$ and edge set $\cE$
such that:
\begin{itemize}\itemsep=0pt
\item  $\cV$ is bipartite, that is, there is a partition of the vertex set $\cV  = V \cup \bar V$, such that for any
element $l\in\cE$, then $ l = \{v,\bar v\}$ where $v\in V$ and $\bar v\in\bar V$. Their cardinalities
satisfy $|\cV| = 2|V| = 2|\bar V|$.
\item  The edge set is partitioned into $D+1$ subsets $\cE = \bigcup_{i  =0}^{D} \cE^i$, where $\cE^i$ is the subset
of edges with color $i$.
\item  It is $(D+1)$-regular (i.e.\ all vertices are $(D+1)$-valent) with all edges
incident  to a given vertex having distinct colors.
\end{itemize}
\end{definition}

We shall call the elements  $v\in V$ ($\bar v \in \bar{V}$) positive (negative) vertices and draw them with the colors
clockwise (anti-clockwise) turning.  The bipartition induces an orientation on the edges, say from $v$ to $\bar v$.
As is to be expected, we shall only be dealing in with connected graphs in the sequel.
Finally,  an important point on notation, we denote by $\widehat i,\widehat j,\dots, \widehat k$ the following
complement in the set of colors $ \{0,\dots, D \} \setminus \{ i,j, \dots, k \} $.

\begin{remark}
 A colored model with real f\/ields can be def\/ined as:
\[
  d\nu  =  \prod_{i} d\mu_{C^i}(\phi^i) e^{-S}   ,\qquad
  S  =  \lambda \sum_{n_i \in X^i } K_{n_1 \dots n_D} \prod_{i=0}^D \phi^i_{n_i}   .
\]
Its graphs are colored, but their vertex set is not bipartite.
\end{remark}

\section{A brief review of (colored) matrix models}

   There are many excellent reviews \cite{Di_Francesco:1993nw} of random matrix models in the literature. This section is not intended to be one. Rather
we want to brief\/ly recall the most prominent features of matrix models, which will then be generalized one by one to colored
tensor models. In the case of matrix models the colors do not play an important role, being merely a decoration. However,
having in mind the generalization to higher dimensions we shall include them in the discussion below.

The particular matrix model in which we are interested is a perturbed Gaussian measure for three $N\times N$  non-hermitian colored matrices,
 with colors $i=\{0,1,2\}$ (extensively analyzed in the literature
\cite{difrancesco-countingRT,difrancesco-rect,difrancesco-coloringRT}):{\samepage
%\begin{alignat*}{4}
%& M^0_{ {\vec n}_0},   	&&				&& {\vec n}_0  =  \big(n^{02}, n^{01}\big) , & \\
%& M^1_{ {\vec n}_1},    	&& \qquad\textrm{where}\qquad &&	{\vec n}_1  =  \big(n^{10},n^{12}\big) , & \\
%& M^2_{ {\vec n}_2}, 	&&	&& 			{\vec n}_2  =  \big(n^{21},n^{20}\big)   .&
%\end{alignat*}
\begin{gather*}
M^0_{ {\vec n}_0}, \qquad M^1_{ {\vec n}_1},\qquad M^2_{ {\vec n}_2},
\end{gather*}
where ${\vec n}_0  =  \big(n^{02}, n^{01}\big)$, ${\vec n}_1  =  \big(n^{10},n^{12}\big)$, ${\vec n}_2  =  \big(n^{21},n^{20}\big)$.}

We will detail below the independent identically distributed~(i.i.d.)~colored matrix model, having covariance and vertex kernels:
\begin{gather*}
  C^i_{ {\vec n}_i, \vec{ \bar n }_i }  =  \prod_{j\neq i} \delta_{ n^{ij} {\bar n}^{ij} }   , \qquad
  K_{ \vec n_0, \vec n_1 \vec n_2 }  = \frac{1}{\sqrt N}  \prod_{i<j}  \delta_{n^{ij}, n^{ji} } ,
\qquad K_{ \vec {\bar n}_0, \vec {\bar n}_1 \vec {\bar n}_2 } = \frac{1}{\sqrt N}
\prod_{i<j}  \delta_{ \bar n^{ij}, {\bar n}^{ji} }   .
\end{gather*}
The reader can convince oneself that the above is just a rather cumbersome way to describe the partition function:
\begin{gather} \label{eq:mataction1}
 Z = e^{ F} = \int \prod_i [dM^i (dM^i)^{\dagger} ]  e^{ - \sum_i \tr [ M^i (M^i)^{\dagger}]  -\frac{\lambda}{\sqrt{N}} \tr [M^0M^1M^2 ]
  - \frac{\bar \lambda}{ \sqrt N } \tr [ (M^0)^{\dagger} (M^1)^{\dagger}(M^2)^{\dagger}] }   ,
\end{gather}
which can also be written, after the change of variables $M = \sqrt{N} T $, in the more familiar form:
\begin{gather} \label{eq:mataction2}
 Z = e^{F} = \int \prod_i [dT^i (dT^i)^{\dagger} ]  e^{ -N \Big{(} \sum_i \tr [ T^i (T^i)^{\dagger}]  + \lambda \tr [T^0T^1T^2 ]
  + \bar \lambda \tr [ (T^0)^{\dagger} (T^1)^{\dagger}(T^2)^{\dagger}] \Big{)} } \; .
\end{gather}
We will denote $E$ the free energy per degree of freedom $E = N^{-2} F$.

\begin{figure}[htb]
\centering
 \includegraphics{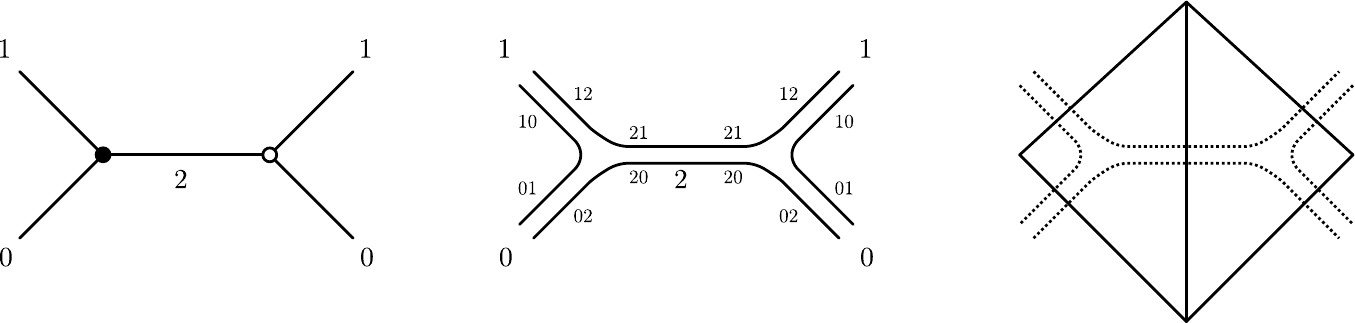}
\caption{Colored ribbon graphs.}\label{fig:ribbongraphs}
\end{figure}

 {\it Graphs:}
The graphs of a matrix model are {\it ribbon graphs} comprising of ribbon lines and ribbon vertices. The sides of the ribbons are called strands,
 and closed strands are called faces. In the colored matrix model above, the ribbon lines have a color: $0$, $1$ or $2$. The strands (which correspond
to the matrix indices~$n^{ij}$) have two colors:~$ij$. For instance, the strand colored by~$12$ is shared by the ribbon lines of color~$1$ and~$2$
at a vertex. The colors of the strands are conserved along the ribbon lines, hence the faces are indexed by couples of colors, see Fig.~\ref{fig:ribbongraphs}. Due to the presence of the colors the ribbon graph representation is, as a matter of fact, redundant. Indeed, one can
represent a graph using only point vertices and colored lines. Every such graph can be uniquely mapped onto a ribbon graph by adding faces for
the cycles made of lines of only two colors.

For every graph one can construct a dual triangulation by drawing a triangle dual to every ribbon vertex and an edge dual to every ribbon line.
The faces of the ribbon graph correspond to the vertices of the triangulation, see again Fig.~\ref{fig:ribbongraphs}. Thus, every graph is dual
 to an orientable surface. The free energy $F$ is the partition function of connected surfaces.

{\it Observables:}
A convenient set of observables in matrix models are the so called {\it loop observables}.  These are traces of products of matrices corresponding
to external faces of connected open ribbon graphs. The colors impose some restrictions on these observables.  As an example, the reader can convince
oneself that a matrix $M^i$ can only be followed by either~$(M^i)^{\dagger}$ or by~$M^{(i+1 \mod 3) }$, hence:
\begin{gather*}
\textrm{Valid:} \quad\begin{cases}   \tr[M^0 (M^0)^{\dagger}]  ,\\  \tr[M^0 M^1M^2  ]  ,\\   \tr[ M^0 (M^0)^{\dagger} (M^1)^{\dagger} M^1]   ,
\end{cases}\qquad
\textrm{Invalid:} \quad \tr[M^0 (M^1)^{\dagger} (M^0)^{\dagger} M^1 ]   .
\end{gather*}

{\it Amplitudes and the $1/N$ expansion:}
As all lines connect a $\lambda$-vertex with a $\bar \lambda$-vertex, a closed connected graph has an even number of vertices
$|\cV| = 2p$. The vertices are trivalent, thus the number of lines $|\cE|$ computes to $|\cE| = \frac{3}{2} |\cV| = 3p $.
The amplitude of a ribbon graph may be readily computed in terms of the numbers $|\cV|=2p$, $|\cE|$ and $|\cF|$ (the number of faces of a graph).
Starting from~\eqref{eq:mataction1} (of course, the same result is obtained, starting from~\eqref{eq:mataction2}),
we obtain a contribution~$N^{-1/2}$ per vertex and a contribution $N$ per face, hence:
\begin{gather*}
 A(\cG) = (\lambda \bar \lambda)^p N^{ - p + \cF } = (\lambda \bar \lambda)^p N^{\cV - \cE + \cF } = (\lambda\bar \lambda)^p N^{2-2g}   ,
\end{gather*}
with $g$ the genus of the graph. As the genus is an nonnegative integer, the free energy~$F$ of a~matrix model supports a~$1/N$ expansion.
The leading order graphs are those with genus $g=0$, that is, planar graphs corresponding to spherical surfaces. Higher genus surfaces
are suppressed in powers of~$1/N$~\cite{Brez,mm, Kaz}.

 {\it Continuum limit:}
The planar sector of the 3-colored matrix model can be analytically sol\-ved~\cite{difrancesco-countingRT,difrancesco-rect,difrancesco-coloringRT}.
After def\/ining the constant $t$ by $ t = (\lambda \bar \lambda)^{1/3}$, one can rewrite the derivative of the free energy of the model (after a change
of variables $ A  = \lambda^{1/3} T $) in parametric form as:
\[
 t \frac{dE}{dt} = U = \frac{z^3}{t^2} (1-3z)   ,\qquad  t= z (1-2z)   ,
\]
which is solved by:
\[
  z = \frac{ 1 - \sqrt{1-8t} } {4}   ,\qquad U =  - \frac{3}{4} + \frac{3}{8 t } - \frac{1}{32 t^2} + \frac{(1-8 t)^{3/2}}{32 t^2}   .
\]
When $t$ approaches the critical value $t_c = \frac{1}{8}$, the free energy of the matrix model (i.e.\ the partition function of connected surfaces)
 exhibits a critical behavior:
\[
 E_{\rm sing} \sim (t_c - t)^{5/2}   .
\]
The average area of the connected surfaces, $ \langle 2p \rangle \sim t\partial_t \ln E \sim \frac{1}{t-t_c}$ diverges, hence,
large surfaces (triangulated by many triangles) dominate.

 {\it Loop equations:}
The integral on two colors, say $1$ and $2$ is Gaussian and can be explicitly performed to express the partition function of the colored
three-matrix model as an integral over only one matrix:
\[
 Z  = \int [dTdT^{\dagger}] e^{- \tr\bigl[ T^{\dagger}T + \ln (1- \lambda \bar \lambda T^{\dagger} T) \bigr]}
=
 \int [dTdT^{\dagger}] e^{ - \tr[ T^{\dagger}T] + \sum_{p=1}^{\infty} \frac{(\lambda \bar \lambda)^p}{p} \tr\bigl[ (T^{\dagger}T)^p \bigr] } \; .
\]
The model becomes the most general model for one (non-hermitian) matrix by replacing the coef\/f\/icients
$\frac{ (\lambda \bar \lambda)^p }{p}$ by an independent $t_p$ for each of the operators $ \tr\bigl[ (T^{\dagger}T)^p \bigr] $
in the ef\/fective action for the last color:
\[
 Z   = \int e^{- N \tr V(T^{\dagger} T )}   ,\qquad V = \sum_{p=1}^{\infty} t_p  \; (T^{\dagger} T)^p  .
\]
The Schwinger--Dyson equations (SDE) of this one-matrix model are:
\begin{gather*}
 0 =  \int [dM] \, \frac{\delta}{\delta M_{ab}} \big( [(MM^{\dagger})^n M]_{ab} \; e^{-N\tr[V(MM^{\dagger})] }\big) \\
\phantom{0}{} =  \Big{\langle} \sum_{k=0}^n [(MM^{\dagger})^{k}]_{aa} [(M^{\dagger}M)^{n-k}]_{bb} \Big{\rangle}
  -N \Big{\langle} \sum_{j=1} j   t_j [(MM^{\dagger})^n M]_{ab} [M^{\dagger}(MM^{\dagger})^{j-1}]_{ba} \Big{\rangle}   ,
\end{gather*}
which, summing over $a$ and $b$, become:
\[
 \Big{\langle} \sum_{k=0}^n \text{Tr}[(MM^{\dagger})^k]\text{Tr}[(MM^{\dagger})^k]^{n-k} \Big{\rangle}
 -N \sum_{j} j\; t_j \Big{\langle} \text{Tr}[(MM^{\dagger})^{n+j}]\Big{\rangle} =0   .
\]
Every insertion of an operator $\tr[(MM^{\dagger})^j]$ in the correlation function can be re-expressed as a~derivative
of $V(MM^{\dagger})$ with respect to~$t_j$. Consequently, the SDEs can be written alternatively~as:
\begin{gather*}%\label{eq:virmat}
L_n Z  =  0  , \qquad \text{for} \quad n\ge 0   ,
\end{gather*}
 where
 \begin{gather*}
 L_{n}  =  N^2 \delta_{0,n} - \frac{2}{N} \frac{\partial }{\partial t_n} +
\frac{1}{N^2}\sum_{k=1}^{n-1} \frac{\partial^2}{\partial t_{k}\partial t_{n-k}} +
  \sum_{j=1}^{\infty} j  t_j \frac{\partial }{\partial t_{n+j}}   ,
\end{gather*}
where the derivatives w.r.t.\ $t_j$, with  $j\le 0$ are understood to be omitted.
A direct computation (involving some relabeling of discrete sums) shows~\cite{Fukuma:1990jw}
that the $L_n$'s respect the commutation relations of (the positive operators of) the Virasoro algebra:
\[
 [L_m,L_n]= (m-n)   L_{m+n} \qquad \text{for}  \quad m,n\ge 0  .
\]
Note that as we only deal with $L_m$, $m\ge 0$, we certainly do not obtain the central charge term.

\section{Topology of colored graphs}\label{sec:topology}

Edge-colored graphs have been extensively used in topology~\cite{FG, Lins} to study manifold crystallization.
We shall present in this section the topology of $(D+1)$-colored graphs, but without restricting to manifold
topologies. The colors encode enough topological information to construct a $D$-dimensional cellular complex,
rather than just the na\"ive 1-complex of a graph. An intuitive picture of the complex associated to a graph is
presented in Remark~\ref{rem:complex}. Essential to the construction of this $D$-dimensional graph complex are the $d$-cells
 (or $d$-bubbles as we term them), for all $d=0,\dots, D$.  This nested cellular structure is precisely the information encoded
in the colors. With this at our disposal, one can subsequently def\/ine a rather rich cellular (co-)homology upon the graph complex.

To aid the reader's navigation through this rather technical section, let us give a f\/lavor of the subsequent subsections.

{\it Cellular structure and pseudo-manifolds:} We shall def\/ine the $d$-cells of the graphs, which are encoded by the colors, and explain
their nested structure.  The aim of this subsection is to show that a $(D+1)$-colored graph is dual to a $D$-dimensional simplicial pseudo-manifold.

{\it Homology and homotopy:} The nested structure of the $d$-cells permits a def\/inition of colored boundary operators and their related colored
homology groups.  We then supply a presentation of the f\/irst homotopy group.  This subsection stands somewhat alone with respect to the development of the rest.

{\it Boundary graphs:}  Here, we examine open rather than closed $(D+1)$-colored graphs.  These open graphs come equipped with a boundary.
We show that these boundaries are themselves colored graphs, but this time the colors are associated to the vertices.
We shall need these def\/initions when we later examine graph factorizations
(below) and the embedded matrix models (Section~\ref{sec:embed}).

{\it Combinatorial moves:}  This and the following two subsections contribute indispensable know\-ledge for a  thorough analysis of the $1/N$-expansion
 of tensor models (Section~\ref{sec:tensomodels}), their critical behavior (Section~\ref{sec:critiid}) and the underlying quantum equations of
motion (Section~\ref{sec:bubbleq}). We explain $k$-dipole creation and its inverse ($k$-dipole contraction).  These are graph manipulations
which preserve certain combinatorial features of the graph, although not necessarily its topology. This allows us to def\/ine the notion of
combinatorial equivalence.   We conclude by highlighting 1-dipole contraction, which plays a signif\/icant role in Section~\ref{sec:tensomodels}.

{\it Jackets and degree:} Jackets are Riemann surfaces embedded in a specif\/ic way inside the $(D+1)$-colored graphs.  We provide the rules for
 their identif\/ication and show that the number of such surfaces, for a given graph, is f\/ixed by its dimension~$D$.   Just as the scaling of matrix
model graphs is controlled by the genus of the Riemann surface associated to that graph,  we discover later that a generalized concept, known as
degree, controls the scaling of higher-dimensional tensor model graphs.  We def\/ine the degree in this subsection; it is simply the sum of the
genera of all the jackets of a graph.

{\it Topological equivalence:}  Although not all $k$-dipole creations/contractions preserve topology, there is a subset that possess this property,
that is, they are homeomorphisms. With this in hand, we are f\/inally at the stage at which we can provide one of the most important results used in
the analysis of tensor models, namely, if the degree of a graph is zero, then it is a sphere.

{\it Graph factorization:}  We conclude this section by developing partitions on the set of jackets for any graph. An element of the partition is
known as a PJ-factorization and each one has a~pleasant property: in order to know the degree of the graph,  one needs to know merely the genera
of the jackets within any one PJ-factorization.   Interestingly, this factorization places a~spotlight on a constraint satisf\/ied by the degree
 (it cannot take just any value).   An understanding of this topics is instrumental for grasping the manner in which matrix models are embedded
 within the tensor model structure (Section~\ref{sec:embed}) and for the analysis of classical solutions of tensor models (Section~\ref{sec:solutions}).

\subsection{Cellular structure and pseudo-manifolds}

At the outset, we need to construct $d$-bubbles \cite{color,lost} for the graph complex $\cG$.

\begin{definition}\label{def:bubble}
The {\it $d$-bubbles} of a graph are the maximally connected subgraphs comprising of edges with $d$ f\/ixed colors.
\end{definition}

The $d$-bubbles are denoted by  $\cB^{i_1\dots i_d}_{ (\rho) }$, where the color indices
are ordered $i_1<i_2<\cdots <i_d$ and~$\rho $ labels the various connected components with the same colors.
We denote $\cB^{[d]}$ the number of $d$-bubbles of the graph.
Note that the 0-bubbles are the vertices of~$\cG$, the 1-bubbles are the edges of~$\cG$. The $2$-bubbles are the faces of~$\cG$.

\begin{figure}[htb]
\centering
 \includegraphics{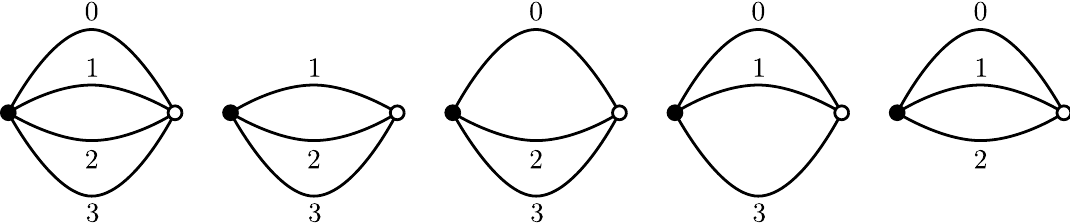}
\caption{$3$-bubbles of a graph in $D=3$.}
\label{fig:exemplubub}
\end{figure}

A graph in $D=3$ and its $3$-bubbles are presented in Fig.~\ref{fig:exemplubub}.
The $3$-bubbles are indexed by the colors of their lines, namely from left to right $123$, $023$, $013$ and $012$.
The $2$-bubbles are the subgraphs consisting each of two lines, of colors $01$, $02$, $03$, $12$, $13$ and $23$.
The $1$-bubbles, as already mentioned are the lines $0$, $1$, $2$ and $3$, while the $0$-bubbles are the vertices.

Now, while the def\/inition of $d$-bubbles seems sensible, we should check they possess a $D$-dimensional cellular complex structure.
Constructing the dual {\it finite abstract simplicial comp\-lex}~\cite{rus} facilitates this analysis.
It goes without saying that the graph complex and the dual complex are the
same topological space.  As a quick note, we write $\cH \subset \cG$, if $\cH$ is a~subgraph of~$\cG$.  Then, to build up
the dual complex~\cite{lost}, we f\/irst assemble all the $D$-bubbles of~$\cG$ into a set~$A$:
  \[
   A = \Bigl\{ \cB^{\widehat i}_{(\rho)} \;| \;  \forall \, \rho ,\   i \in \{ 0,\dots, D \}  \Bigr\} \; .
  \]
We consider the subsets of $A$ indexed by the various $(D+1-d)$-bubbles of $\cG$, $d\in \{1,\dots, D+1\}$:
   \[
    \sigma^{ \cB^{ \widehat i_1  \widehat i_2\dots \widehat i_d}_{ (\kappa) } } =
    \Bigl\{ \cB^{\widehat i_k}_{(\rho)} \; | \; \cB^{ \widehat i_1 \widehat i_2\dots \widehat i_d}_{ (\kappa) }
           \subset \cB^{\widehat i_k}_{(\rho)}   , \  k\in \{1,\dots, d\}  \Bigr\}   .
   \]
Note that for a  given bubble $\cB^{ \widehat i_1 \widehat i_2\dots  \widehat i_d}_{ (\kappa) }$ and for each $k\in\{1,\dots,d\}$,
there exists a unique $D$-bubble
 $\cB^{\widehat i_k}_{(\rho)} \supset \cB^{ \widehat i_1 \widehat i_2\dots  \widehat i_d}_{ (\kappa) }$, namely, the maximal
connected component (in $\cG$) obtained by starting from $\cB^{ \widehat i_1 \widehat i_2\dots  \widehat i_d}_{ (\kappa) }$ and
adding lines of all colors except $i_k$.  Thus, the cardinality of
$\sigma^{ \cB^{ \widehat i_1  \widehat i_2\dots \widehat i_d}_{ (\kappa) } }$ is $d$.  Any subset
  $\tau  \in \sigma^{ \cB^{ \widehat i_1  \widehat i_2\dots \widehat i_d}_{ (\kappa) } } $ is indexed by a choice
 of subset $S  \subset \{ 1,\dots, d \} $:
   \[
    \tau =  \Bigr\{ \cB^{\widehat i_k}_{(\rho)} \;|
           \; \cB^{ \widehat i_1 \widehat i_2\dots \widehat i_d}_{ (\kappa) }
           \subset \cB^{\widehat i_k}_{(\rho)}   , \   k \in  \{ 1,\dots, d \} \setminus  S  \Bigl\}   ,
   \]
and so $\tau = \sigma^{\cB^{ \widehat i_1  \widehat i_2\dots \widehat i_{d-|S|}}_{ (\xi) }}$, where
$\cB^{ \widehat i_1  \widehat i_2\dots \widehat i_{d-|S|}}_{ (\xi) }$  is the unique subgraph obtained by adding
the colors~$i_s$, $s\in S$, to the subgraph
$\cB^{ \widehat i_1 \widehat i_2\dots  \widehat i_d}_{ (\kappa) }$.

Crucially, the sets $\sigma^{\cB^{ \widehat i_1  \widehat i_2\dots \widehat i_{d}}_{ (\kappa) }}$ are
the $(d-1)$-simplices of a f\/inite abstract simplicial complex, as the collection (multi set)
  \[
\Delta = \Big\{ \sigma^{ \cB^{\hat i_1\hat i_2\dots \hat i_d}_{ (\kappa) } } \; | \;
\cB^{\hat i_1\hat i_2\dots \hat i_d}_{ (\kappa) } \subset \cG\Big\}   ,
  \]
is such that $\forall\,  \sigma \in \Delta$, $\tau \subset \sigma\Rightarrow \tau \in \Delta$.
The cardinality of $\sigma \in \Delta$ is $d$ (it corresponds to a~$(D+1-d)$-bubble) and so its dimension is~$d-1$.
In fact, since $\Delta$ is {\it non-branching}, {\it strongly connected} and {\it pure}, it is a~{\it $D$-dimensional simplicial
pseudo-manifold} \cite{lost}.\footnote{For the sake of self-containment, we provide here a concise explanation of the above comment. A~{\it $D$-dimensional simplicial pseudo-manifold} is a f\/inite abstract simplicial
complex with the following properties:
\begin{itemize}\itemsep=0pt
\item{\it non-branching}: each $(D-1)$-simplex is a face of precisely two $D$-simplices;
\item{\it strongly connected}: any two $D$-simplices can be joined by a ``strong chain'' of $D$-simplices in which each pair of neighboring simplices have a common $(D-1)$-simplex;
\item{\it pure} (that is, {\it dimensional homogeneity}): each simplex is a face of some $D$-simplex.
\end{itemize}
}

\begin{remark}\label{rem:complex}
The vertices of the graph correspond to the $D$-simplices of the simplicial complex. The half-lines of a vertex represent
the $(D-1)$-simplices bounding a $D$-simplex and have a color. Any lower-dimensional sub-simplex is colored by the colors of the $D-1$
 simplices sharing it. In Fig.~\ref{fig:complex} we sketched the dual complex in $D=3$ dimensions. The vertices are dual to tetrahedra.
 A triangle (say $3$) is dual to a line (of color $3$) and separates two tetrahedra.
An edge (say common to the triangles $2$ and $3$) is dual to a face (2-bubble of colors~$2$
and~$3$). A~vertex (say common to the triangles~$0$,~$2$ and~$3$) is dual to a~$3$-bubble
(of colors~$0$,~$2$ and~$3$).
\begin{figure}[htb]
\centering
 \includegraphics{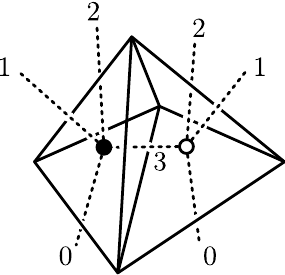}
\caption{The dual complex in $D=3$.}
\label{fig:complex}
\end{figure}
\end{remark}

\begin{remark}
One can give a representation of a colored graph which includes the strands.
Recall that the strands represent the faces of the graph and thus are identif\/ied by pairs of colors. Every line of color $i$ is represented
by $D$ parallel strands with colors $ij$, $j\neq i$. Inside every colored vertex the strand~$ij$ is the strand common to the half-lines~$i$
and $j$ incident at the vertex. Therefore, the ribbon representation of matrix models graphs becomes a stranded representation for graphs made
 of lines and vertices.  This is depicted in Fig.~\ref{fig:strandstrands}.
\begin{figure}[htb]
\centering
 \includegraphics{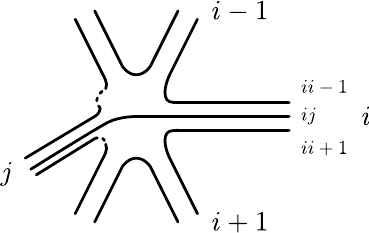}
\caption{Stranded representation.}
\label{fig:strandstrands}
\end{figure}

This representation is often used in the literature. Being redundant and very cumbersome, we shall not use it at all
in this review.
\end{remark}

\begin{remark}
The graphs of the real colored model are also simplicial pseudo-manifolds. Indeed, the fact that the vertex set
is bipartite not played an important role so far.
This will be the case, up to trivial generalizations, for many of the results
we present in the sequel.  The bipartite condition is equivalent to the orientability of the
pseudo-manifold~\cite{caravelli}.
\end{remark}

\subsection{Homology and homotopy}

This section details the topology of the graph complex. It can be skipped at a f\/irst read, especially by the
reader interested in the critical behavior of tensor models, as it does not play a prominent role in the latter.
The homology of the topological space def\/ined by a graph is studied by means of a colored homology
def\/ined for the graph complex~\cite{color}.

\begin{definition}
The {\it $d$'th chain group} is the group f\/initely generated by the $d$-bubbles:
\[
 \alpha_d=\sum_{{\cal B}^{i_1 \dots i_d }_{(\rho)} \subset \cG } c^{i_1 \dots i_d }_{(\rho)}
   {\cal B}^{i_1 \dots i_d }_{(\rho)}   , \qquad c^{i_1 \dots i_d }_{(\rho)} \in \mathbb{Z}    .
\]
\end{definition}

 The chain groups def\/ine homology groups via a boundary operator,

\begin{definition}\label{def:bound}
The {\it $d$'th boundary operator $\partial_d$} acting on a $d$-bubble  ${\cal B}^{i_1 \dots i_d }_{(\rho)}$  is:
\begin{itemize}\itemsep=0pt
\item for $d\ge 2$,
\begin{gather*} %\label{eq:boundary}
\partial_d( {\cal B}^{i_1 \dots i_d }_{(\rho)} )=
\sum_{q=1}^d (-)^{q+1} \sum_{ \cB^{i_1 \dots \widehat{i_q} \dots i_d }_{(\kappa)} \subset {\cal B}^{i_1 \dots i_d }_{(\rho)} }
    \cB^{i_1 \dots \widehat{i_q} \dots i_d }_{(\kappa)}   ,
\end{gather*}
which associates to a $d$-bubble the alternating sum of all $(d-1)$-bubbles formed by subsets of its vertices.

\item for $d=1$, since the edges $\cB^i_{(\rho)}$ connect a positive vertex $v$ to a negative one $\bar v$:
\[
 \partial_d \cB^i_{(\rho)}= v - \bar v   .
\]
 \item for $d=0$, $\partial_0 v = \partial_0 \bar v =0$.
\end{itemize}
\end{definition}

The colored boundary operators def\/ine a homology as $\partial_{d-1}\circ \partial_{d}=0$ \cite{color} and thus we def\/ine
the {\it $d$'th colored homology group} to be $H_d \equiv \rm{ker}(\partial_d)/\rm{Im}(\partial_{d+1})$.  We list some
properties of the maximal and minimal colored homology groups in the lemma below~\cite{color}.

\begin{lemma}
 Denoting  the total number of $d$-bubbles by $\cB^{[d]}$ $($hence $\cB^{[0]}$ its the number of vertices, $\cB^{[1]}$ is
 the number of lines, etc.$)$:
  \begin{gather*}
  \begin{array}{@{}lll}
 \displaystyle   \rm{ker}(\partial_0)  =  \bigoplus_{\cB^{[0]}}{\mathbb Z}  , \qquad &   \displaystyle  \rm{Im} (\partial_1)  =
 \bigoplus_{\cB^{[0]}-1} {\mathbb Z}  , \qquad &  \textrm{hence}\quad H_0 ={\mathbb Z}  ,
 \\
  \displaystyle  \rm{ker} (\partial_{D})   =  {\mathbb Z}  ,  \qquad &  \rm{Im}(\partial_{D+1})  = 0  ,  \qquad   & \textrm{hence}\quad H_D={\mathbb Z}   .
\end{array}
  \end{gather*}
\end{lemma}

 At the same time, some more useful information falls into our lap:
\[
\rm{Im} (\partial_D) =  \bigoplus_{\cB^{[D]}-1} \Z , \qquad  \rm{ker} (\partial_1)  =
  \bigoplus_{\cB^{[1]} -\cB^{[0]} + 1}\Z .
\]

\begin{remark} A f\/inite presentation of the {\it fundamental group} of the graph $\cG$ is obtained by associating a generator
$h_l$ to all edges $l\in\cG$ (apart from those edges lying on a tree in $\cG$, which we f\/ix to the identity~$e$)
and a relation to all faces $\cB^{ij}_{(\rho)}$:
\[ %\label{eq:relations}
{\cal R}_{\cB^{ij}_{(\rho)}}= \prod_{ l \in \cB^{ij}_{(\rho)}   }^{\rightarrow}   h_{l}^{\epsilon(l)} =1  ,
\]
where the product  follows the boundary of the face $\cB^{ij}_{(\rho)} $ in question and $\epsilon(l)$ is $+1$ if our
direction around the boundary agrees with the orientation of $l$ and $-1$ if not. One often comes across tensor models
 that weight graphs according to this characteristic.
\end{remark}

\begin{example}
We present some examples of graphs and their associated homology groups. The black (white) vertices are the positive
clockwise turning (negative anti-clockwise turning)
vertices.
\begin{figure}[htb]
\centering
\begin{minipage}{0.32\textwidth} \centering\includegraphics{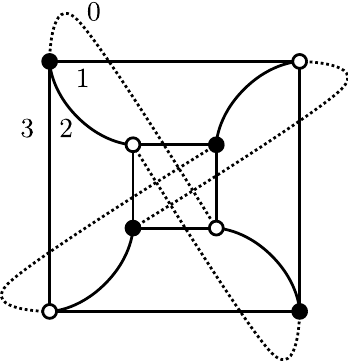}\end{minipage}
\begin{minipage}{0.32\textwidth}\centering \includegraphics{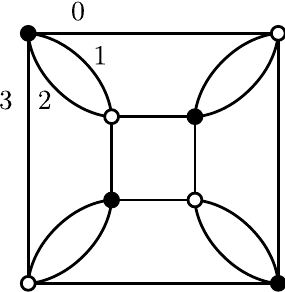} \end{minipage}
\begin{minipage}{0.32\textwidth}\centering \includegraphics{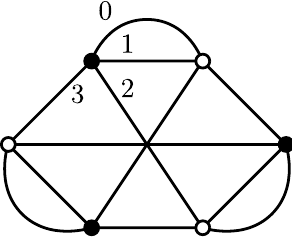}\end{minipage}
\caption{Examples of 3-dimensional graphs.}
\label{fig:graph1}
\end{figure}

\begin{itemize}\itemsep=0pt
\item Consider the leftmost 4-colored graph in Fig.~\ref{fig:graph1}.  Its homology groups are:
\[
 H_0=\Z  , \quad  H_1= \Z_2  , \qquad  H_2 =0  , \qquad  H_3 =\Z  ,
\]
matching those of $\mathbb{R}P^3$. In fact, one can show that the graph represents  $\mathbb{R}P^3$, but this requires
 more ef\/fort than merely computing homology groups.

\item A second example is given by the central graph in Fig.~\ref{fig:graph1}. Its homology groups are:
\[
 H_0=H_3 = \Z  , \qquad H_1=H_2=0   ,
\]
which are those of the 3-sphere $\mathbb{S}^3$ (we shall see later that this graph represents a sphere).

\item For the f\/inal example, we take the rightmost graph in Fig.~\ref{fig:graph1}.  Its homology groups are:
\[
  H_0= \Z   , \qquad  H_1=\Z \oplus\Z  , \qquad  H_2=0   , \qquad  H_3=\Z   .
\]
This graph represents a pseudo-manifold with two isolated singularities. In 3-dimensions, the topological singularities
 arise just at the vertices of the simplicial complex and we f\/ind in this case that
the {\it links}\footnote{The link of a vertex in a simplicial complex is the simplicial complex
$\text{lk}_{\Delta}(V) = \{ \sigma \in \Delta | V\notin \sigma \text{ and } V \cup \sigma \in \Delta \}$.} of two vertices
 are homeomorphic to the torus.
\end{itemize}
\end{example}

\subsection{Boundary graphs}\label{ssec:boundary}

Up to now we dealt with vacuum graphs, that is, closed $(D+1)$-colored graphs. Non-vacuum graphs arise when one evaluates
observables and have a natural interpretation of $D$-dimensional pseudo-manifolds with boundary.
As such they play an important role in colored tensor models.
The boundary is itself a
 $(D-1)$-dimensional pseudo-manifold.  To begin, let us describe these non-vacuum graphs:
\begin{definition}
 An {\it open $(D+1)$-colored graph} is a graph $\cG$ satisfying some additional constraints:
\begin{itemize}\itemsep=0pt
\item It is bipartite, that is, there is a partition of the vertex set $\cV  = V \cup \bar V$, such that for any
element $l\in\cE$, then $ l = \{v,\bar v\}$ where $v\in V$ and $\bar v\in\bar V$.

\item The positive vertices are of two types $V  = V_{\rm int} + V_\partial$, where $V_{\rm int}$ is the set of  $(D+1)$-valent
{\it internal} vertices and the elements of $V_\partial$ are 1-valent {\it boundary} vertices.  A similar distinction
holds for negative vertices.

\item  The edge set is partitioned into $D+1$ subsets $\cE = \bigcup_{i  \in \Z_{D+1}} \cE^i$, where $\cE^i$ is the subset
of edges with color $i$. Furthermore, each $\cE^i = \cE^i_{\rm int} \cup \cE^i_{\rm ext}$, such that {\it internal} edges $\cE^i_{\rm int}$
join two internal vertices, while {\it external} edges $\cE^i_{\rm ext}$ join an internal vertex to a boundary vertex.

\item All edges incident to a $D+1$ valent vertex have distinct colors.
\end{itemize}
\end{definition}

 We consider only connected open graphs, from which we construct the boundary graph~\cite{PolyColor} as follows.

\begin{definition}
 The {\it boundary graph $\cG_{\partial}$} of an open $(D+1)$-colored graph $\cG$ comprises of:

   \begin{itemize}\itemsep=0pt

    \item the vertex set $\cV_\partial = V_{\partial}\cup \bar V_{\partial}$.   We stress that it is not bipartite with
 respect to this splitting.  The vertices inherit the color from the external edges of $\cG$ upon which they lie, so that
a more appropriate partition is $\cV_\partial = \bigcup_{i \in\Z_{D+1}} \cV_\partial^i$, where $\cV_\partial^i$ denotes
the set of boundary vertices with color~$i$.

    \item the edge set $\cE_\partial = \mathop{\bigcup_{i\neq j \in\Z_{D+1}}} \cE^{ij}_\partial$, where
 $l^{ij} = \{v, w\} \in \cE^{ij}_\partial$ exists if there is a bi-colored path from $v$ to $w$ in $\cG$ consisting of colors
 $i$ and $j$.  Thus, the lines $\cE^{ij}_\partial$ inherit the colors of the path in $\cG$.
   \end{itemize}
\end{definition}

\begin{figure}[htb]
\centering
\includegraphics[width=13cm]{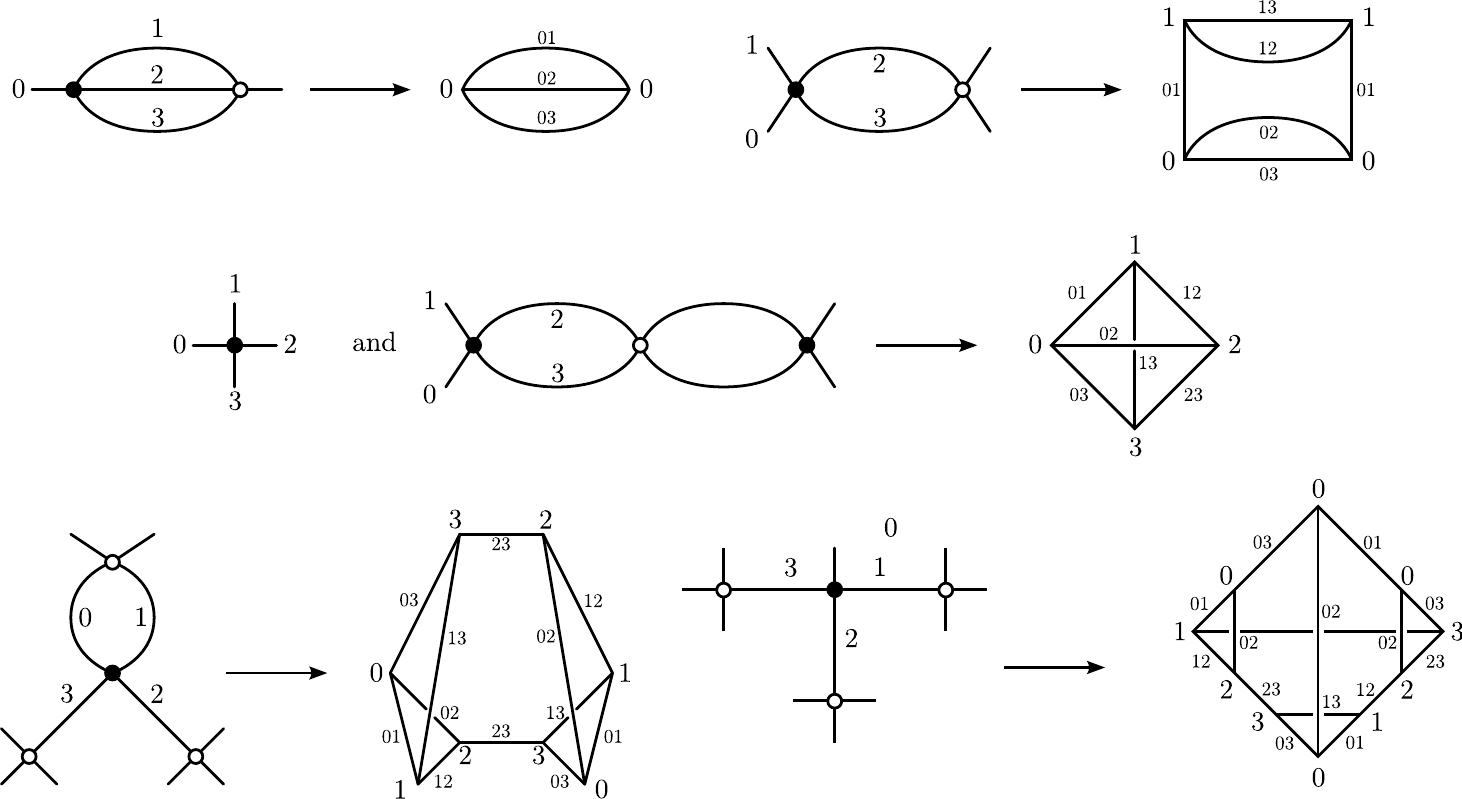}
\caption{Open 4-color graphs and their corresponding boundary graphs.}
\label{fig:boundgr}
\end{figure}

These boundary graphs, possess a number of additional properties. The line $l^{ij} \in \cE_\partial^{ij}$  can only
exist if $ l^{ij}  = \{v^i,w^i\}$ or $\{v^i,w^j\}$ or $\{v^j,w^i\}$ or $\{v^j,w^j\}$, where $v^i, w^i \in \cV_\partial^i$
and $v^j, w^j\in\cV_\partial^j$.   Each boundary vertex is $D$-valent and for $v^i\in \cV_\partial^i$, the incident boundary
edges are $l^{i j}$ where $j \neq i$.  Several examples for $D=3$ are presented in Fig.~\ref{fig:boundgr}.

\begin{remark}
As it has colored vertices and bi-colored edges, the boundary graph $\cG_{\partial}$ is a~priori very dif\/ferent from the
initial graph $\cG$. The boundary graph can have several connected components.  Each component has a  cellular complex
 structure. For $d\ge 1$ the {\it boundary  $d$-bubbles} $(\cB_{\partial})^{i_1\dots i_{d+1}}_{ (\rho) }$ are the maximally
connected components of $\cG_\partial $ formed by boundary vertices $v^{i_a}$ and boundary edges $l^{i_bi_c}$, where
 $i_a,i_b, i_c \in \{i_1,\dots, i_{d+1} \} $. Following step-by-step the constructions for the vacuum graphs, but taking into
account that the boundary $d$-bubbles have $d+1$ colors, one can show that each connected component of $\cG_\partial$ is dual
to a simplicial complex (and is a~pseudo-manifold). In fact, the simplicial complex dual to $\cG_\partial$ is the boundary of
the simplicial complex dual to $\cG$.  Consequently, one can study its colored homology.
\end{remark}

A particular subclass of boundary graphs are those obtained from open $(D+1)$-colored graphs, such that all external but none of the
internal lines have color $D$. In this case, the open graph possesses exactly one internal $D$-bubble.  Additionally, all vertices in the
$\cG_{\partial}$  have the same color, $D$ and all edges in $\cG_{\partial}$ correspond to the existence of a bi-colored
path in $\cG$ joining the two vertices.  Hence, all edges are of the form $l^{Dk}$ for an internal edge of color $k$.
In fact, if one deletes the label $D$ from all the vertices and edges of the boundary graph $\cG_{\partial}$, one f\/inds
that it is identical to the graph obtained from $\cG$ by deleting the external edges.

\subsection{Combinatorial moves: $k$-dipole reductions}

The colored graphs support a class of combinatorial moves, termed $k$-dipole moves \cite{Gur3,Gur4,GurRiv}, which have a well controlled ef\/fect
on their bubble structure. In a subsequent section, we shall see that under a further assumption, the moves implement
homeomorphisms of the topological space associated to the graph.
These moves are crucial to understanding the structure of arbitrary terms in the large $N$ expansion of colored tensor models.
However, the leading order of the expansion can be well understood without going into the details of this section.

\begin{definition} A {\it $k$-dipole} $d_k$ is a subset of $\cG$ comprising of two vertices $v,\,\bar v$ such that:
\begin{itemize}\itemsep=0pt
\item $v$ and $\bar v$ share $k$ edges colored by $i_1, \dots, i_k\in \Z_{D+1}$;
\item $v$ and $\bar v$ lie in distinct $(D+1-k)$-bubbles: $B^{\hat{i}_1\dots \hat{i}_{k}}_{(\alpha)}\neq B^{\hat{i}_1\dots
\hat{i}_{k}}_{(\beta)}$.
\end{itemize}
 \end{definition}

We say that $d_k$ {\it separates} the bubbles $B^{\hat{i}_1\dots \hat{i}_{k}}_{(\alpha)}$ and
$B^{\hat{i}_1\dots \hat{i}_{k}}_{(\beta)}$.
Yet more important is how we manipulate the graph structure with respect to these subsets.

\begin{figure}[htb]
\centering
 \includegraphics[width= 6cm]{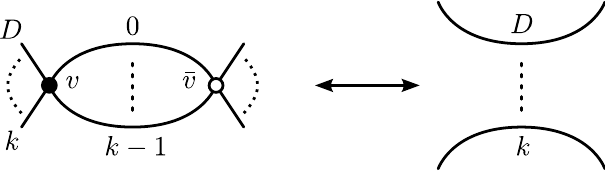}
\caption{The process of $k$-dipole contraction/creation in the case $i_p =  p-1$ for all $p \in \{1,\dots, k\}$.}
\label{fig:1canc}
\end{figure}

\begin{definition}The process of {\it $k$-dipole contraction}:
\begin{itemize}\itemsep=0pt
\item deletes the vertices $v_1$ and $v_2$;
\item deletes the edges $i_1,\dots i_k$;
\item connects the remaining edges {\it respecting coloring}, see Fig.~\ref{fig:1canc}.
\end{itemize}
\end{definition}

The inverse of $k$-dipole contraction is called {\it $k$-dipole creation}\footnote{The contraction/creation of a $D$-dipole is
superf\/luous and can always be traded for the contraction/creation of a $1$-dipole.}. As a point of clarif\/ication, we note that it is certainly {\it not} the $k$-dipole creation results case that selecting arbitrarily $D+1-k$ distinctly colored strands, snipping them and inserting the partial subgraph on the left hand side of Fig.~\ref{fig:1canc}.  It is rather the case that we much carefully choose the strands such that the resulting $(D+1-k)$-bubbles are distinct. We denote $\cG/d_k$ the graph obtained from $\cG$ by contracting~$d_k$.

\begin{definition}\label{def:combeq}
Two graphs as said to be {\it combinatorially equivalent}, denoted $\sim^{(c)}$, if they are related by a sequence of $k$-dipole
 contractions and creations.
 \end{definition}

 In particular, $\cG \sim^{(c)}\cG/d_k$.

We now analyze the ef\/fect a $k$-dipole contraction has on the graph.  Consider a vacuum graph~$\cG$ and denote by $\cV_{\cG}$,
 $\cE_{\cG}$, $\cF_{\cG}$ its  vertex, line
and  face sets respectively, such that $ |\cV_{\cG}|=2p$. Under a $k$-dipole contraction, $\cG \rightarrow \cG/d_k$ and the
 cardinalities of their various sets are related by:
\begin{gather}
|\cV_{\cG/d_k}| = |V_{\cG}|  - 2  , \qquad |\cE_{\cG/d_k}| = |\cE_{G}| - (D+1) ,\nonumber\\
 |\cF_{\cG/d_k}| =\dsty |\cF_{\cG}| - \frac{k(k-1)}{2} - \frac{(D+1-k)(D-k)}{2}  .\label{eq:facesdipole}
\end{gather}
In words, the number of vertices decreases by two and the number of lines by $D+1$ (one per color). For the faces, note that
prior to contraction, they come in four types: the~$k(k-1)/2$ faces  that are comprised exclusively of lines in the dipole;
 the $k(D+1-k)$ faces that contain exactly one line in the dipole; the
$ (D+1-k) (D-k) $ faces that do not contain any line in the dipole, but contain one of the two vertices~$v$ or~$\bar v$; and
the rest, which do not concern us here.  After contraction of the $k$-dipole, the faces made up exclusively by lines in the
 dipole are erased, while the faces containing either~$v$ or~$\bar v$ but no line in the dipole are
merged pairwise. The $k(D+1-k)$ faces containing exactly one line in the dipole will lose one line, but their number is
conserved.

In subsequent sections, 1-dipole contraction will play a signif\/icant role and so let us examine this case in more detail.
 Consider a $1$-dipole $d_1$ of color~$i$. We can see from above formulae that $\cG/d_1$ has two less vertices.  Interestingly,
 $\cG/d_1$ also has one less $D$-bubble than $\cG$ of colors $\widehat{i}$, while the number of $D$-bubbles of colors
 $\widehat{j}, \; j\neq i$ is left unchanged. Thus, the quantity $p - \cB^{[D]}$ is conserved by 1-dipole contraction.
Also, the {\it connectivity} of the graph remains the same. So now let us contract a maximal number of $1$-dipoles in $\cG$ so that it
 reduces to some graph $\cG_f$ with $2p_f \ge 2$ vertices. With a little inspection, one notices that $\cG_f$ possesses
exactly one $D$-bubble for each color $i$, and so, $\cB_{\cG_f}^{[D]}=D+1$.  We have arrived at the important inequality:
\begin{gather}\label{eq:smeche}
p- \cB_\cG^{[D]} = p_f - \cB^{[D]}_{\cG_f} \quad\implies\quad  p_f-1 = p + D - \cB_\cG^{[D]} \ge 0  .
\end{gather}

\subsection{Jackets and degree}

In order to gain a better understanding of colored graphs, one would like to def\/ine some simpler graphs which capture only
some of the information encoded in the colors.  A f\/irst class of subgraphs is already available: the bubbles. The bubbles
are, however,
themselves colored graphs.  Hence although they have less colors, they are still relatively dif\/f\/icult to handle. A~second
class of simpler graphs is given by the jackets~\cite{BOpcont3,Gur3,Gur4,GurRiv}. The main advantage of the jackets is that they are merely ribbon graphs
(i.e.\ they comprise just of  vertices,  edges and faces), like the ones generated by matrix models graphs.
The jackets contain all the vertices and all the lines of~$\cG$ but only some of its faces.
As they are needed in order to def\/ine the `degree', they play a~crucial role in the large~$N$ expansion of colored tensor models.
As we shall see, the jackets are Riemann surfaces embedded in the cellular complex.

\begin{definition}
 A colored {\it jacket} $\cJ$ is a 2-subcomplex of $\cG$, labeled by a $(D+1)$-cycle $\tau$, such that:
 \begin{itemize}\itemsep=0pt
 \item $\cJ$ and $\cG$ have identical vertex sets, $\cV_\cJ = \cV_\cG$;
 \item $\cJ$ and $\cG$ have identical edge sets, $\cE_\cJ = \cE_\cG$;
 \item the face set of $\cJ$ is a subset of the face set of $\cG$: $\cF_{\cJ} = \{f\in \cF_{\cG}\;|\;
f = (\tau^q(0),\tau^{q+1}(0)), q\in \Z_{D+1} \}$.
 \end{itemize}
 \end{definition}

It is evident that $\cJ$ and $\cG$ have the same connectivity. In actual fact, a given jacket is independent of the overall
orientation of the cycle, meaning that the number of jackets is in one-to-two correspondence with $(D+1)$-cycles.  Therefore,
 the number of independent jackets is $D!/2$ and the number of jackets containing a given face is $(D-1)!$.\footnote{It is,
 however, sometimes more transparent to over count the distinct jackets by a factor of two associating them one to one with
cycles. For example, on can count that from the $D!$ cycles of $D+1$ colors, $(D-1)!$ will contain the pair $ij$ and $(D-1)!$
the pair $ji$.}

The jacket has the structure of a {\it ribbon graph}.  Note that each edge of $\cJ$ lies on the boundary of two of its faces,
thus corresponds to a ribbon line in the ribbon graph. As we said, the ribbon lines separate two faces, $(\tau^{-1}(i),i)$
and $(i,\tau(i))$ and inherit the color $i$ of the line in $\cJ$.  Ribbon graphs are well-known to correspond to Riemann
surfaces, and so the same holds for jackets.  Given this, we can def\/ine the {\it Euler characteristic} of the jacket as:
$\chi(\cJ) = |\cF_\cJ| - |\cE_\cJ| + |\cV_\cJ| = 2 - 2g_\cJ$, where $g_{\cJ}$ is the {\it genus} of the
jacket\footnote{A momentary ref\/lection reveals that the jackets necessarily represent orientable surfaces.}.

\begin{figure}[htb]
\centering
 \includegraphics[width=12cm]{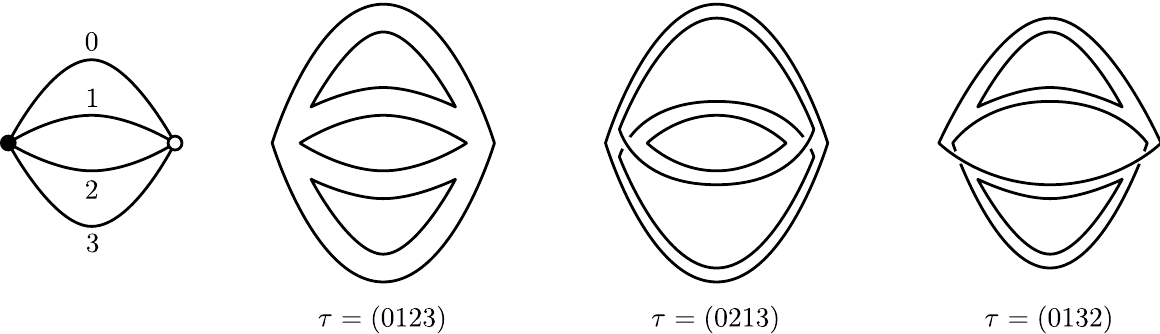}
\caption{A vacuum 4-colored graph and its jackets.}
\label{fig:exemplujacket1}
\end{figure}

In $D=2$, the (unique) jacket of a $(2+1)$-colored graph is the graph itself.  In $D=3$, an example of a graph and its
jackets (and their associated cycles) is given in Fig.~\ref{fig:exemplujacket1}. For instance the leftmost jacket
corresponding to the cycle $\tau = (0123)$ contains only the faces~$01$,~$12$,~$23$ and~$30$.

For a $(D+1)$-colored graph $\cG$, its $D$-bubbles are $D$-colored graphs $\cB_{(\rho)}^{\widehat i}$.  Thus, they also possess
jackets, which we denote by $\cJ_{(\rho)}^{\widehat i}$.  It is rather elementary to construct the $\cJ^{\widehat{i}}_{(\rho)}$
 from the~$\cJ$.   Let us construct the ribbon graph $\cJ^{\widehat i}$ consisting of vertex, edge and face sets:
\begin{gather*}
\cV_{\cJ^{\widehat i}} = \cV_{\cJ},  \qquad \cE_{\cJ^{\widehat i}} = \cE_{\cJ}\setminus \cE^i, \qquad %\textrm{and}\quad
\cF_{\cJ^{\widehat{i}}}
 = \big(\cF_\cJ  \setminus \{(\tau^{-1}(i), i), (i,\tau(i))\}\big) \cup \big\{ (\tau^{-1}(i), \tau(i)) \big\} ,
\end{gather*}
that is having all vertices of~$\cG$, all lines of $\cG$ of colors dif\/ferent from $i$ and some faces.
Given that the face set of $\cJ$ is specif\/ied by a $(D+1)$-cycle $\tau$, the f\/irst thing to notice is that the face set of
 $\cJ^{\widehat i}$ is specif\/ied by a $D$-cycle obtained from $\tau$ by deleting the color $i$.   The ribbon subgraph~$\cJ^{\widehat{i}}$
is the union of several connected components, $\cJ^{\widehat{i}}_{(\rho)}$. Each $\cJ^{\widehat{i}}_{(\rho)}$ is a jacket of
 the $D$-bubble $\cB^{ \widehat{i} }_{ (\rho) }$. Conversely, every jacket of~$\cB^{ \widehat{i} }_{ (\rho) }$ is obtained
from exactly $D$ jackets of~$\cG$.  To realize this, consider a jacket~$\cJ^{\widehat{i}}_{(\rho)}$. It is specif\/ied by a
 $D$-cycle (missing the color~$i$).  On can insert the color $i$ anywhere along the cycle and thus get $D$ independent
 $(D+1)$-cycles.

More generally, the $d$-bubbles are $d$-colored graphs and they also possess jackets which can be obtained from the jackets
of $\cG$.

Consider once again Fig.~\ref{fig:exemplujacket1}.   Applying our procedure to the jacket $(0123)$ leads to the three
 jackets $(123)$, $(023)$ and $(012)$. Each of these jackets corresponds to a bubble of Fig.~\ref{fig:exemplubub} and
is a~\mbox{$2+1$} colored graph.

\begin{definition}We def\/ine:
    \begin{itemize}\itemsep=0pt
    \item The {\it convergence degree} (or simply {\it degree}) of a graph $\cG$ is $\omega(\cG)=\sum_{\cJ} g_{\cJ}$,
    where the sum runs over all the jackets $\cJ$ of $\cG$.
    \item The {\it degree} of a $k$-dipole $d_k$ separating the bubbles $\cB^{\widehat{i}_1 \dots \widehat{i}_k}_{(\alpha)}$
   and $\cB^{\widehat{i}_1 \dots \widehat{i}_k}_{(\beta)}$, denoted $\omega (d_k)$,  is the smallest of the
    two degrees $ \omega(\cB^{\widehat{i}_1 \dots \widehat{i}_k}_{(\alpha)})$ and $
    \omega(\cB^{\widehat{i}_1 \dots \widehat{i}_k}_{(\beta)})  $.
   \end{itemize}
\end{definition}

The degree is a number one can readily compute starting from a graph. As it will play a major role in
tensor models, we list below a number of properties of the degree. For understanding the leading order in the
large $N$ expansion one needs to recall from the reminder of this section that the degree is a positive number,
and equation~\eqref{eq:faces} relating the number of faces of a graphs with its degree.

First, the degree of a graph and of its bubbles are not independent.

\begin{lemma}\label{lem:degjackets}
   The degrees of a graph $\omega(\cG)$ and its $D$-bubbles $\omega(\cB^{\widehat{i}}_{(\rho)})$ respect:
   \[
    \omega(\cG) = \frac{(D-1)!}{2} \big( p+D-\cB^{[D]} \big) + \sum_{i;\rho} \omega\big(\cB^{\widehat{i}}_{(\rho)}\big)   .
   \]
    In particular, by  \eqref{eq:smeche}, $p+D-\cB^{[D]} \ge 0$, thus
     $\omega(\cG)=0 \Rightarrow \omega(\cB^{\widehat{i}}_{(\rho)}) = 0$ $\forall\, i,\rho$ and $ p+D-\cB^{[D]} =0$.
\end{lemma}

\begin{proof}
Consider a jacket $\cJ\subset\cG$.   The number of vertices and lines of $\cJ$ are: $|\cV_{\cJ}| = |\cV_{\cG}|=2p$ and
 $|\cV_{\cJ}| = |\cE_{\cG}|=(D+1)p$, respectively. Hence, the number of faces $\cJ$ is $|\cF_{\cJ}| = (D-1) p + 2-2g_{\cJ}$.
Taking into account that $\cG$ has $\frac{1}{2} D!$ jackets and each face belongs to $(D-1)!$ jackets:
\begin{gather}\label{eq:faces}
 |\cF_{\cG}| = \frac{D (D-1)}{2} p + D - \frac{2}{(D-1)!} \sum_{\cJ} g_{\cJ}
= \frac{D (D-1)}{2} p + D - \frac{2}{(D-1)!} \omega(\cG)   .
\end{gather}
Each of the $D$-bubbles $\cB^{\widehat{i}}_{(\rho)}$ (with $|\cV_{ \cB^{\widehat{i}}_{(\rho)} }| = 2 p^{ \widehat{i}}_{(\rho)}$)
is a $D$-colored graph and  thus an analogous formula to  \eqref{eq:faces} holds for each $D$-bubble:
\begin{gather}\label{eq:faces2}
 |\cF_{ \cB^{\widehat{i}}_{(\rho)} }| = \frac{(D-1)(D-2)}{2} p^{ \widehat{i}}_{(\rho)}
+ (D-1) - \frac{2}{(D-2)!} \omega\big(\cB^{\widehat{i}}_{(\rho)}\big)   .
\end{gather}
Each vertex of $\cG$ contributes to $D+1$ of its $D$-bubbles and each face to $D-1$ of them. Thus,
$\sum_{i;\rho} p^{\widehat{i}}_{(\rho)} = (D+1)p$ and $\sum_{i;\rho} |\cF_{ \cB^{\widehat{i}}_{(\rho)} }| = (D-1)|\cF_\cG|$.
  Summing over $D$-bubbles in  \eqref{eq:faces2} and dividing by $D-1$ yields:
\begin{gather}\label{eq:faces3}
 |\cF_{\cG}| = \frac{(D-2)(D+1)}{2} p + \cB^{[D]} - \frac{2}{(D-1)!}  \sum_{i;\rho}  \omega\big(\cB^{\widehat{i}}_{(\rho)}\big)   .
\end{gather}
One equates \eqref{eq:faces3} to  \eqref{eq:faces} to prove the lemma.
\end{proof}

  Second, due to  \eqref{eq:facesdipole}, the degree of a graph changes under a $k$-dipole contraction.

\begin{lemma}\label{lem:degreecontr}
   The degree of $\cG$ and $\cG/d_k$ are related by:
   \[
    \omega(\cG) = \frac{(D-1)!} {2} \big (  (D+1)k - k^2 - D \big )  + \omega(\cG/d_k)   .
   \]
 \end{lemma}

One can track in more detail the ef\/fect of a $1$-dipole contraction on the degrees of the $D$-bubbles of a graph.
Let us denote by  $d_1$ a $1$-dipole of color $i$ separating the two $D$-bubbles $\cB^{\widehat{i}}_{(\alpha)}$ and
$\cB^{\widehat{i}}_{(\beta)}$. For $j\neq i$ or $\{j=i, \; \rho\neq \alpha,\beta\}$ we denote
$\cB^{\widehat{j}}_{(\rho)}$ (resp. $\cB^{\widehat{j}}_{(\rho)}/d_1$) the $D$-bubbles
before (resp.\ after) contraction of $d_1$. The two $D$-bubbles $\cB^{\widehat{i}}_{(\alpha)}$ and
$\cB^{\widehat{i}}_{(\beta)}$ are merged into a single bubble $\cB^{\widehat{i}}_{(\alpha)\cup(\beta)}/d_1$
after contraction.

\begin{lemma}\label{lem:degbub}
   The degrees of the bubbles before and after contraction of a $1$-dipole respect:
   \begin{gather*}
    \omega\big( \cB^{\widehat{j}}_{(\rho)}/d_1\big) = \omega\big( \cB^{\widehat{j}}_{(\rho)}\big)   ,  \qquad  i\neq j \quad
        \textrm{or}  \quad   \{i=j, \; \rho\neq \alpha,\beta\}, \\
  \omega\big( \cB^{\widehat{i}}_{(\alpha)\cup(\beta)}/d_1\big) = \omega\big(\cB^{\widehat{i}}_{(\alpha)} \big) +
     \omega\big( \cB^{\widehat{i}}_{(\beta)} \big)   .
   \end{gather*}
\end{lemma}

\begin{proof}
Only the  $D$-bubbles containing one (or both) of the end vertices $v$ and $\bar v$ of $d_1$ are af\/fected by the contraction.

For $j\neq i$,  using formulae analogous to~\eqref{eq:facesdipole} for $D$-colored graphs, one sees that any jacket
$\cJ^{\widehat{j}}_{(\rho)}/d_1$ of $\cB^{\widehat{j}}_{(\rho)}/d_1$ has $2$ vertices less, $D$ lines less and $D-2$
faces less than the corresponding jacket $\cJ^{\widehat{j}}_{(\rho)}$ of $\cB^{\widehat{j}}_{(\rho)}$. Therefore,
 $g_{\cJ^{\widehat{j}}_{(\rho)}/d_1}= g_{\cJ^{\widehat{j}}_{(\rho)}}$ and the degree is conserved.

   Any jacket $\cJ^{\widehat{i}}_{(\alpha)\cup(\beta)}/d_1 $ of the $D$-bubble $\cB^{\widehat{i}}_{(\alpha)\cup(\beta)}/d_1$, is
obtained by gluing two jackets $ \cJ^{\widehat{i}}_{(\alpha)}$ and $\cJ^{\widehat{i}}_{(\beta)}$ of
$ \cB^{\widehat{i}}_{(\alpha)}$ and $\cB^{\widehat{i}}_{(\beta)}$. It follows  $\cJ^{\widehat{i}}_{(\alpha)\cup(\beta)}/d_1 $
has $2$ vertices  less, $D$ lines  less, $D$ faces  less and~$1$ connected component  less than the two jackets
$ \cJ^{\widehat{i}}_{(\alpha)}$ and $\cJ^{\widehat{i}}_{(\beta)}$. Hence
$g_{\cJ^{\widehat{i}}_{(\alpha)\cup(\beta)}/d_1 }=g_{\cJ^{\widehat{i}}_{(\alpha)}} +g_{\cJ^{\widehat{i}}_{(\beta)}}$, and the
lemma follows.
\end{proof}

An important consequence of Lemma \ref{lem:degjackets} is that two graphs related by a 1-dipole contraction have
the same degree.

\subsection{Topological equivalence}

We now include the topology in the picture. We shall utilize a fundamental result from combinatorial topology \cite{FG,Lins}:

\begin{theorem} Two pseudo-manifolds dual to $\cG$ and $\cG/d_k$ are homeomorphic if one of the bubbles
$\cB^{\widehat{i}_0\dots \widehat{i}_{k-1}}_{(\alpha)} $ or $\cB^{\widehat{i}_0\dots \widehat{i}_{k-1} }_{ (\beta) }$
 separated by the dipole is dual to a {\it sphere} $S^{D-k}$.
\end{theorem}

 This allows us to propose another equivalence relation on the set of colored graphs.

\begin{definition}
Two graphs, $\cG_i$ and $\cG_f$, are said to be {\it topologically equivalent}, denoted~$\sim^{(t)}$, if they are
related by a sequence of dipole contraction and creation moves satisfying an additional property:  for any dipole
move in the sequence, at least one of the bubbles
separated by the dipole is a sphere.
\end{definition}

It is in principle very dif\/f\/icult to check whether  a graph is a sphere or not. However we can establish the
following partial result, crucial for the large $N$ expansion, including at leading order, of colored tensor models.

\begin{lemma} \label{lem:sph}
 If $\omega(\cG)=0$ then $\cG$ is dual to a sphere $S^D$. The reciprocal holds in $D=2$.
\end{lemma}

\begin{proof} We use induction on $D$. In $D=2$, $\cG$ is a ribbon graph and its degree equals its genus.
 For $D\ge 3$, as $\omega(\cG)=0$, Lemma~\ref{lem:degjackets}  implies that $\omega(\cB^{\widehat{i}}_{(\rho)})=0$ and,
using the inductive hypothesis, all the bubbles $\cB^{\widehat{i}}_{(\rho)} $ are dual to spheres $S^{D-1}$. Any
$1$-dipole $d_1$ will separate a sphere, so that $\cG/d_1 \sim^{(t)} \cG$ and by Lemma~\ref{lem:degreecontr},
 $\omega(\cG/d_1)=\omega(\cG)=0$.  We iteratively contract a full set of $1$-dipoles to reduce $\cG$ to a f\/inal
 graph $\cG_f$, such that $\cG_f \sim^{(t)}\cG$, $\omega(\cG_f)=0$ and $\cG_f$ does not posses any 1-dipoles. It
follows that $\cG_f$ has $\cB^{[D]}_f=D+1$ remaining $D$-bubbles (one for each colors $\widehat{i}$) and, by
Lemma~\ref{lem:degjackets}, $p_f+D-\cB_f^{[D]}=0$. We conclude that $p_f=1$ and $\cG_f$ represents the coherent
identif\/ication of two $D$-simplices along their boundary i.e.\ it is dual to a sphere~$S^D$.
\end{proof}

A last result we shall need is a lower bound for the degree of a graph as a function of the degrees of its
$D$-bubbles with f\/ixed colors $\cB^{\widehat{D}}_{(\rho)}$.

\begin{lemma}\label{lem:ddegg}
Let $\cG$ be a $D+1$ colored graph and $\cB^{ \widehat{D} }_{(\rho)}$ its $D$-bubbles
with colors $\widehat{D}$. Then
\[
 \omega(\cG) \ge D \sum_{\rho} \omega \big(\cB^{\widehat{D} }_{(\rho)} \big)   .
\]
\end{lemma}

\begin{proof}
Consider a jacket $\cJ$ of $\cG$. By eliminating the color $D$ in its associated cycle we obtain
a~cycle over $0,\dots,D-1$ associated to a~jacket $\cJ^{\widehat{D}}_{(\rho)}$ for each of its bubbles.
As ribbon graphs, $ \cJ^{\widehat{D}}_{(\rho)} $ are in one-to-one correspondence with disjoint subgraphs of~$\cJ$.
 One obtains these subgraphs by deleting the lines of color $D$ and joining the strands $(\pi^{-1}(D),D)$
and $(D,\pi(D))$ in mixed faces corresponding to $(\pi^{-1}(D),\pi(D))$
in $ \cJ^{\widehat{D}}_{(\rho)} $ \cite{GurRiv}. Consequently:
\[
 g_{\cJ} \ge \sum_{\rho } g_{\cJ^{\widehat{D}}_{(\rho)}  }   .
\]
As we already mentioned, every jacket $ \cJ^{\widehat{D}}_{(\rho)} $ is obtained as subgraph of exactly $D$ distinct \mbox{jackets~$\cJ$} (corresponding to inserting the color $D$ anywhere in the cycle associated to
$ \cJ^{\widehat{D}}_{(\rho)} $). Summing over all jackets of $\cG$ we obtain:
\begin{gather*}
 \sum_{J} g_{\cJ} \ge D \sum_{\rho } \sum_{ \cJ^{\widehat{D} }_{ (\rho) } } g_{\cJ^{\widehat{D}}_{(\rho)}  }   .\tag*{\qed}
\end{gather*}
  \renewcommand{\qed}{}
\end{proof}

\subsection{Graph factorization}\label{ssec:factor}

Earlier, we introduced jackets, which are instrumental in the def\/inition of the convergence degree.  Later, we shall see that the amplitudes of many colored tensor
models capture information about these jackets rather than higher-dimensional subspaces.
Let us recall that the tensor models generate Feynman graphs using identical building blocks, namely the interaction vertices.  Interestingly, it emerges
that there is a seed for the jackets directly within this building block and thus in the tensor model action itself.  Identifying this basic structure will
be our purpose in this section; we shall utilize it later to provide an interesting perspective on the colored tensor models in question.

Our f\/irst port of call is the interaction vertex.  We usually picture this as a single vertex with $D+1$ distinctly colored half-edges emanating from it.
Obviously, this is a viable graph in its own right for the interior of a manifold with boundary.  It is more convenient here to treat it as such and to
formulate our analysis in terms of its boundary graph.  Using the rules set out in Section~\ref{ssec:boundary}, we note that the boundary graph
has $D+1$ vertices, each
of which is connected to all the others by a single boundary edge.  In other words, the boundary graph is the boundary graph of a $D$-simplex.  On top
of that, each of the boundary vertices possesses a distinct color (inherited from the corresponding colored half edge of the interior), while each boundary edge is labeled by
two colors (inherited from the corresponding interior face).

The introduction of this boundary graph allows us to perform a more concise investigation.  Its power lies in the fact that any result found for the
boundary graph can be easily translated back to the interior since, in this case, the vertex and edge sets of the boundary are in one-to-one correspondence
with the edge and face sets of the interior, respectively.

To keep the following manipulations as  succinct as possible, we shall also need some elementary def\/initions from graph theory.

\begin{definition} Some graph theory def\/initions:
\begin{itemize}\itemsep=0pt
\item A {\it complete graph} is one in which every pair of distinct vertices is connected by a unique edge. Moreover, we shall be interested in complete graphs with distinctly colored vertices. We shall denote such a colored complete graph with $n$ vertices by~$\cK_{n}$.
\item A {\it $k$-factor} $\cI_k$ is a k-regular spanning subgraph of $\cG$.  In other words, its vertex set coincides with
 that of $\cG$ and all vertices are $k$-valent.
\item A {\it $k$-factorization} of $\cG$ partitions the edge set of $\cG$ into disjoint $k$-factors\footnote{For small values of $k$, these $k$-factors often come under dif\/ferent names.
\begin{itemize}\itemsep=0pt
\item A {\it matching} in a graph is a set of edges without common vertices.  A {\it perfect matching} is one which contains
all vertices of the graph. Thus, a perfect matching and a 1-factor are equivalent concepts.
\item A {\it cycle} is an alternating sequence of points and edges, $\{v_0, e_{01}, v_1, \dots, v_{n}, e_{n0}\}$, such that
 $e_{ii+1} = v_iv_{i+1}$ and all vertices are distinct. We denote it by $\tau = (v_0,v_1,\dots, v_n)$, leaving the edges
implicit.  A {\it Hamiltonian cycle} is a cycle which contains every vertex of $\cG$ and thus it is a 2-factor.
\item An {\it edge decomposition} of a graph is a partition of its edges into subgraphs.  A {\it Hamiltonian decomposition}
is an edge decomposition consisting of Hamiltonian cycles and so it is a 2-factorization.\end{itemize}}.

\item The {\it graph subtraction} operation is based on the set-theoretic one.  In particular, we shall be interested in
$\cG-\cI_1$, the graph $\cG$ with the edges of a 1-factor removed from its edge set.
\end{itemize}
\end{definition}

The boundary of a colored $D$-simplex is the colored complete graph $\cK_{D+1}$.  Furthermore, a~2-factor corresponds to a~$(D+1)$-cycle in the
 vertex set of the boundary that is a cycle in the set of colors.  Thus, on construction of the Feynman graph this 2-factor determines a jacket.
  In a moment, we shall also need information about the face subset of a generic tensor model graph~$\cG$ that is determined by a 1-factor of~$\cK_{D+1}$.
For this, we shall introduce a new object in~$\cG$, called a {\it patch}, corresponding to the image of this 1-factor.

\begin{definition}
A colored {\it patch} $\cP$ is a 2-subcomplex of $\cG$, for odd $D$, labeled by a 1-factor~$\cI_1$ in~$\cK_{D+1}$, such that:
 \begin{itemize}\itemsep=0pt
 \item $\cP$ and $\cG$ have identical vertex sets, $\cV_\cP = \cV_\cG$;
 \item $\cP$ and $\cG$ have identical edge sets, $\cE_\cP = \cE_\cG$;
 \item the face set of $\cP$ is a subset of the face set of $\cG$: $\cF_{\cP} = \{f\in \cF_{\cG}\;|\; f = (ij)\in\cI_1 \}$.
 \end{itemize}
\end{definition}

Unlike the jackets, a patch does not correspond to a Riemann surface embedded in the cellular complex.  On the contrary, any two faces in a patch are either
disjoint or intersect in a f\/inite number of (zero-dimensional) points.

 Before going any further, we recall a rather well-known result that subsequently plays a~sig\-nif\/icant role. It relates to a permissible factorization
of $\cK_{D+1}$ and as such, it is the simplest and most generic.

\begin{proposition}\label{prop:wal}
Say $n\in \N$. There exists a $2$-factorization of $\cK_{2n+1}$ and $\cK_{2n+2} - \cI_1$.
\end{proposition}

\begin{proof} We employ an explicit construction due to Walecki \cite{walecki} in order to prove this proposition.   For
$\cK_{2n+1}$, label the vertex set by $\Z_{2n+1}$ and construct the 2-factor:
\[
(2n,0,1,2n-1,2,2n-2,3,2n-3,\dots,n-1,n+1,n).
\]
Next act on this cycle with the permutation $\rho = (0,1,2,\dots, 2n-1)(2n)$.  This generates another 2-factor disjoint
 from the f\/irst.  Repeating this action another $n-2$ times generates $n$ disjoint 2-factors and thus a 2-factorization.
Perhaps an illustration for $n=3$ shall serve to clarify matters, see Fig.~\ref{walodd}:
  We form a $2n$-gon  with the vertex labeled by $2n$ at the center.  We can think of the action of $\rho$ as rotating
the labels (passive) or equivalently rotating the cycle (active).
\begin{figure}[htb]
\centering
\includegraphics[scale = 0.8]{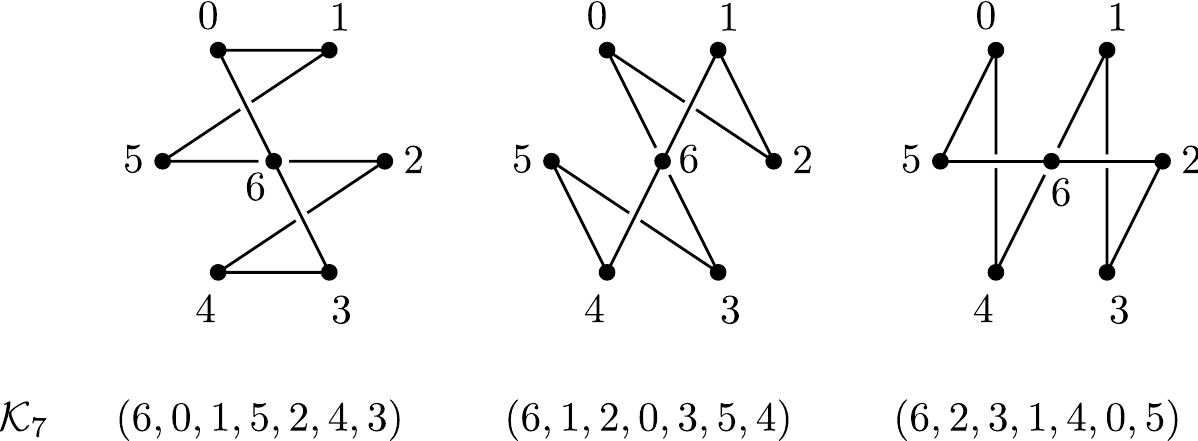}
\caption{\label{walodd} 2-factorization of $\cK_7$.}
\end{figure}

For  $\cK_{2n+2}$, we label the vertex set by $\Z_{2n+2}$ and construct the 2-factor:
\begin{gather*}
\left(2n+1, 0,1,2n-1,2,2n-2,3,2n-3, \dots, \dfrac{n}{2}, 2n, \dfrac{3n}{2}, \dots  , n-1,n+1, n\right)
\end{gather*}
 for $n$ even,
\begin{gather*}
\left(2n+1, 0,1,2n-1,2,2n-2,3,2n-3, \dots, \dfrac{3n+1}{2}, 2n, \dfrac{n+1}{2}, \dots  , n-1,n+1, n\right)
\end{gather*}
 for $n$ odd.
Once again by applying $\rho = (0,1,2,\dots, 2n-1)(2n)(2n+1)$, one generates $n$ disjoint 2-factors.  But in this case, it
does not exhaust the edge set of $\cK_{2n+2}$.  There is a 1-factor left over:
\begin{gather*}
\cI_1 = \{(2n,2n+1)\}\cup \{(x, x+n) : x \in \Z_{n}\}.\tag*{\qed}
\end{gather*}  \renewcommand{\qed}{}
\end{proof}

Note that in our case, the vertices of $\cK_{D+1}$ are distinguished by a color.  Therefore,  the 2-factors of this labeled $K_{D+1}$ are in
 one-to-one correspondence with the jackets of a~$(D+1)$-colored graph $\cG$, that is, there are $\frac{D!}{2}$ of them.   For odd $D$, we shall
also need to take into account the patches.  For a labeled $\cK_{D+1}$, there are $D!!$ 1-factors and thus the same number of patches.

Let us use the factorization result above to formulate an idea of face-set factorization for a~graph~$\cG$.
\begin{definition}
A {\it PJ-factorization} $\cPJ_\cG$ of a $(D+1)$-colored graph $\cG$ is:
 \begin{itemize}\itemsep=0pt
\item for even $D$, a subset of its jackets corresponding to a 2-factorization of $\cK_{D+1}$;
\item for odd $D$, a patch
corresponding to a 1-factor $\cI_1$, along with the subset of its jackets corresponding to a 2-factorization of $\cK_{D+1} - \cI_1$.
\end{itemize}
\end{definition}
A PJ-factorization has the useful properties that all its elements have disjoint face sets, but yet it contains all
the faces of $\cG$.  Note that for even $D$, it contains $\frac{D}2$ jackets, while for odd $D$, it contains $\frac{D-1}{2}$
jackets and a patch.

Another result follows for a labeled $D$-simplex $\cK_{D+1}$:
\begin{lemma}\label{lem:pjfactor}
Consider the complete graph $\cK_{D+1}$ with distinguished vertices.  One may place a~partition on the set of $2$-factors of $\cK_{D+1}$ such that:
\begin{itemize}\itemsep=0pt
\item For even $D$, one may partition the set of all $2$-factors into $(D-1)!$ $2$-factorizations.  Moreover, there are $D+1$ such partitions.
\item For odd $D$, one may partition the set of all $2$-factors into $D [(D-2)!]$ $2$-factorizations.  Once again there are $D+1$ such partitions.
\end{itemize}
\end{lemma}

\begin{proof}
For even $D$,  consider the 2-factorization constructed in Proposition~\ref{prop:wal}.  We work from the passive stance, where the cycle through the
vertices is f\/ixed and the labels are permuted. In Proposition~\ref{prop:wal}, we started from an initial cycle and considered the action of cyclic subgroup of the permutation group $S_{D+1}$ generated by the element $(0,1,2,\dots, D-1)(D)$.  This subgroup f\/ixes the label $D$ and it has $D$
elements (although due to a symmetry in the initial 2-factor, the action of this subgroup generates just $D/2$ distinct 2-factors).
There are $(D+1)[(D-1)!]$ left cosets with respect to this subgroup, each of which corresponds to a 2-factorization of $\cK_{D+1}$.
Of these left cosets,  $(D-1)!$ f\/ix a given label, say $D$,  and all contain distinct 2-factors.  Thus, this is a partition on the set of 2-factors
and there are $D+1$ such partitions.

For odd $D$, the situation is bit more involved but one follows essentially the same argument.  Here, there are two vertex labels f\/ixed and a cyclic subgroup
of $S_{D+1}$ is used to generate the rest of the 2-factors.   This subgroup has $D-1$ elements.  Then, the $(D+1)D[(D-2)!]$ left cosets are such
 $D[(D-2)!]$ of them f\/ix a given label, say $D$, and contain distinct 2-factors.  So, once again one has a partition on the set of 2-factors
and there $D+1$ such partitions.

Note that for odd $D$, the left-over 1-factor is not the same for each 2-factorization, but it is uniquely determined once one has all
the 2-factors in the factorization.
\end{proof}

Translating this to a graph $\cG$, we get analogous statements for its set of jackets and patches.
\begin{proposition}
Consider a $(D+1)$-colored graph $\cG$.  One may place a partition on the set of jackets of $\cG$ such that:
\begin{itemize}\itemsep=0pt
\item For even $D$, one may partition the set of all jackets into $(D-1)!$ PJ-factorizations.  Moreover, there are $D+1$ such partitions.
\item For odd $D$, one may partition the set of all jackets into $D [(D-2)!]$ PJ-factorizations.  Once again there are $D+1$ such partitions.
\end{itemize}
\end{proposition}

What is of more interest to us here is a def\/inition of the Euler characteristic of a PJ-factorization.  For even $D$, the factorization contains only
 jackets, which are Riemann surfaces and thus there is an natural extant def\/inition.  For odd $D$, one must extend this to deal with patches.
We shall now def\/ine the Euler characteristic of a PJ-factorization:
\begin{gather}\label{eq:factorchar}
\chi(\cPJ_\cG) =
\begin{cases}
\displaystyle  \sum_{\cJ\in{\cPJ_{\cG}}} \chi(\cJ) & \textrm{for even}\; D, \quad {\rm where} \quad \chi(\cJ)
 =  |\cV_\cP| -  |\cE_\cP| + |\cF_\cP|, \\
\displaystyle \sum_{\cJ\in{\cPJ_{\cG}}} \chi(\cJ) + \chi(\cP)& \textrm{for odd}\; D,  \quad {\rm where} \quad \chi(\cP)
 = \frac{1}{2} |\cV_\cP| - \frac{1}{2} |\cE_\cP| + |\cF_\cP| .
\end{cases}\!\!\!\!\!
\end{gather}

Note that $\chi(\cPJ_\cG)$ is an integer and $\chi(\cPJ_\cG) = D - \frac{2}{(D-1)!} w(\cG)$.  This puts some constraint on the values of $w(\cG)$:
\[
w(\cG) = \frac{(D-1)!}{2} (\textrm{an integer}) .
\]
Additionally, we notice that to obtain the degree of a graph we need only topological information attached to a subset of the jackets (and perhaps a~patch),
that is, those in a~single PJ-factorization.   This def\/inition reveals a nice view of Lemma \ref{lem:degjackets} relating the degrees of a~graph and its $D$-bubbles:
\begin{gather}\label{eq:char}
\chi(\cPJ_\cG) = - p + \frac{1}{D-1} \sum_{i;\rho}\chi\big(\cPJ_{\cB^{\hat{i}}_{(\rho)}}\big)   ,
\end{gather}
which stresses the nested structure existing among the various bubbles of a graph. As a word of caution, however, we mention that a single
 PJ-factorization of a graph does not automatically provide a PJ-factorization for all its $D$-bubbles. In passing from a graph to one of
its $D$-bubbles, one deletes a given color.  This color gets deleted from the cycles determining the jackets in the PJ-factorization.  These cycles are
 no longer a factorization of the $\cK_{D}$ generated from $\cK_{D+1}$ by removing the vertex and edges containing the chosen color.

Let us focus brief\/ly on the case when $D = 3$, where we can readily utilize this machinery to uncover some further results about the topology of the 4-colored graphs.

\begin{proposition}\label{prop:spht}
If $D=3$ and $\cG$ possesses a spherical jacket then $\cG$ is spherical.
\end{proposition}

Note that one can give several alternative proofs of this result.

\begin{proof}
If $w(\cG) = 0$, we may apply Lemma \ref{lem:sph} and we are done.  Therefore, we consider the case where $w(\cG) \neq 0$.

Since $\cG$ possesses a spherical jacket, there is an ordering of the colors around the vertex that provides a planar representation of the graph.
 Deleting a single color to obtain the $3$-bubbles of $\cG$, preserves the planarity of this representation. Thus, each 3-bubble possesses a spherical jacket.
 Since 3-bubbles are Riemann surfaces, they are spheres  and $w(\cB^{\hat i}_{(\rho)}) = 0$.

Since all 1-dipoles separate spheres, we may iteratively contract a full set of 1-dipoles to get a graph $\cG_1\sim^{(t)}\cG$ such that $\cG_1$ has just four
3-bubbles, one for each color.  Now, pick the PJ-factorization of $\cG_1$ that contains the spherical jacket.  Then, $\chi(\cPJ_{\cG_1}) = 2+ \chi(\cP)$ and
along with the formula \eqref{eq:char}, this gives us:
\begin{gather}\label{eq:patchface}
2 + |\cF_\cP|  - p  = 2 +  \chi(\cP) = -p + \frac{1}{2} \sum_{i = 0}^3 \chi(\cPJ_{\cR^{\hat i}_{(1)}}) = -p + 4 \quad\implies \quad |\cF_\cP| = 2 .
\end{gather}
Let the spherical jacket be labeled by the cycle $(abcd)$.  Then, equation~\eqref{eq:patchface} states that after all 1-dipoles are contracted, we are left
 with a graph having one face of type~$(ac)$ and one of type~$(bd)$.  As an illustration of the planar jacket helps at this point, see Fig.~\ref{fig:patchthree}.

\begin{figure}[h]
\centering
\includegraphics[scale = 1.2]{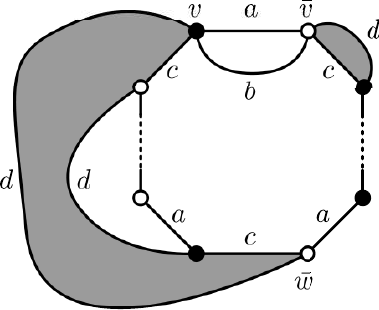}
\caption{A possible graph $\cG_1$.}\label{fig:patchthree}
\end{figure}

We have drawn explicitly the face of type $(ac)$ and left most of the lines of color~$b$ and~$d$ implicit. We shall now show that we can iteratively contract
a set of 2-dipoles until we arrive at the spherical graph with two vertices, denoted~$\cG_{f}$.

Planarity of the jacket implies that of the faces of types $(ab)$ or $(bc)$, there are at least two which have only two edges.  We have drawn one of these,
of type $(ab)$,  at the top of the f\/igure.  Showing the lines $d$ emanating from the vertices $v$ and $\bar{v}$, one can see that the face $(ab)$ separates
two distinct faces of type $(cd)$ and so it is a 2-dipole.  The only catch would be if $\bar{w} = \bar{v}$.  In that case, the graph must be $\cG_f$, else it would
 have more than one face of type $(bd)$.  We contract this 2-dipole and arrive at~$\cG_f$ or at a graph with the same properties as~$\cG_1$. Therefore, it contains
 a 2-dipole and we may iterate this procedure until we arrive at~$\cG_f$. To conclude, $\cG$ is spherical.
\end{proof}

We present here another result which relates the jackets to important objects in 3-dimensional topology:
\begin{definition}
A {\it Heegaard splitting} of a compact connected oriented 3-manifold $\cM$ is an ordered triple $(\Sigma, \cH_1, \cH_2)_\cM$ consisting of a
 compact connected oriented surface $\Sigma$ and two handlebodies $\cH_1$, $\cH_2$ such that $\partial\cH_1 = \partial\cH_2 = \Sigma$. $\Sigma$
is known as the {\it Heegaard surface} of the splitting.
\end{definition}

\begin{proposition}
If $\cG$ is a manifold, then its jackets $\cJ$ are Heegaard surfaces.
\end{proposition}

In fact, this result (see~\cite{Ryan:2011qm}) provides an alternative route to proving Proposition~\ref{prop:spht}.  If $\cG$ possesses a spherical
 jacket, then all its 3-bubbles are spheres.  This implies that it is a manifold (rather than a pseudo-manifold) and its jackets are Heegaard surfaces.
 A well-known result in 3-dimensional topology is that a 3-manifold possessing a spherical Heegaard surface must be a~sphere~\cite{hatcher-book}.

\section{Tensor models}\label{sec:tensomodels}

   We have discussed so far the topological and cellular structure of colored graphs. These graphs arise due to the coloring of the f\/ields, and thus the results
hold for $(D+1)$-colored tensor models irrespective of their precise details, that is, the f\/ields, their arguments, the vertex kernels, the covariances.  From
now on, we shall start making specif\/ic choices for
 the action of our tensor model.  The amplitudes of graphs and the physical interpretation of our results depend strongly on the particularities of the model.

Once again, we present a digest of forthcoming subsections.  First and foremost, this section as a whole details the~$1/N$ expansion(s) appropriate for colored tensor models.

{\it Introducing colored tensor models:}
We introduce the independent identically distributed (i.i.d.) probability measure for colored tensor f\/ields
upon which we perform essentially all our subsequent analysis. The reader should keep in mind that similar
results hold for many other choices of measure. We will present in the conclusion a more detailed discussion on the
degree of generality of our results.

{\it Boundary graphs as observables:}
We mention as a quick note how the boundary graphs of Section~\ref{ssec:boundary} arise as moments of this probability measure.

{\it Amplitude:}
We write down the amplitude associated to each graph in the i.i.d.~model.  We highlight the importance of choosing the correct scaling
(with respect the to parameter $N$) for the coupling constants of the model. We note that the amplitude of a graph depends solely on its degree and moreover,
in such a way that the leading order graphs are necessarily $D$-spheres.

{\it Combinatorial $1/N$ expansion:}
In this subsection, we introduce the f\/irst of two $1/N$ expansions.   The idea is that for a given number of vertices, one can identify a f\/inite
set of graphs, combinatorial core graphs, which are `simplest' with respect to a combinatorial criterion.
Every core graph indexes an inf\/inite class of graphs, all having the same amplitude, thus represents a~term in the~$1/N$ expansion of the
free energy of our model. Core graphs at higher order (i.e.\ with more vertices) are suppressed in powers of~$1/N$.

{\it Topological $1/N$ expansion:}
Here, we repeat the process of the previous subsection with the restriction that we are only allowed to use topology-preserving moves.
This increases the number of core graphs in the topological case and leads to topological core equivalence classes according to
which the expansion is ordered.

{\it Redundancies in the $1/N$ expansion:}
Despite the many useful properties of this topological expansion, it is not as powerful as in the 2-dimensional case, mainly because the i.i.d.~amplitude
is not sensitive to many aspects of higher-dimensional topology.  Since it merely depends on the degree of the graph, it only captures
combinatorial properties.  We outline specif\/ic superf\/luities of the topological expansion in the i.i.d.~scenario.

{\it First terms of the $1/N$ expansion:}
We present some explicit examples of core graphs at low orders.

\subsection{Introducing colored tensor models}

   We will always chose the domain of def\/inition of our random functions to be (several copies of) a compact Lie group $G$.
Generally, however, it is more convenient to work with Fourier transformed f\/ields, that is, f\/ields dependent on the (discrete)
 representation space of the Lie group in question.  We will denote by $N$ a (large) cutof\/f in the representations, the range
of the discrete indices. Thus, generically our colored f\/ield is a colored random tensor of some rank $r$,   $\phi^i_{\vec n_i}$
 with $\vec n_i=\{ n^{a_1},\dots, n^{a_r} \}$. The crucial feature of the colored models is that one can assign weights to the
Feynman graphs according to their cellular structure. One might wish, perhaps capriciously, to place a weight $N$ just on
the $d$-bubbles of colors $i_1, \dots, i_d$.   This can be achieved simply by assigning
an index $ n^{i_1 \dots i_d}_{i_k} $ (resp. $\bar n^{i_1 \dots i_d}_{i_k}$) to the
f\/ield $\phi^{i_k}$ (resp. $\bar \phi^{i_k}$) for $k\in \{1,\dots, d\}$ and utilizing the following covariances and vertex kernel:
\begin{gather*}
C^{i}  =  1 \!\!\qquad \textrm{for} \quad i\notin\{i_1,\dots, i_d\},\qquad
 C^{i}_{n^{i_1 \dots i_d}_{i}, \bar n^{i_1 \dots i_d}_{i}  }  =
 \delta_{ n^{i_1 \dots i_d}_{i} , \bar n^{i_1 \dots i_d}_{i}  } \!\!\qquad \textrm{for} \quad i \in \{i_1,\dots, i_d\}
\end{gather*}
 and
 \[
 K_{\{n_{i_k}\}} = \sum_{n=1}^N \prod_{k=1}^d \delta_{n,n^{i_1 \dots i_d}_{i_k}  }.
 \]
More generally, one can weight several cells (either of the same dimension or not) at the same time, use dif\/ferent cutof\/fs
for the various indices and so on.

We shall concentrate in the sequel on a particular model which assigns equal weights $N$ to {\it all} the faces (two cells)
 of our graph. It is the straightforward generalization of a matrix model to higher dimensions. Since a line belongs to $D$
faces, our tensors are rank $D$ tensors and we denote them  $\phi^i_{\vec n_i}$, where
 $\vec n_i =  (n^{ii-1},\dots, n^{i0},\, n^{iD}, \dots, n^{ii+1}) $, with
the index $n^{ij}$ associated to the face with colors $i$ and $j$. The tensors have {\it no symmetry} properties in their indices,
 that is, $ \phi^i_{\vec n_i}$,~$\bar \phi^i_{ \bar {\vec n}_i}$~are independent random variables for each choice of their indices.
 Furthermore, we chose a~trivial covariance and vertex kernel:
\[
  C^i_{\vec n_i,  \bar {\vec  n}_i } = \prod_{j\neq i} \delta_{n^{ij},\bar n^{ij}}   ,
\qquad K_{ \vec n_0\dots \vec n_D } = \frac{1}{N^{ D(D-1)/4 }}
 \prod_{i<j} \delta_{n^{ij},n^{ji}}   .
\]
This is the {\it independent identically distributed~$($i.i.d.$)$~colored tensor model} in $D$ dimensions~\cite{Gur3,Gur4,GurRiv}. The probability measure
we shall deal with in the sequel is therefore:
\begin{gather}
 d\nu  =  \prod_{i, \vec n_i} \frac{d\phi^i_{\vec n_i} d\bar\phi^i_{ \vec{\bar n}_i}} {2\pi}
  e^{-S}   ,\nonumber\\
 S (\phi,\bar\phi)  =  \sum_{i=0}^{D} \sum_{\vec n_i, \vec {\bar n}_i  } \bar \phi^i_{\vec {\bar n}_i}
  \delta_{ \vec {\bar n}_i , \vec n_i } \phi^i_{\vec n_i}  +
\frac{\lambda}{ N^{D(D-1)/4} } \sum_{\{\vec n\}} \prod_{i=0}^D \phi^i_{ \vec n_i } +
\frac{\bar \lambda}{ N^{D(D-1)/4} } \sum_{\{\vec {\bar n} \}}
\prod_{i=0}^D \bar \phi^i_{ \vec {\bar n}_i }   ,\label{eq:iid}
\end{gather}
where $\sum_{\vec n}$ denotes the sum over all indices $n_{ij}$ from $1$ to $N$.
Note that rescaling $\phi^i_{\vec n_i} = N^{-D/4} T^i_{\vec n_i}$ leads to:
\[
 S (\bar T ,T  ) = N^{D/2} \left(
\sum_{i=0}^{D} \sum_{\vec n} \bar T^i_{\vec n_i} T^i_{\vec n_i}  +
\lambda \sum_{\{\vec n\}} \prod_{i=0}^D T^i_{ \vec n_i } +
\bar \lambda \sum_{\{\vec {\bar n}\}} \prod_{i=0}^D \bar T^i_{ \vec{\bar n}_i } \right)  .
\]

The scaling of the coupling constant with respect to the large parameter $N$ in~\eqref{eq:iid} will be explained in Section~\ref{sec:iidampli}. We shall discuss at length the physical interpretation of this model in Section~\ref{sec:dtcontlim}.
The notions required for the developing the leading order in the large $N$ expansion are the bubbles, the degree, and
the relation between the number of faces of a graph and its degree in equation~\eqref{eq:faces}.

\subsection{Boundary graphs as observables}

Tensor models with face weights possess a very convenient property: their observables are indexed by the boundary graphs.
 More specif\/ically, the observables are the connected correlations:
\[
 \Big \langle  \phi^{i_1}_{\vec n_{i_1} } \cdots \phi^{i_q}_{\vec n_{i_q} }
\bar \phi^{j_1}_{\bar {\vec n}_{j_1} } \cdots \bar \phi^{j_q}_{\bar {\vec n}_{j_q} }
 \Big \rangle_c.
\]
As the indices are conserved along the faces, an observable is non-zero if and only if every index $n^{i_1k}$ is paired
either with an index $\bar n^{i_1k}$ or with an index $n^{ki_1}$. Representing each f\/ield as a (colored) boundary vertex,
and each such identif\/ication
as a (bi-colored) boundary line represents the observable as a boundary graph. Conversely, starting from a boundary graph
one builds an observable by considering a product of f\/ields, one for each boundary vertex, with indices identif\/ied according
to the boundary lines.

\subsection{Amplitude}\label{sec:iidampli}

Consider an arbitrary connected $D+1$-colored graph $\cG$ with amplitude specif\/ied by the i.i.d.\ model.  Notably,
the $\delta_{n, \bar n }$ functions associated to the boundary of the faces compose with each other until one is left
with a exactly one redundant summation per face.  Thus, given the scaling of the coupling constant in \eqref{eq:iid},
the amplitude for a given $\cG$ is:
\[ %\label{eq:ampliiid}
A(\cG) = (\lambda \bar\lambda)^{p} \; N^{|\cF_\cG| - p \frac{D(D-1)}{2}} =
(\lambda \bar\lambda)^{p} \; N^{D - \frac{2}{(D-1)! } \omega(\cG) } ,
\]
where we used \eqref{eq:faces} to arrive at the second equality.    For the i.i.d.~model, we conclude that the convergence degree
$\omega(\cG)$ indexes the behavior of the amplitude with respect to the large parameter~$N$.\footnote{Hence, the degree
acts in higher dimensions as the genus did in matrix models.}  In particular, this justif\/ies the scaling of the coupling
 constant in~\eqref{eq:iid}. Indeed, as
we shall see in the sequel, there exists an inf\/inite family of graphs of degree~$0$. If one does not use the appropriate
 rescaling of the coupling constant, either this family of graphs is increasingly divergent at increasingly higher orders
(hence the theory makes no sense), or it, along with all other graphs, is increasingly suppressed at increasingly higher
 orders (and in this case the theory is trivial, having
at most a f\/inite number of divergent graphs. The only scaling which leads to a sensible non-trivial theory is the one chosen
 in~\eqref{eq:iid}, for which one has inf\/inite families of uniformly divergent graphs.

The perturbative series pertaining to free energy and correlation functions of the i.i.d.~model are an expansion in the degree:
\begin{gather*}
  F(\lambda,\bar\lambda) = \ln(Z) = \sum_{\omega=0}^{\infty} C^{[\omega]}(\lambda ,\bar\lambda)
   N^{D - \frac{2}{(D-1)!}\omega},\qquad
  C^{[\omega]}(\lambda ,\bar\lambda) =
\sum_{\cG :  \omega(\cG)  =  \omega } \frac{ 1 } {s(\cG)} (\lambda \bar\lambda)^{|\cV_{\cG}|/2}    ,
\end{gather*}
with $s(\cG)$ \looseness=-1 some symmetry factor. The coef\/f\/icient $C^{[\omega]}(\lambda, \bar\lambda)$ is the sum over the set of graphs
 of degree $\omega$. The fundamental dif\/ference between the two-dimensional case and the general case is that, whereas
the genus of a surface is a topological invariant, the degree {\it is not} in higher dimensions. This leads to two
distinct expansions,
 the {\it combinatorial~$1/N$ expansion} and the {\it topological~$1/N$ expansion}, both of which we shall present in detail
 in a moment.
Crucially, the leading order in both expansions is given by graphs of degree~$0$, which we know, from Lemma~\ref{lem:sph},
 are spheres.

The reminder of this section is dedicated to establishing the full series indexed by the degree. For the purpose
of the critical behavior in the large~$N$ limit however this study is not immediately important, as in the large~$N$ limit all but the graphs of degree zero are suppressed.

\subsection{Combinatorial $1/N$ expansion}

The aim is to give an alternative characterization of $C^{[\omega]}(\lambda, \bar\lambda)$ that can be used
to index the series. From Lemma~\ref{lem:degreecontr}, the degree of a graph is {\it invariant}
under arbitrary $1$-dipole contractions, that is, combinatorially equivalent graphs (see Def\/inition~\ref{def:combeq}) have
 the same degree.  For any graph $\cG$ we reduce a maximal number of $1$-dipoles to obtain a simpler graph with the same
amplitude as~$\cG$~\cite{Gur4}.

 {\it Combinatorial bubble routing:}
\begin{itemize}\itemsep=0pt
\item {\it Designate a root.}  For a given color $i$, we pick one of the $D$-bubbles $\cB^{\widehat{i}}_{(\rho)} $
as a root $\cR^{\widehat{i}}_{(1)}$ bubble. The total number of roots of a graph is $\cR^{[D]} = D+1$.

\item {\it Identify $\widehat i$-connectivity graph.}   We associate to the bubbles $\widehat{i}$ of $\cG$ an
{\it $\widehat{i}$-connectivity graph}. Its vertices represent the various bubbles $\cB^{\widehat{i}}_{(\rho)} $. Its
 lines are the lines of color $i$ in $\cG$.  They either start and end on the same bubble $ \cB^{\widehat{i}}_{(\alpha)} $,
 known as {\it tadpole lines} in the connectivity graph, or they do not. A particularly simple way to picture the
$\widehat{i}$-connectivity graph is to draw $\cG$ with the lines $j\neq i $ much shorter than the lines $i$.

\item {\it Choose a tree.}
We choose a rooted tree $\cT^{i}$ in the $\widehat{i}$-connectivity graph, such that its root is $\cR^{\widehat{i}}_{(1)}$. We refer
to the rest of the lines of color $i$ as {\it loop lines}.

\item {\it Contract.} All the $\cB^{[\widehat{i}]} - 1 $ lines of $\cT^i$ are 1-dipoles and we contract them.
We end up with a connectivity graph with only one vertex corresponding to the root bubble $\cR^{\widehat{i}}_{(1)}$.
The remaining lines of color $i$ cannot be contracted further (they are tadpole lines in the connectivity graph). The number
 of the $D$-bubbles of the other colors is unchanged
under these contractions.

\item {\it Repeat.}  We iterate the previous three points for all colors starting with $D$. The routing tree $\cT^{j}$ is
 chosen in the graph
obtained {\it after} contracting $\cT^{j+1}, \dots, \cT^D $.  The number of bubbles of colors $q>j$ are constant under
 contractions of $1$-dipoles of color $j$, hence the latter {\it cannot} create new 1-dipoles of color $q$.  Reducing
a full set of 1-dipoles indexed by $D+1$ routing trees $\cT^0, \dots, \cT^{D} $  we obtain a graph in which all bubbles
are roots. This is called a  {\it combinatorial core graph}.

\end{itemize}

\begin{definition}\label{def:corecomb} A {\it combinatorial core graph at order $p$}, denoted $\cG^{(c)}_p$, is a
$(D+1)$-colored graph with $2p$ vertices, such that for all colors $i$, it has a unique $D$-bubble $\cR^{\widehat{i}}_{(1)}$.
\end{definition}

In fact, note that we already used the combinatorial core graphs three times: f\/irst to prove~\eqref{eq:smeche}, second
to prove Lemma~\ref{lem:sph} and third to prove Proposition~\ref{prop:spht}. The important feature of combinatorial core graphs is that
 their degree (and consequently amplitude)
admits a bound with respect to the number of vertices $2p$.  By Lemma~\ref{lem:degjackets}:
    \[
    \omega(\cG^{(c)}_p) = \frac{(D-1)!}{2}  \big(  p+D-\cR^{[D]} \big )  + \sum_{i;\rho} \omega\big(\cR^{\widehat{i}}_{(\rho)}\big)
    \ge \frac{(D-1)!}{2} (p-1)  .
   \]

 {\it Combinatorial core equivalence classes:} The combinatorial core graph one obtains by the above routing
 procedure is {\it not} independent of the routing trees. The same graph leads to several combinatorially equivalent
 core graphs, all at the same
order $p$ and all possessing the same amplitude.  We denote such equivalence by $\cG^{(c)}_p \simeq^{(c)} \cG^{(c)}_p{}'$
and call two such core graphs {\it combinatorially core equivalent}.

An arbitrary graph $\cG$ routes to a unique combinatorial core equivalence class, $[\cG^{(c)}_p]$.  There exists a relation
 between their respective amplitudes:
\[ %\label{eq:famcomb}
 A(\cG)= (\lambda \bar\lambda)^{ \cB^{[D] } - D-1 } A \big(\cG^{(c)}_{p}\big)   ,
  \qquad p= \frac{|\cV_{\cG}|}{2} - \bigl(\cB^{ [D]  }- D-1  \bigr)   .
\]
Thus, the $1/N$ expansion of the free energy of a $(D+1)$-colored tensor model can be recast in combinatorial core
equivalence classes:
\[ %\label{eq:free}
 F(\lambda, \bar\lambda) = \sum_{p=1}^{\infty}
\sum_{  \omega\ge \frac{(D-1)!}{2} ( p-1 ) } \left[   \sum_{[\cG^{(c)}_p]   :  \omega(\cG^{(c)}_p) = \omega}
 C^{[\cG^{(c)}_p]}(\lambda,\bar\lambda) \right]  N^{D - \frac{2}{ (D-1)! } \omega }   ,
\]
where  $C^{[\cG^{(c)}_p]}( \lambda,\bar\lambda )$ counts all the graphs
routing (via a combinatorial bubble routing) to the equivalence class~$[\cG^{(c)}_p]$. The crucial feature of the
expansion in combinatorial core classes is that it can be evaluated order by order.
One needs to draw all core graphs at order~$p$, compute their amplitudes, and divide them into combinatorial
equivalence classes. The last step, identifying the combinatorial equivalence classes, is potentially
dif\/f\/icult, but one needs to deal with this problem only a {\it finite} number of times  to
write all terms up to a given order.

\subsection{Topological $1/N$ expansion}\label{sec:topo1/N}

\looseness=1
To write a series taking into account not only the amplitude but also the topology of the graphs one must rely on
topological equivalence rather than combinatorial equivalence. This means that, instead of contracting all $1$-dipoles,
 one should contract only $1$-dipoles separating spheres~\cite{Gur4}. Of
course it is out of the question to reduce all such $1$-dipoles for an arbitrary graph (one would need to deal with
a very dif\/f\/icult problem for every graph~$\cG$). We will modestly only reduce the $1$-dipoles of degree~$0$. Not
only such contractions preserve the topology (from Lemma~\ref{lem:sph}), but also they
lead to a well def\/ined~$1/N$ expansion. Unsurprisingly the f\/inal series will have some degree of redundancy
which we will detail later.

 {\it Topological bubble routing:}
\begin{itemize}\itemsep=0pt
\item {\it Designate a root.}  Consider a color $i$.  If there exist $D$-bubbles of non-zero degree, that
 is $\omega(\cB^{\widehat{i}}_{(\rho)}) \ge 1 $,
we set all of them as roots $\cR^{\widehat{i}}_{(\mu)} $. Otherwise, we choose one of them as the sole root
 $\cR^{\widehat{i}}_{(1)}$.
No matter what the scenario, we call $\cR^{\widehat{i}}_{(1)} $ the principal root and $\cR^{\widehat{i}}_{(\mu)}, \; \mu >1 $
{\it branch roots}.   Iterating for all colors, we identify all the roots of $\cG$.  We denote by $\cR^{[\widehat{i}]}$ the
 number of roots of color $\widehat{i}$ and  $\cR^{[D]} = \sum_i \cR^{[\widehat{i}]}$.

\item {\it Identify $\widehat{i}$-connectivity graph.} As in the combinatorial case.

\item {\it Choose a tree.}  We choose a rooted tree $\cT^{i}$ in the $\widehat{i}$-connectivity graph. Consider any branch
root $\cR^{\widehat{i}}_{(\mu)}$, $\mu>1$.  It is represented by a vertex in $\cT^{(i)}$.  We represent by a dashed
line, the line incident to this vertex and belonging to the unique path in $\cT^{i}$ connecting~$\cR^{\widehat{i}}_{(\mu)} $
to the principal root
$\cR^{\widehat{i}}_{(1)} $.
All the other lines in $\cT^{i}$ are represented as solid lines. An example is given in
Fig.~\ref{fig:treebub}.
\begin{figure}[htb]
\centering
 \includegraphics[scale = 2]{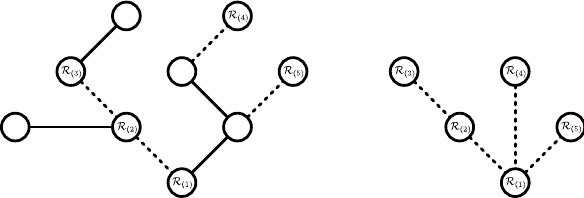}
\caption{A tree $\cT^3$ in the $012$ connectivity graph.}
\label{fig:treebub}
\end{figure}

\item {\it Contract.} All the $\cB^{[\widehat{i}]} - \cR^{[\widehat{i}]} $ solid lines in $\cT^i$ are 1-dipoles of
degree $0$ and we contract them.  We arrive at an $\widehat i$-connectivity graph with vertices corresponding to the roots
$\cR^{\widehat{i}}_{(\mu)}$. The remaining lines of color $i$ cannot be contracted further.   They are either tadpole
 lines or dashed lines.  The tadpole lines are not 1-dipoles and the dashed lines now separate two roots and so are
 not 1-dipoles of degree 0.
Neither the number nor the topology of the bubbles of the other colors is changed under these contractions.

\item {\it Repeat.} We iterate for all colors starting with $D$.  The routing tree $\cT^{j}$ is chosen in the graph
obtained {\it after} contracting $\cT^{j+1}, \dots, \cT^D $.  The degrees $\omega(\cR^{\widehat{q}}_{(\rho)})$, $q>j$
are unaf\/fected by the contractions of $1$-dipoles of color $j$ (by Lemma~\ref{lem:degbub}), hence the latter {\it cannot}
create new 1-dipoles of degree 0 and color~$q$. Reducing a full set of 1-dipoles of degree $0$ indexed by $D+1$ routing
 trees $\cT^0, \dots, \cT^{D}$, we obtain a graph in which  all bubbles are roots.  This is called a {\it topological core graph}.
\end{itemize}

\begin{definition}\label{def:coretop}  A {\it topological core graph at order $p$}, denoted $\cG^{(t)}_p$, is a
$(D+1)$-colored graph with $2p$ vertices such that for each color $i$:
   \begin{itemize}\itemsep=0pt
    \item either $\cG$ has a unique $D$-bubble $\cR^{\widehat{i}}_{(1)}$ of degree $\omega(\cR^{\widehat{i}}_{(1)})=0$;
    \item or all bubbles $\cR^{\widehat{i}}_{(\rho)}$, have degree $\omega(\cR^{\widehat{i}}_{(\rho)}) \ge 1$.
   \end{itemize}
\end{definition}

Just as for combinatorial core graphs, the degree and amplitude of topological core graphs admit a bound
in the number of vertices.  By Lemma \ref{lem:degjackets}:
\[
    \omega\big(\cG^{(t)}_p\big) = \frac{(D-1)!}{2} \big (  p+D-\cR^{[D]} \big )  + \sum_{i;\rho} \omega\big(\cR^{\widehat{i}}_{(\rho)}\big)   ,
\]
The def\/inition of a topological core graph implies:
$ \sum_{i;\rho} \omega(\cR^{\widehat{i}}_{(\rho)}) \ge \cR^{[D]}- (D+1)$, hence:
\[ %\label{eq:degtopol}
   \omega\big(\cG^{(t)}_p\big) \ge  \left( \frac{(D-1)!}{2} -1 \right )  \big (  p+D-\cR^{[D]} \big )  +p-1 \ge p-1   ,
\]
   where for the last inequality we used again~\eqref{eq:smeche}.

 {\it Topological core equivalence classes:} The core graph one obtains by the above routing procedure is
again not independent of the routing trees.  The same graph leads to several topologically equivalent core graphs,
all at the same
order $p$ and all possessing the same amplitude. We denote such an equivalence by $\cG^{(t)}_p \simeq^{(t)} \cG^{(t)}_p{}'$
and call two such graphs {\it topologically core equivalent}.

An arbitrary graph $\cG$ routes to a unique topological core equivalence class, $[\cG^{(t)}_p]$.  There exists a relation
 between their respective amplitudes:
\[ %\label{eq:famtop}
 A (\cG) = (\lambda \bar\lambda)^{ \cB^{[D] } - \cR^{[D]} } A \big(\cG^{(t)}_{p} \big)  ,
  \qquad p= \frac{\cN_{\cG}}{2} - \bigl(\cB^{[D]  }- \cR^{[D]  } \bigr)   .
\]
Thus, the $1/N$ expansion of the free energy of a $(D+1)$-colored tensor model can be recast in topological core equivalence
classes:
\begin{gather}\label{eq:topoiid}
 F(\lambda, \bar\lambda) = \sum_{p=1}^{\infty}
\sum_{  \omega \ge p-1  } \left[  \sum_{[\cG^{(t)}_p] , \; \omega(\cG^{(t)}_p)=\omega}
 C^{[\cG^{(t)}_p]}(\lambda,\bar\lambda) \right]  \; N^{D - \frac{2}{ (D-1)! } \omega } \; ,
\end{gather}
where $C^{[\cG^{(t)}_p]}( \lambda,\bar\lambda )$ counts all the graphs
routing (via a topological bubble routing) to $[\cG^{(t)}_p]$. The classes $[\cG^{(t)}_p]$ can again be listed order by
order.  As before, identifying the topological equivalence classes at order $p$ is dif\/f\/icult, but one needs to deal
with this problem only a f\/inite number of times.

\subsection{Redundancies in the $1/N$ series}

    The topological expansion \eqref{eq:topoiid} groups together only graphs with both the same topology and amplitude.
 However, the scaling with $N$ only separates graphs of dif\/ferent amplitudes,  as captured by the combinatorial expansion.
  Classifying further the graphs with the same amplitude into {\it topological} core equivalence classes is a choice. While
this leads to a well def\/ined series, which one can compute order-by-order, the topological expansion has a certain degree
of redundancy:
\begin{itemize}\itemsep=0pt
\item  At a given order $p$, there exist graphs $ \cG^{(t)}_p$, $\cG^{(t)}_p{}'$ with the same amplitude but which are
not topologically core equivalent since they are not topologically equivalent.  In formulae:
$A( \cG^{(t)}_p) = A( \cG^{(t)}_p{}')$ but $\cG^{(t)}_p \not\simeq \cG^{(t)}_p{}'$ since
 $ \cG^{(t)}_p \nsim \cG^{(t)}_p{}'$. Although the scaling with $N$ does not distinguish two such graphs, we obviously
group them in dif\/ferent terms in~\eqref{eq:topoiid}.

\item  At a given order $p$, there exist graphs $ \cG^{(t)}_p$, $\cG^{(t)}_p{}'$ which are topologically equivalent
but not topologically core equivalent since they do not have the same amplitude. In formulae:
 $\cG^{(t)}_p \sim \cG^{(t)}_p{}'$ but $\cG^{(t)}_p \not\simeq \cG^{(t)}_p{}'$ since
$A(\cG^{(t)}_p)\neq A(\cG^{(t)}_p{}')$.  In other words, the same topology can appear in several distinct
 topological core equivalence classes at a given order.

\item  For two distinct orders $p<q $ there exist graphs $ \cG^{(t)}_p$, $\cG^{(t)}_q$ which are
both topologically equivalent and have the same amplitude but are not topologically core equivalent since they are at
dif\/ferent orders.  In formulae:  $\cG^{(t)}_p \sim \cG^{(t)}_q$ and $A(\cG^{(t)}_p)=A(\cG^{(t)}_q)$
but $\cG^{(t)}_p \not\simeq \cG^{(t)}_q$ since $p\neq q$.
 For instance, if $\cG^{(t)}_q$ can be obtained from $\cG^{(t)}_p$ by creating $1$-dipoles of non-zero degree which
 separate spheres\footnote{Note that, as the degree of the $D$-bubbles is additive under  $1$-dipole creations,
$q\le p+ \omega(\cG_p)$.}.

\item  Any topology appears at arbitrary order. For all topological core graphs $\cG^{(t)}_p$ one obtains
       topologically equivalent core graphs (but diminished in power counting) by creating an arbitrary number of
       $k$-dipoles of degree 0 for $k\ge2$, which separate spheres.
\end{itemize}

As topology in higher dimensions is a very involved subject, it is to be expected that the f\/inal~$1/N$ expansion is
somewhat dif\/f\/icult to handle. Counting all graphs which route to a~given core graph is relatively straightforward
(by performing all possible creations of $1$-dipoles of degree~$0$). However, one must pay attention to double
 counting problems when grouping together the graphs routing to a
topological core equivalence class. Fortunately, at low orders, the topological core equivalence classes have a
 single representative, and the over counting problem is absent. In fact, later in this review, we shall explicitly
sum the leading order graphs and analyze the critical behavior of the leading order contribution.

To conclude this topic, let us consider for a brief moment the 3-dimensional case.  A bubble
 $\cB^{\widehat{i}}_{(\rho)}$ of a $(3+1)$-colored graph~$\cG$ represents a sphere if and only if
 $\omega( \cB^{\widehat{i}}_{(\rho)})=0$.   Hence, a $1$-dipole~$d_1$ separates a sphere if and only if~ $\omega(d_1)=0$. This eliminates most of the redundancies of the~$1/N$ expansion.

\subsection{First terms of the $1/N$ expansion}\label{sec:firstterms}

\begin{figure}[htb]
\centering
 \includegraphics{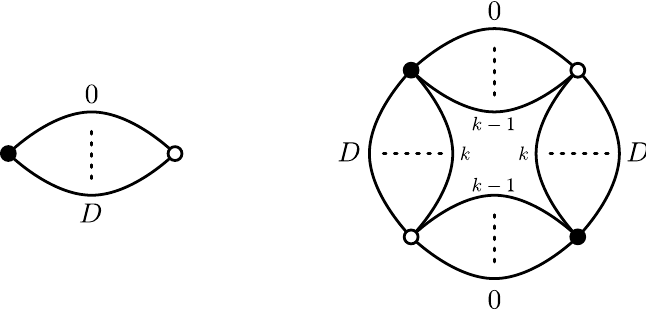}
\caption{Core graphs at $p=1$ and $p=2$.}\label{fig:t1t2}
\end{figure}

To give us a feel for the expansions describes above, let us calculate the f\/irst few terms.  Note that {\it all} combinatorial
core graphs are also topological core graphs. The topological core graphs up to $p=2$ are represented in Fig.~\ref{fig:t1t2}.
 In these cases, since they only have one $D$-bubble~$\cB^{\widehat{i}}$, for each color~$i$, they are also
combinatorial core graphs.

At $p=1$, we have a unique topological/combinatorial core graph, denoted $\cG^{(c)}_1=\cG^{(t)}_1$. It is the
unique topological/combinatorial core graph of degree $0$ and thus it represents a sphere. For reasons which will
 become clear below we call this graph the super-melon. As the degree of a graph is invariant under arbitrary
 $1$-dipole contractions, the
combinatorial class~$[\cG^{(c)}_1]$ is identical with the topological class~$[\cG^{(t)}_1]$.

At $p=2$, we have the topological/combinatorial core graphs $\cG^{(c)}_{2;k}=\cG^{(t)}_{2;k}$, with
 $2\le k \le \lfloor\frac{D+1}{2}\rfloor$. This time the combinatorial and topological classes
$[\cG^{(c)}_{2;k}]$, $[\cG^{(t)}_{2;k}]$ are dif\/ferent. All these core graphs are dual to spheres $S^D$
(thus, $\cG^{(t)}_1\sim^{(t)} \cG^{(t)}_{2;k}$).

 The amplitudes of the various graphs are:
\[
 A(\cG_1) = N^{D}, \qquad A (\cG_{2;k}) = N^{D-(D+1)k+k^2+D}   .
\]
The two $1/N$ expansions of the free energy are:
\begin{gather}
 F(\lambda,\bar\lambda)  =  C^{[\cG^{(c)}_1]}(\lambda,\bar\lambda)   N^{D} +
   \sum_{k=2}^{ \lfloor\frac{D+1}{2}\rfloor } C^{[\cG^{(c)}_{2;k}]}(\lambda,\bar\lambda)    N^{D-(D+1)k+k^2+D}
   +O \bigl( N^{D - 2 } \bigr)  ,  \nonumber\\
 F(\lambda,\bar\lambda)  =  C^{[\cG^{(t)}_1]}(\lambda,\bar\lambda)   N^{D} +
   \sum_{k=2}^{ \lfloor\frac{D+1}{2}\rfloor } C^{[\cG^{(t)}_{2;k}]}(\lambda,\bar\lambda)    N^{D-(D+1)k+k^2+D}
  +O \bigl( N^{D - \frac{2}{(D-1)!} 2 } \bigr)  .\label{eq:final}
\end{gather}
Note that the combinatorial expansion has a better a priori bound on the remainder terms than the topological
 expansion. This should come as no surprise, as the topological classes $[\cG^{(t)}_{2;k}]$ contain fewer graphs
 than the combinatorial classes $[\cG^{(c)}_{2;k}]$.
At f\/irst sight, the bound on the remainder terms is not very tight. The f\/irst explicit correction with respect
to the dominant behavior in~\eqref{eq:final} comes from $\cG_{2,2}$. To ensure that all terms with amplitude
 larger or equal than $\cG_{2,2}$ have
been taken into account,  one needs to check combinatorial core graphs up to $p_{\max,(c)}=D-1$ in the combinatorial
expansion and topological core graphs up to $p_{\max, (t)}=1 +\frac{(D-1)!}{2} (D-2)$ in the topological expansion.

\section{Embedded matrix models}\label{sec:embed}

We return here to the graph factorizations we developed in Section~\ref{sec:topology}.  Our earlier disposition on graph factorization
 was intended to highlight the relationship between factors of labeled complete graphs $\cK_{D+1}$ and subsets of $(D+1)$-colored graphs $\cG$.
 We saw that 1- and 2-factors translated into patches and jackets, respectively.     In particular, a factorization of $\cK_{D+1}$ in terms of
 1- and 2-factors determined a PJ-factorization of $\cG$.

 The colored version of $\cK_{D+1}$ makes a reappearance when we examine the tensor model~\eqref{eq:iid}.  Irrespective of the index structure,
each interaction term clearly generates a vertex with D+1
 distinctly colored incident lines.    The boundary of such a graph component is a colored $\cK_{D+1}$.  Thus, at the level of the action,
 we have the same picture as earlier:  a single $(D+1)$-valent vertex and its boundary graph.

 The colored tensors are associated to the edges of the interior and thus the vertices of the boundary graph.  Our choice of index structure
rests on which substructures we would like to weight in our tensor model.  In the i.i.d.~action, each tensor has $D$ indices and a given tensor
 is contracted with each of the other $D$ tensors using one of these indices.  Given that the tensors are associated to the vertices of the
boundary graph, these contractions weight the edges in the boundary graph.  As a check, from the counting above,  we have $D(D+1)/2$ pairs
of contracted indices, which coincides with the number of edges in~$\cK_{D+1}$.

 {\it Matrix models and jackets:}
A 2-factor of $\cK_{D+1}$ is a $(D+1)$-cycle in the vertex set of $\cK_{D+1}$ and thus picks out $D+1$ edges.  Translating this into tensor
language, this cycle picks out $D+1$ pairs of contracted indices in the interaction term and each tensor contains one element of two pairs.
 Neglecting all other indices for the moment, this has the character of a $(D+1)$-colored matrix model.  To illustrate this, let us
examine the 2-factor $(0,2,1,3,4,\dots, D)$.  It picks out the edges $(02)$, $(21)$, $(13)$, $(34)$, and so on up to $(D0)$.
  Then, the interaction term in the action takes the form:
\begin{gather*}
 S_{\rm int}(\phi,\bar\phi)  =
\frac{\lambda}{ N^{D(D-1)/4} } \sum_{n}  \phi^0_{ n_{D0}n_{02} } \phi^1_{ n_{21}n_{13} }  \phi^2_{ n_{02}n_{21} } \phi^3_{ n_{13}n_{34} }
  \prod_{i=4}^D \phi^i_{ n_{D-1D}n_{DD+1} }  + \textrm{c.c.}\\
\phantom{ S_{\rm int}(\phi,\bar\phi) }{}  =  \frac{\lambda}{ N^{D(D-1)/4} } \tr_{\cJ} (\phi^0\phi^1\phi^2\phi^3\phi^4\cdots\phi^D) + \textrm{c.c.}  ,
\end{gather*}
 where we left all other indices implicit and introduced some new notation $\tr_{\cJ}$ meaning the trace over the indices associated to the edges picked out by the 2-factor.  As we have written it, this subset of the tensor model is a matrix model in its own right.  As a result, it generates
a Riemann surface none other than the jacket~$\cJ$ determined by the 2-factor.  This is a general rule: jackets are weighted by matrix models
 embedded inside the tensor model.

{\it Vector models and patches:}
 For odd $D$, a 1-factor is a pairwise association of vertices of~$\cK_{D+1}$ and thus picks out $(D+1)/2$ edges.  Translating this into tensor
 language, the 1-factor couples the $D+1$ tensors into $(D+1)/2$ pairs.  Neglecting all other indices for the moment, this has the character of
a set of coupled vector models.  Consider the 1-factor specifying the edges $\{(i\,{i+ (D+1)/2}) : i \in \Z_{(D+1)/2}\}$.
 Then, the interaction term takes the form:
 \begin{gather*}
 S_{\rm int}(\phi,\bar\phi)  =
\frac{\lambda}{ N^{D(D-1)/4} } \sum_{n}   \phi^0_{ n_{0 \frac{D+1}2} } \cdots  \phi^{\frac{D-1}2}_{ n_{ \frac{D-1}2} D}
 \phi^{\frac{D+1}{2}}_{ n_{0\frac{D+1}{2}} } \cdots  \phi^{D}_{n_{\frac{D-1}{2} D}}  + \textrm{c.c.}\\
\phantom{S_{\rm int}(\phi,\bar\phi)}{}
=   \frac{\lambda}{ N^{D(D-1)/4} }    \bigl(\phi^0\cdot \phi^{\frac{D+1}{2}} \bigr) \cdots   \bigl( \phi^{\frac{D-1}{2}}\cdot \phi^{D} \bigr)+ \textrm{c.c.}\\
\phantom{S_{\rm int}(\phi,\bar\phi)}{}
 =  \frac{\lambda}{ N^{D(D-1)/4} } \tr_{\cP} \big(\phi^0\phi^1\phi^2\cdots\phi^D\big) + \textrm{c.c.}  ,
\end{gather*}
 where in the second line we see its manifestation as coupled vectors and in the third line, we have introduced another piece of notation
 $\tr_\cP$ meaning the trace over the indices associated to the edges picked out by the 1-factor.  These vectors are, once again, embedded
inside the tensor structure.  This vector structure does not generate a Riemann surface this time, but it still generates the faces of the
patch~$\cP$ determined by the 1-factor in question.  This is another general rule: patches are weighted by vector models embedded inside the tensor model.

 We now possess enough tools to translate the concept of PJ-factorization into the tensor model setting.  As we have stated, a PJ-factorization
 is determined by a factorization of edge set of~$\cK_{D+1}$ into 2-factors (and, for odd~$D$, a single 1-factor).  In our tensor model, contracted
pairs of indices correspond to edges in~$\cK_{D+1}$, so that we can factorize them into 1- and 2-factors.  Thus, for even $D$, we f\/ind that indices
of the interaction term are exhausted by~$D/2$ 2-factors, while for odd~$D$, they are exhausted by~$(D-1)/2$ 2-factors and a 1-factor.

This factorization allows us to rewrite the action as:
\begin{gather}\label{eq:pjfactorize}
S (\phi,\bar\phi) = \sum_{i=0}^{D} \tr( \bar \phi^i\phi^i)  +
\frac{\lambda}{ N^{D(D-1)/4} } \tr_{\cPJ}\left(\prod_{i=0}^D\phi^i\right) +
\frac{\bar \lambda}{ N^{D(D-1)/4} }\tr_{\cPJ}\left(\prod_{i=0}^D\bar\phi^i\right)    ,
\end{gather}
where $\tr_{\cPJ}$ denotes the matrix trace over the indices associated to the disjoint 2-factors and the scalar
 product over the indices associated to the possible 1-factor.  Once a particular PJ-factorization is chosen,
 this partition of degrees of freedom among the jackets is preserved by amplitudes. Remember that:
\[
A(\cG)  =  (\lambda\bar\lambda)^p \, N^{D- \frac{2}{(D-1)!}w(\cG)} =  (\lambda\bar\lambda)^p\, N^{\chi(\cPJ_\cG)} ,
\]
using equation~\eqref{eq:factorchar}. This amplitude is a remarkably simple generalization of the 2-dimensional case.  We note that each jacket in
the factorization contributes a factor~$N^{\chi(\cJ)}$, exactly what one would expect from the matrix model associated to Riemann surfaces.
For odd dimensions, there is also a patch which contributes a factor~$\chi(\cP)$.  While such coupled vector models are not so common in the literature,
the amplitude~$\chi(\cP)$ is appropriate to such a model.

\begin{remark}
In Section~\ref{sec:tensomodels}, we have developed a $1/N$-expansion such that the sphere dominates in the large $N$ limit (due to the fact
 that it is the only topology that contributes at $w(\cG) = 0$).  In other words, all the jackets have zero genus.  But this reasoning was formulated
treating all jackets in a rather egalitarian manner.  Now that we have chosen a particular factorization, we see that the dominant conf\/igurations
correspond to having only spherical jackets and in certain cases, a patch such that $\chi(\cP)$ is maximal.  In using a particular factorization,
we have chosen to highlight certain surfaces.  One could imagine making this distinction more pronounced, for example, by weighting the dif\/ferent
surfaces by dif\/ferent scaling parameters.  In such a~scenario, the tensors $\phi^i$ are no longer maps with the domain $(\Z_N)^D$, but rather with
the domain $\times_{i}\, \Z_{N_{\cJ_i}} \times \Z_{N_\cP}$, where $\cJ_i$  and $\cP$ are respectively the jackets and the (possible) patch in a
PJ-factorization.  We could then scale each of the jackets independently.  In this case, no longer do the conf\/igurations with $w(\cG) = 0$
 necessarily dominate.  We could for example make just $N_\cJ$ for just one jacket large, keeping the other relatively small.  In this case,
all conf\/igurations with one spherical jacket would be equally dominant.  In Proposition~\ref{prop:spht}, we showed that all these conf\/igurations
were still spheres, at least in $D=3$.  Thus, these generalizations are of interest as they are likely to exhibit dif\/ferent critical behavior to
the simplest i.i.d.~model (see Section~\ref{sec:critiid}).
\end{remark}

\begin{remark}
The identif\/ication of embedded matrix and vector models~\cite{Ryan:2011qm, jprtoappear} provides a~f\/irst glimpse of an currently active research avenue
in that it hints at how
 one might adapt the plethora of powerful multi-matrix model techniques to the higher-dimensional setting~\cite{Di_Francesco:1993nw}. Although it is
unlikely that these
are utilizable in the most generic cases, there is tantalizing evidence that they can be used to solve models exactly in specif\/ic regimes~\cite{jprtoappear}.
\end{remark}

\section{Critical behavior}\label{sec:critiid}

We are f\/inally ready to examine the critical behavior of tensor models. We shall see in this section that the colored tensor models  undergo a phase
transition to a continuum theory of large spaces. The topology of the space thus obtained is
spherical\footnote{Its precise geometry, however, is still to be understood.}.
The leading order in the~$1/N$ expansion consists of a subclass of colored triangulations of the $D$-sphere.
We will now address the combinatorics of the dominant triangulations.
As usual, we include an abridged version of the forthcoming developments here.

{\it The dominant order:}
 Our f\/irst task is to identify the combinatorics of the leading order graphs (which are all spherical).  It emerges that all graphs have a
 characteristic `melonic' structure. The  leading order graph with the lowest number of vertices is the `super-melon' presented in
Fig.~\ref{fig:supermelon}. All other leading order graphs are obtained by repeatedly inserting $D$-bubbles with two vertices (i.e.\ two vertices
connected by $D$ lines) arbitrarily on lines.   As we explain in a short prologue to this subsection,  it is possible and more convenient to deal
explicitly with 2-point graphs and translate to vacuum graphs at the end.

{\it From melons to trees: }
The melonic graphs are in one-to-one correspondence with the so-called colored rooted $(D+1)$-ary trees.  We present
their def\/inition and a detailed map between the two combinatorial species.

{\it Resumming the dominant series:}
Having recast the dominant order as a series indexed by a species of trees (and possessing a f\/inite radius of convergence), we present no
 less than three ways to perform the arithmetic:  a direct method; a method using the $(D+1)$-ary tree measure from Section~\ref{sec:Gaussmeasure}; and a method using Cayley's theorem.

{\it Critical exponent:}
We extract the critical value of the coupling constant and critical exponent associated to the free energy (the
`string' susceptibility exponent~$\gamma$).

 {\it Melons as branched polymers:}
We outline the interpretation of the $(D+1)$-colored i.i.d.~tensor model as the generator of dynamical triangulations.   This allows us to visualize the problem
from a more geometrical and hopefully physical standpoint. Translating our dominant melonic graphs into this formalism, we f\/ind that they
display a remarkable resemblance to the sector of dynamical triangulations known as branched polymers.  They both have large positive
 curvature and moreover their entropy exponents match. However, the spectral and Hausdorf\/f dimensions of melonic graphs are still to be determined before
concluding on the precise relation with branched polymers.

{\it Large volume limit:}
We conclude this section with a description of the large volume limit. This is obtained by sending the length (a parameter introduced in the
map to the dynamical triangulation picture) of the individual simplices to zero and the coupling constant of our tensor model
to its critical value.  We highlight the subtlety inherent between taking this limit and the continuum limit.  For the latter we must also keep the renormalized cosmological
constant f\/ixed.  We f\/inish with some interesting current research directions.

To get to the melon rolling, so to speak, the free energy and the 2-point function of the i.i.d.\ model are related by a Schwinger--Dyson (SD) equation. (This is in fact true for a generic tensor model.) It turns out
 that in order to count the leading order graphs, it is easier to count the 2-point graphs and recover the free energy using this equation.  Due to the conservation of the indices along the strands, the 2-point function is necessarily
connected and has the index structure:
\begin{gather}\label{eq:2point}
\langle \bar{\phi}^i_{ \bar {\vec n}_i}\,\phi^i_{ {\vec n}_i}\rangle_{\rm c} =
  \delta_{ \bar {\vec n}_i, {\vec n}_i} \,G(\lambda,\bar{\lambda})   ,
\end{gather}
where $\delta_{ \bar { \vec n}_i, {\vec n}_i} $ denotes $ \prod_{k\neq i} \delta_{ \bar n_{ik} n_{ik} }$.
\begin{proposition} \label{relation GF}
The full connected $2$-point function is:
\[
G(\lambda,\bar{\lambda}) = 1 + \lambda\,\partial_\lambda E(\lambda,\bar{\lambda})   ,
\]
where $E = N^{-D} F$ is the free energy per degree of freedom of the model.
\end{proposition}

The proof of this, by standard techniques can be found for instance in~\cite{Bonzom:2011zz}.
In words, this SD equation tells us that vacuum graphs can be obtained from 2-point graphs
by reconnecting the two external edges (and conversely, cutting any edge in a
vacuum graph generates a 2-point graph).

\subsection{The dominant order: the world of melons} \label{sec:melons}

\begin{figure}[htb]
\centering
 \includegraphics[scale = 1.2]{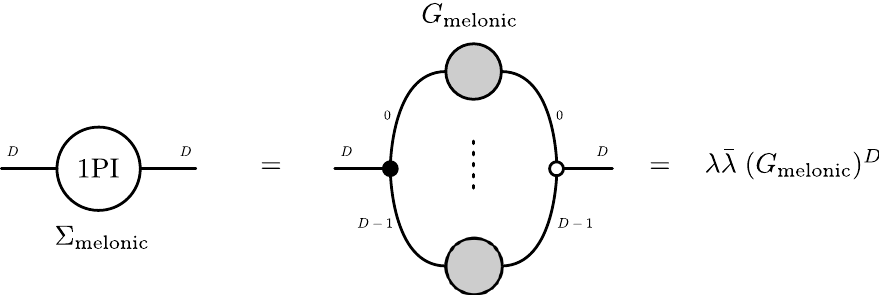}
\caption{The 1PI 2-point function and the full connected 2-point function for melonic graphs.}
\label{fig:melonsig}
\end{figure}

We present below the precise construction of the leading order melonic graphs. As the construction is somewhat convoluted
it is useful to f\/irst state the end result. Call $\Sigma$ the self energy of the model (that is the 1 particle irreducible
amputated two point function). At leading (melonic) order $\Sigma_{\rm melonic}$ factors into the convolution of $D$ connected two point
functions $G_{\rm melonic}$, one for each color. This is represented in Fig.~\ref{fig:melonsig}. We will now establish this result.

\begin{figure}[htb]
\centering
\subfigure[The {\it super-melon} graph.]{\quad\qquad
 \includegraphics[scale=1]{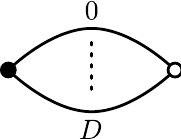}\qquad\quad
  \label{fig:supermelon}
     }
 \qquad\qquad
 \subfigure[A $D$-bubble with two vertices.]{\quad\qquad
 \includegraphics[scale = 1]{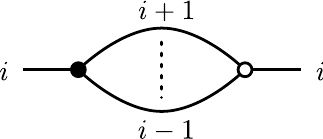}\qquad\quad
  \label{fig:redmelon}
     } \caption{The super-melon and $D$-bubble elimination.}
\end{figure}

 Our construction relies on eliminating $D$-bubbles with two vertices. A $D$-bubble with two vertices
 $\cB^{\widehat{i}}_{(\rho)}$ is a
$D$-dipole\footnote{This is the case unless it is a bubble of the {\it super-melon} graph depicted in the Fig.~\ref{fig:supermelon}.}. Thus, by Lemma~\ref{lem:degbub}, the degree and topology of a graph $\cG$ are identical to those
of the graph $\cG/\cB^{\widehat{i}}_{(\rho)}$ obtained by replacing $\cB^{\widehat{i}}_{(\rho)} $
with a line of color $i$ (see Fig.~\ref{fig:redmelon})

For $D\ge 3$, it can be shown \cite{Bonzom:2011zz} that a leading order 2-point graph must possess a $D$-bubble with
exactly two vertices. First one proves:

\begin{lemma} \label{lemma:face-2vertices}
 If $D \ge 3$ and $\cG$ is a vacuum graph with degree $0$, then $\cG$ has a face with exactly two vertices.
\end{lemma}

\begin{proof} Since the graph $\cG$ is of degree 0, from equation~\eqref{eq:faces} we can ascertain that its
 face set has cardinality  $|\cF_\cG| = \left(\frac{D (D-1)}{2} p + D\right) $. Let us denote by $|\cF_{\cG,s}|$, the
number of faces with $2s$ vertices (every face must have an even number of vertices). Then:
\begin{gather} \label{sumF_s}
|\cF_{\cG,1}| + |\cF_{\cG,2}| + \sum_{s\geq3}|\cF_{\cG,s}| =  \frac{D (D-1)}{2}  p + D   .
\end{gather}
Let $2 p^{ij}_{(\rho)}$ be the number of vertices of the $\rho$-th face with colors~$ij$. On the one hand, we count the
 total number of vertices by summing the numbers of vertices per face.  On the other hand, each vertex contributes to
 $D(D+1)/2$ faces.  This leads us to:
\begin{gather} \label{eq1}
 \sum_{\rho, i<j} p^{ij}_{(\rho)} =\begin{cases}
\displaystyle |\cF_{\cG,1}| + 2 |\cF_{\cG,2 }|+ \sum_{s\ge 3} s  |\cF_{\cG,s}| ,\\
\displaystyle   \frac{D(D+1)}{2}  p  .
\end{cases}
\end{gather}
We solve \eqref{sumF_s} for the number $|\cF_{\cG,2}|$ of faces
 with four vertices, and insert the result in \eqref{eq1} to get:
\[
|\cF_{\cG,1}| = 2 D + \sum_{s\ge 3} (s-2) |\cF_{\cG,s}| +  \frac{D(D-3)}{2}  p   .
\]
Notice that on the right hand side, the f\/irst two terms yield a strictly positive contribution
for any $D\geq 2$, whereas the third term changes sign at $D=3$. Thus, we conclude that:
\begin{gather*}
|\cF_{\cG,1}| \geq 1\qquad \text{if} \quad D\geq3  .\tag*{\qed}
\end{gather*}  \renewcommand{\qed}{}
\end{proof}

Using Lemma \ref{lemma:face-2vertices} and the fact that all jackets of a leading order graph are planar, one can
 prove \cite{Bonzom:2011zz}:

\begin{proposition} \label{lemma:melon}
 If $D \ge 3$ and $\cG$ is a leading order $2$-point graph, then it contains a $D$-bubble with exactly two vertices.
\end{proposition}

This result reveals some properties of leading order vacuum graphs.  Consider a leading order vacuum graph $\cG$. It is
either the super-melon graph of Fig.~\ref{fig:redmelon} or it contains a $D$-bubble with exactly two vertices which can
be eliminated. This elimination procedure preserves colorability, degree and topology and so we
 obtain a leading order vacuum graph with two less vertices.   Thus, the new graph satisf\/ies the same conditions:
either it is the super-melon graph or it possesses a $D$-bubble with two vertices.  In the latter case, we iterate
the procedure. It follows that the leading order vacuum graphs reduce to the {\it super-melon} graph after sequentially
eliminating their $D$-bubbles. When counting the leading order graphs it is more useful, however, to take the reversed
point of view. One starts with the
super-melon graph and inserts $D$-bubbles with two vertices on its lines.

Now let us move back to dealing exclusively with leading order connected amputated 2-point graphs.
\begin{definition}
A {\it melon} graph $\cM$ is a leading order 1-particle irreducible (1PI) amputated 2-point graph.
\end{definition}
Not all leading order connected amputated 2-point graphs are 1PI (see for instance Fig.~\ref{fig:t5}),
but one may build a generic leading order 2-point graph
from those possessing this property.  Thus, it is possible to introduce the following class of graphs.

\begin{definition}
A {\it melonic} graph $\cG$ is a graph in which all 1PI amputated 2-point subgraphs are melon graphs.
\end{definition}
All leading order amputated 2-point graphs are melonic graphs (although not necessarily melon graphs, as they are not
necessarily 1PI), as is the case for the example drawn in Fig.~\ref{fig:t5}. Additionally, we shall denote the
 {\it set of melons} of~$\cG$ by~$S_\cG$.

The intuitive picture is that a melonic graph itself comprises of melons within melons. The $D$-bubble with only two vertices
is obviously the smallest melon, while for the leading order 1PI amputated 2-point graph, the largest melon is the graph itself.

At leading order, the connected 2-point function and its related free energy take the form:
\[
G_{\rm melonic} = \sum_{p=0}^{\infty}    G_p   (\lambda\bar\lambda)^p  ,   \qquad
E_{\rm melonic} = \sum_{p=0}^{\infty}    E_p  (\lambda\bar\lambda)^p  ,
\]
where $G_p$ (resp.~$E_p$) is the number of leading order 2-point (resp.\ vacuum) $(D+1)$-colored melonic graphs with~$2p$
vertices.  We want to f\/ind a closed form for $G_{\rm melonic}$ (and hence $E_{\rm melonic}$) by resumming the above series.

 {\it Direct solution:}
We now extract some juice from the world of melons.
Let $\Sigma_{\rm melonic}$ be the melonic 1PI 2-point function (with external edges of color $D$). Any contribution
 to $\Sigma_{\rm melonic}$ can be obtained by arbitrary insertions of melonic graphs on the interior edges of a $D$-bubble.
 (We do not allow insertions on the active exterior edge of the root melon, since this would cause the graph to lose its
 1PI property).  Thus,  $\Sigma_{\rm melonic}$ necessarily has the structure represented in Fig.~\ref{fig:melonsig}.
Taking into account~\eqref{eq:iid} and~\eqref{eq:2point}, it can be written as  a function of the melonic 2-point
 function $G_{\rm melonic}$:
\begin{gather*}
\Sigma_{\rm melonic}  =  \frac{(\lambda\bar{\lambda})}{N^{D(D-1)/2}}
\sum_{\bar n \notin \bar {\vec n}_D, \, n\notin \vec n_D }
\prod_{i=0}^{D-1} \langle \bar \phi^i_{\bar {\vec n}_i} \phi^i_{\vec { n}_i} \rangle \\
\phantom{\Sigma_{\rm melonic}}{}
 =
\frac{(\lambda\bar{\lambda})}{N^{D(D-1)/2}} \sum_{\bar n \notin \bar {\vec  n}_D, \, n\notin \vec n_D }
\prod_{i=0}^{D-1} \delta_{\bar {\vec  n}_i \vec n_i}  (G_{\rm melonic})^D
  =   (\lambda\bar{\lambda})\; \delta_{\bar {\vec {n}}_D \vec n_D}\; (G_{\rm melonic})^D   .
\end{gather*}

One gets a closed equation for $G_{\rm melonic}$ by recalling that the sum of the geometric series of 1PI amputated
 2-point functions yields the connected 2-point function: $G_{\rm melonic} = (1-\Sigma_{\rm melonic})^{-1}$. Hence,
\begin{gather} \label{melonic eq}
G_{\rm melonic} = 1+ (\lambda\bar{\lambda}) (G_{\rm melonic})^{D+1}.
\end{gather}
The condition that at leading order in the large $N$ limit the 1PI two point functions factors into contributions from
the connected two point functions is at the core of all the studies of the critical behavior of tensor models
performed so far.

This equation is well-known in the literature (\cite[Exercise~2.7.1]{goulden-combinatorial},
\cite[p.~200]{graham-concrete}, \cite[Proposition~6.2.2]{stanley-enumerative}) appearing in various problems
of enumeration~\cite{heubach-catalan}. Similar equations appear in the study of branched polymers. Consequently, the
leading order in $1/N$ shares the same combinatorics with the branched polymers, hence many of their
statistical properties. Such conf\/igurations have been observed in numerical studies of dynamical triangulations.
However, as already mentioned, there is a major point of divergence between branched polymers and our melonic graphs. Namely,
the intrinsic Hausdorf\/f and spectral dimensions of the melons are yet to be computed, and as we will explain below,
one has no a priori reason to expect that they will equal those of the branched polymers.

The solution of equation~\eqref{melonic eq} which goes to 1 when $(\lambda\bar{\lambda})$ goes to zero can
be written as a~power series in $(\lambda\bar{\lambda})$ with coef\/f\/icients the $(D+1)$-Catalan numbers.

\begin{proposition} \label{prop:melonic 2pt}
The melonic $2$-point function admits the following expansion:
\begin{gather*} %\label{D+1 catalan}
G_{\rm melonic}(\lambda, \bar{\lambda}) = \sum_{p=0}^\infty C^{(D+1)}_p  (\lambda\bar{\lambda})^p,
\qquad \text{with}\quad C^{(D+1)}_p = \frac1{(D+1)p+1} \binom{(D+1)p+1}{p} ,
\end{gather*}
the $(D+1)$-Catalan numbers.
\end{proposition}

The usual Catalan numbers correspond to the case $D=1$. Note that the case $D=3$ can be solved explicitly:
\begin{gather} \label{G 3D}
G_{{\rm melonic}|D=3} = \frac{  (1+4v)^{1/4} - \left[2-(1+4v)^{1/2} \right]^{1/2}  }{2 (vg)^{1/4}},
\end{gather}
 where
 \[
v= \frac{g^{1/3}}{ 2^{1/3} } \left[ \left(1+\sqrt{ 1 - \frac{2^8}{3^3} g   } \right)^{1/3} +
\left( 1 -\sqrt{1 - \frac{2^8}{3^3} g } \right)^{1/3} \right]\qquad \text{and}\qquad g =
\lambda\bar{\lambda}  .
\]

We shall see below that the leading order connected 2-point graphs are in
one-to-one correspondence with {\it colored rooted $(D+1)$-ary trees}~\cite{manes-kary-trees}.

\subsection{From melons to trees}\label{sec:meltree}

 We calculate directly the coef\/f\/icients $G_p$.   To do so, we pick a f\/ixed color, $D$ say, for the pair of external legs and count
 the number of melonic graphs at a given order.  Remember that once we list the dominant graphs at a given order, we can
 generate those at the next order by arbitrarily inserting $D$-bubbles with two vertices. In fact, the lowest orders in $p$ can
 be evaluated by a~direct counting of Wick contractions: $G_0 = G_1 =1$, $G_2=(D+1)$ and so on, but since we need all orders,
 we provide a systemic treatment.

 {\it Order $\lambda\bar\lambda$:}
\begin{itemize}\itemsep=0pt
\item The lowest order graph consists in exactly one $D$-bubble with two vertices (and external lines of color $D$). There is
only one Wick contraction leading to this graph, hence $G_1 =1$.

\begin{figure}[htb]
\centering
 \includegraphics[scale = 1.4]{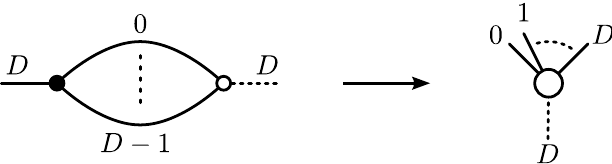}
 \caption{The f\/irst order melonic graph and its corresponding rooted tree.}
  \label{fig:order1}
\end{figure}

\item We represent this graph by the rooted tree with one central vertex decorated with $D+1$ leaves and one root. A {\it leaf}
is a vertex with only one incident edge. These $D+1$ leaves correspond to all the edges touching the positive vertex. The
leaves inherit the colors.
The root corresponds to the external edge of color $D$ touching the negative vertex.  The vertex to which the root belong is
 called the {\it root vertex}.
We consider all edges incident at the positive vertex to be {\it active}, while the external edge incident at the negative
vertex is {\it inactive}. The leafs of the tree inherit this (in)activity.    See Fig.~\ref{fig:order1} for an illustration,
where the inactive line and leaf are represented as dashed lines.
\end{itemize}

 {\it Order $(\lambda\bar\lambda)^2$:}
\begin{itemize}\itemsep=0pt
\item At second order,  $D+1$ graphs contribute.  They arise from inserting a $D$-bubble with two vertices on any of the $D+1$
 active lines of the f\/irst order graph.
Say, we insert the new bubble on the active edge of color $j$.  With respect to the new $D$-bubble, all edges incident at its
 positive vertex are deemed active, while the exterior edge (of color $j$) incident at its negative vertex is deemed inactive.
Each graph has a combinatorial weight $\frac{1}{2!^2}$, but it is produced by $2!^2$ dif\/ferent Wick contractions. These
correspond to the independent relabeling of the positive and negative vertices. Thus, the overall factor is $1$.

\begin{figure}[htb]
\centering
 \includegraphics[scale = 1.3]{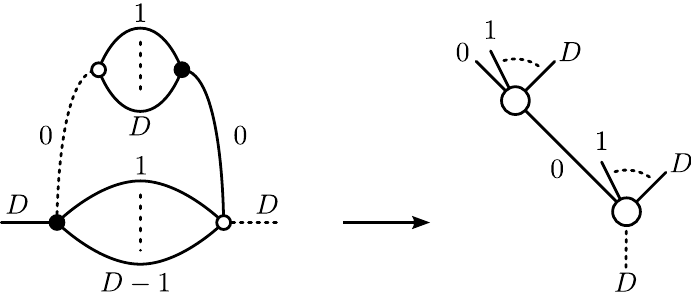}
 \caption{A second order graph and its corresponding tree.}
  \label{fig:order2}
\end{figure}

\item Once again, let us examine the case where we insert the new bubble on the active edge of color $j$. This graph
corresponds to a
tree obtained from the f\/irst order tree by connecting its leaf of color $j$ to a new $(D+2)$-valent vertex. This new
vertex has $D+1$ leaves, one of each color. The remaining line, known as a {\it tree line}, is the line of color $j$
joining the two $(D+2)$-valent vertices. The leaves correspond to the active lines of the new bubble and hence are
themselves active.
We count $D+1$ distinct trees, hence $G_2=D+1$. See Fig.~\ref{fig:order2} for the case $j=0$.  We shall introduce a
 canonical labeling of colored rooted $(D+1)$-ary trees in Section~\ref{sec:algebra}. With regards to this labeling,
each of these trees is a $\{(\;), (j)\}$ tree.

\end{itemize}

 {\it Order $(\lambda\bar\lambda)^{p+1} $:}
\begin{itemize}\itemsep=0pt
\item We obtain the graphs at order $p+1$ by inserting a $D$-bubble with two vertices on any of the active lines of a
 graph at order $p$.
Once again, with respect to the new bubble, all edges incident to its positive vertex are deemed active, while the
exterior edge incident to its negative vertex is deemed inactive.

\item In terms of the trees, we represent this insertion by connecting a $(D+2)$-valent vertex, with $D+1$ active
leaves, to one of the active leaves of a tree at order $p$. The new {\it tree line} inherits the color of the
active edge on which we inserted the $D$-bubble.
 Each of these graphs has a combinatorial weight $\frac{1}{(p+1)!^2}$, while is produced by $(p+1)!^2$ Wick
contractions (corresponding to the relabeling of of positive and negative vertices). Thus, the overall
combinatorial weight  is $1$.
\end{itemize}

To recapitulate,  at a given order $(\lambda\bar\lambda)^{p}$, we obtain contributions (with combinatorial
weight~$1$) from all rooted colored $(D+1)$-ary trees with $p$ vertices (and hence with $Dp+1$ leaves).
We remark that the tree associated to a graph is a colored version of a {\it Gallavotti--Nicolo} (GN) tree~\cite{Gal}.

A tree is associated to a partial ordering. In our case this is an ordering, denoted $\geq$, of the
set $S_\cG$ of melons of the graph. For two melons $\cM_1, \cM_2 \in S_\cG$:
\begin{gather*} %\label{ordered melons}
\cM_1 \geq \cM_2 \qquad {\rm if }\quad \begin{cases}
		\text{either} &	\cM_1 \supset \cM_2 , \\
		\text{or} &  \dsty \exists \,  \cN_{(\rho)}\in S_\cG:  \cG\supset \cM_1\vcup
                               \left( \vec{\bigcup_{\rho}}  \cN_{(\rho)} \right) \supset \cM_2 ,
	\end{cases}
\end{gather*}
where we have def\/ined a $\vcup$ operation on $S_\cG$ such that $\cM_1\vcup \cM_2$  denotes the connection of $\cM_1$ and
 $\cM_2$ using the active external edge of $\cM_1$ and the inactive external edge of~$\cM_2$.  In addition, $\ge$ is transitive.

We may depict this partial ordering very neatly using the tree graph developed earlier.   First, not that we can
identify each melon by the positive vertex at which its active external edge is incident. Since the positive vertices
 are in one-to-one correspondence with $D+2$-valent vertices of the tree, we may label each such vertex of the tree
graph by the associated melon.   Then, the tree graph captures the partial ordering explicitly.  The {\it root melon}
is the melon mapped to the root vertex of the tree.  For a melon $\cM_2$, if another melon~$\cM_1$ lies on its unique
path to the root melon, then $\cM_1\ge \cM_2$.  $\cM_1$ is an {\it ancestor} of $\cM_2$, while conversely~$\cM_2$ is
 a {\it descendant} of~$\cM_1$.

\begin{figure}[htb]
\centering
 \includegraphics[scale = 1]{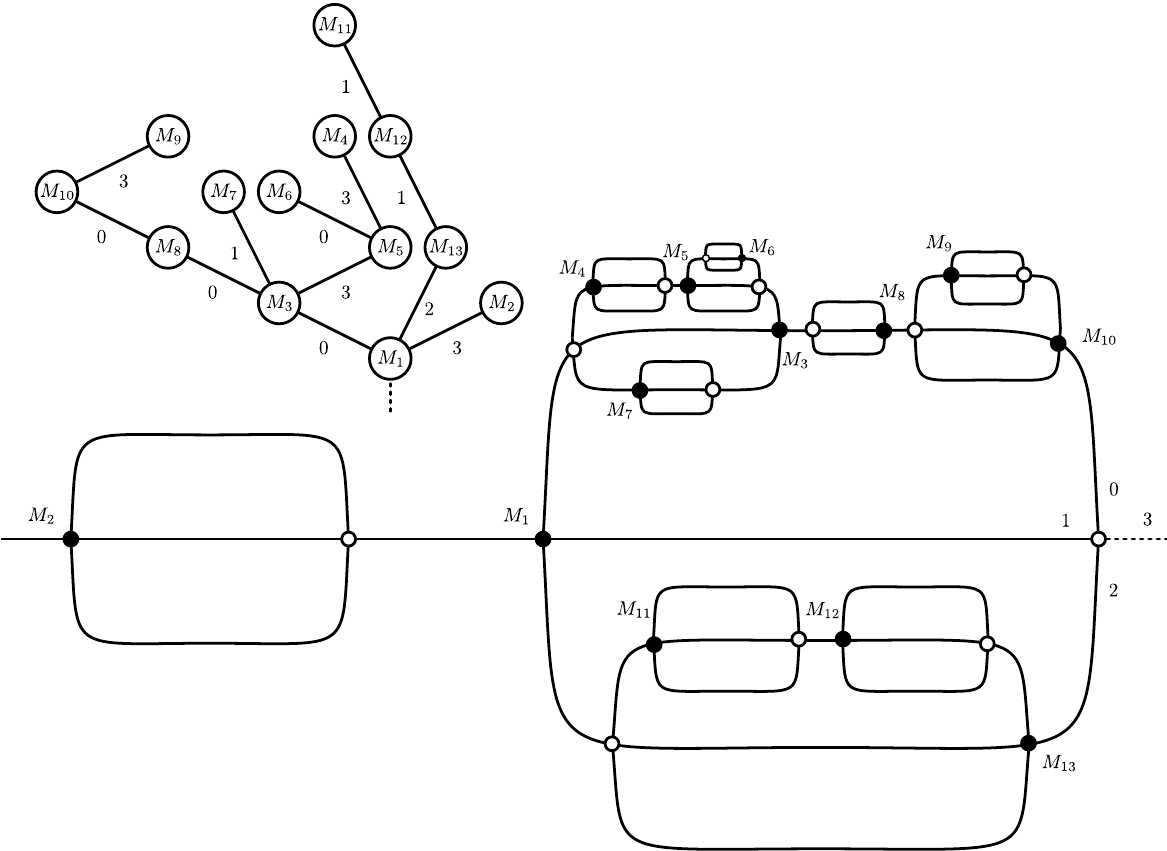}
\caption{A melonic graph and its associated colored GN rooted tree.}
\label{fig:t5}
\end{figure}

An example in $D=3$ is given in Fig.~\ref{fig:t5}.   The root melon is $\cM_1$. Note that
 $\cM_3\supset \cM_4,\cM_5,\cM_6,\cM_7$,
hence it is their ancestor. Also, since $\cM_3 \vcup \cM_8 \vcup \cM_{10}\subset \cG$ and $\cM_9 \subset \cM_{10}$,
the melon $\cM_3$ is the ancestor of $ \cM_4$, $\cM_5$, $\cM_6$, $\cM_7$, $\cM_8$, $\cM_9$, $\cM_{10}$.

The leading order connected vacuum graphs (also called melonic) are obtained by reconnecting the two exterior edges
of a melonic 2-point graph. Their amplitude is $N^D$. If a graph is a~$(D+1)$-colored melonic graph, all its
$D$-bubbles are melonic graphs with $D$ colors. This is easy to see, as the reduction of a~$D$ bubble with two
vertices represents the reduction of a~$(D-1)$-bubble with two vertices for all $D$-bubbles which contain these
two vertices. When reducing the graph to its root melon, one by one all its~$D$-bubbles reduce to $D$-bubbles with
 two vertices.

Moreover, the $D$-ary trees of the $D$-bubbles are trivially obtained from the $(D+1)$-ary tree of the graph by
deleting all lines (and leaves) of color $D$. We shall need below the following obvious fact: given a melonic graph
and one of its $D$-bubbles (say $\cB^{\widehat{D}}_{(1)}$), all the lines of color $D$ connecting  on it either
separate it from a dif\/ferent $D$-bubble (and are tree lines in the associated colored rooted tree) or they connect
 the two external points of a 1PI amputated 2-point subgraph with $D-1$ colors of $\cB^{\widehat{D}}_{(1)}$ (i.e.\
 they connect the two external points of a melon in $\cB^{\widehat{D}}_{(1)}$), in which case they are leaves of
the associated tree.

We shall display below two additional techniques to resum the melonic graphs using this correspondence with
trees, since they are important tools in their own right.

\begin{figure}[htb]
\centering
 \includegraphics[scale = 1.5]{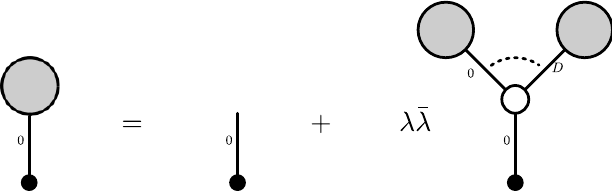}
\caption{The melonic equation represented from
the point of view of the colored rooted trees.} \label{fig:eq_trees}
\end{figure}

 {\it Counting using the $D+1$-ary tree measure:}
From Section~\ref{sec:meltree}, we conclude that $G_{\rm melonic}(\lambda\bar{\lambda})$ is the generating function
of rooted $(D+1)$-ary trees. The equation \eqref{melonic eq} encodes their proliferation.  Its interpretation in term
of trees is straightforward, and depicted in the Fig.~\ref{fig:eq_trees}.
The leading order connected 2-point function expands in colored rooted trees with $p$ (unlabeled) $D+2$-valent
vertices, $Dp+1$ active leaves, and one inactive leaf taken as the root. The result is recovered
using Example~\ref{eq:darymeasure}.

 {\it Counting using Cayley's theorem:}
{\it Cayley's theorem} states that the number ordinary trees on~$n$ labeled vertices, such that the vertex $i$ is $d_i$-valent,
 is: $(n-2)!/\prod_i (d_i -1)! $.
We shall work towards applying this result bit by bit.  First let us denote by~$T_p$, the number of ordinary trees with~$p$
 labeled $(D+2)$-valent vertices and $Dp+2$ labeled 1-valent vertices. Using the above formula, one has:
\[
T_p = \frac{\bigl[ (D+1)p\bigr] !}{\bigl[ (D+1)!\bigr]^p} .
\]
There are two dif\/ferences between ordinary trees and our $(D+1)$-ary trees: $i)$ the vertex-labeling and $ii)$ the line-coloring.
We f\/irst start by coloring the tree. The color of the inactive leaf of the root vertex is f\/ixed by the external color of the
2-point function. We have $(D+1)!$ colorings of the lines of the root vertex. Any $(D+2)$-valent vertex will support $(D+1)!$
colorings of its lines (as the color of the line connecting it towards the root is f\/ixed by its ancestor). Thus we pick up a
 factor $[(D+1)!]^p$ for the colorings of the lines. We then remove the vertex labels.  This brings a
factor $\frac{1}{p!}$ for the $(D+2)$-valent vertices and a factor  $\frac{1}{(Dp+1)!}$ for the active leaves. We conclude that:
\[
G_p = \frac{\bigl[ (D+1)!\bigr]^p}{p! (Dp+1)!}\ T_p = \frac{1}{p!} \frac{1}{(Dp+1)!}  [(D+1)p]! = C_p^{(D+1)}  .
\]
Note that there is one subtle point in using Cayley's theorem: one can not f\/irst relabel the vertices and then color the
lines. Indeed, in order to identify the allowed colorings of the lines one must have distinguished vertices (otherwise one can
not properly count for instance the colorings of leaves touching the same vertex).

\subsection{Critical exponent}

Of course the whole point of this counting was to use the outcome to categorize some critical behavior of the class of
 models in question.  Applying Stirling's formula to the expansion in Proposition \ref{prop:melonic 2pt}, we get the large
order behavior.
\begin{proposition} \label{prop:entropy}
The number of melonic graphs with $2p$ vertices behaves for large $p$ like:
\[
G_p \sim A\ g_c^{-p}\ p^{-3/2} ,
\]
with
\[
g_c = \frac{D^D}{(D+1)^{D+1}},\qquad \text{and}\qquad A = \frac{e}{\sqrt{2\pi}} \frac{\sqrt{D+1}}{D^{3/2}} .
\]
In particular, $G_p$ is exponentially bounded by $e^{-p\ln g_c}$, with $\ln g_c<0$.
\end{proposition}

The $D$-dependent constant $g_c$ is the critical value of the coupling $g\equiv \lambda\bar{\lambda}$. Indeed,
 it is well-known that a series with coef\/f\/icients of the form $g_c^{-p} p^{-\alpha}$ behaves in the neighborhood
of $g_c$ like $(g_c-g)^{\alpha -1}$. Hence, the most singular part of the melonic 2-point function is:
\[
G_{\rm melonic, sing} \sim K\ \left(\frac{g_c -g}{g_c}\right)^{1/2} ,
\]
for some constant $K$.\footnote{This can be checked explicitly from~\eqref{G 3D} in the $D=3$ case.}  The free energy
of melonic graphs is obtained using Proposition~\ref{relation GF}:
\begin{gather*}
G_{\rm melonic} = 1 + \lambda\,\partial_{\lambda} E_{\rm melonic}
\ \Rightarrow \  E_{\rm melonic, sing} \sim K'
\ \left(\frac{g_c -g}{g_c}\right)^{2-\gamma_{\rm melonic}}  , \qquad \gamma_{\rm melonic} = \frac12 .
\end{gather*}
The exponent $\gamma_{\rm melonic}$ is known as the susceptibility, or entropy exponent\footnote{Once an exponential
bound is found for the proliferation of a species, the entropy exponent characterizes the polynomial part of the number
of such objects.}.

Alternatively, one can derive the critical exponent from equation \eqref{melonic eq}, noting that its solution may be written as:
\[
 g= G_{\rm melonic}^{-D} - G_{\rm melonic}^{-D-1}   ,
\]
and that in the neighborhood of the critical point: $ \frac{dg}{dG_{\rm melonic}}(G^c_{\rm melonic}) =0$, so that
the coupling expands as: $g-g_c = (G_{\rm melonic} - G_{\rm melonic}^c  )^2$.

This critical behavior is a key ingredient in the provision of a continuum limit  for random colored melonic triangulations.
But, {\it a priori}, the geometric interpretation of the continuum limit depends on the details of the model under
 consideration. In the following, we use the natural interpretation of the i.i.d.~model as a generator of dynamical triangulations.

\subsection{Melons and branched polymers}\label{sec:dtcontlim}

One can interpret the  i.i.d.~model quite naturally as a model of dynamical triangulations (DT)
\cite{ ambjorn-houches94, ambjorn-revueDT,ambjorn-book, ambjorn-scaling4D, david-revueDT}\footnote{Tensor
models were indeed originally proposed in this context \cite{ambj3dqg,sasa1}.}. In the remainder of this section, we
 switch to notations more familiar in the DT literature.
A melonic graph is dual to a colored triangulation of the $D$-sphere.  We denote the number of $k$-simplices (i.e.\
the number of $(D-k)$-bubbles) by $N_k$.  In this case,  the amplitude of a graph $\cG$ becomes:
\begin{gather}\label{eq:ampligraph}
A(\cG) = e^{\kappa_{D-2}N_{D-2}  -  \kappa_D N_D} ,
\end{gather}
where $\kappa_{D-2}$ and $\kappa_D$ are:
\[
\kappa_{D-2} = \ln N,\qquad \text{and} \qquad
\kappa_D = \frac12\left( \frac12 D(D-1)  \ln N - \ln (g)\right) .
\]
Once again,  $g=\lambda\bar{\lambda}$.  The argument of the exponential takes the form of the Regge action (discrete
 form of the Einstein--Hilbert action for general relativity) on a triangulation with regular $D$-simplices of length,
say,~$a$. Indeed, on a Regge discretization, the curvature is concentrated around the $(D-2)$-simplices and measured
 by the def\/icit angle $\delta(\sigma_{D-2})$ ($2\pi$ minus the sum of the dihedral angles hinged on the
$(D-2)$-simplex $\sigma_{D-2}$) and the total volume is measured by the cosmological term. The Regge action is:
\begin{gather}\label{eq:regge}
S_{\rm Regge} = \Lambda \sum_{\sigma_D} \vol(\sigma_D)
- \frac{1}{16\pi G}\sum_{\sigma_{D-2}} \vol(\sigma_{D-2}) \delta(\sigma_{D-2}) .
\end{gather}
The volume of a regular $k$-simplex is: $\vol(\sigma_k) = \frac{a^k}{k!}\sqrt{\frac{k+1}{2^k}}$,
and \eqref{eq:regge} takes the form \eqref{eq:ampligraph} on a~regular triangulation if we identify~$N$ and~$g$ in terms of the (bare) dimensionful parame\-ters~$G$,~$\Lambda$ and the length~$a$ via:
\begin{gather}
\ln g  =
 \frac{D}{16\pi G}\vol(\sigma_{D-2})\left(\pi(D-1) - (D+1) \arccos \frac1{D}\right) - 2\Lambda \vol(\sigma_D)
\equiv  -2 a^D\ \widetilde{\Lambda} ,  \nonumber \\
\ln N  =  \frac{\vol(\sigma_{D-2})}{8G} .\label{log g}
\end{gather}
Notice that in two dimensions, $\ln g = - 2\Lambda \vol(\sigma_D)$.

The large $N$ limit corresponds to $G\rightarrow 0$. Since $g$ is
kept f\/inite, $\Lambda$ becomes large, positive, and scales like $\Lambda\sim 1/(a^2 G)$. The melonic triangulations
have degree $0$, hence their amplitudes take the form:
\begin{gather*}
N^{-D} A(\cG)  =  g^{N_D/2} = e^{-a^D \widetilde \Lambda N_D}
 =  e^{\frac{D}{32\pi G}\vol(\sigma_{D-2})\bigl(\pi(D-1) - (D+1) \arccos \frac1{D}\bigr) N_D}  e^{-\Lambda\vol(\sigma_D) N_D} ,
\end{gather*}
where we have explicitly split the contribution of the Einstein--Hilbert term and of the cosmological term.
The scalar curvature is the argument of the f\/irst
exponential (up to $1/G$ and irrelevant constants). It grows linearly with the number of $D$-simplices
and is positive (since $\pi(D-1) - (D+1) \arccos \frac1{D}>0$ for $D\geq3$).
 The entropy, that is the logarithm of the number of triangulations with a f\/ixed volume (f\/ixed
number of $D$-simplices), grows linearly with $p$ for melonic graphs, by Proposition \ref{prop:entropy}.

For the melonic family, the numbers of $D$-simplices and $(D-2)$-simplices are not independent.  They are related by:
\[
N_{D-2} = \frac{D(D-1)}{4}N_D + D .
\]
Hence, the large $N$ limit of the colored tensor model projects dynamical triangulations to
 curves (parametrized by $g= \lambda\bar{\lambda}$):
\[
\kappa_D - \frac{D(D-1)}{4}\,\kappa_{D-2} = - \frac12 \ln g > -\frac12 \ln g_c ,
\]
where $g_c$ is given in the Proposition~\ref{prop:entropy}. This gives the critical curve:
\[
\kappa_D^c(\kappa_{D-2}) = \frac{D(D-1)}{4} \kappa_{D-2} - \frac12 \ln g_c .
\]

\begin{figure}[htb]
\centering
 \includegraphics[scale = 0.4]{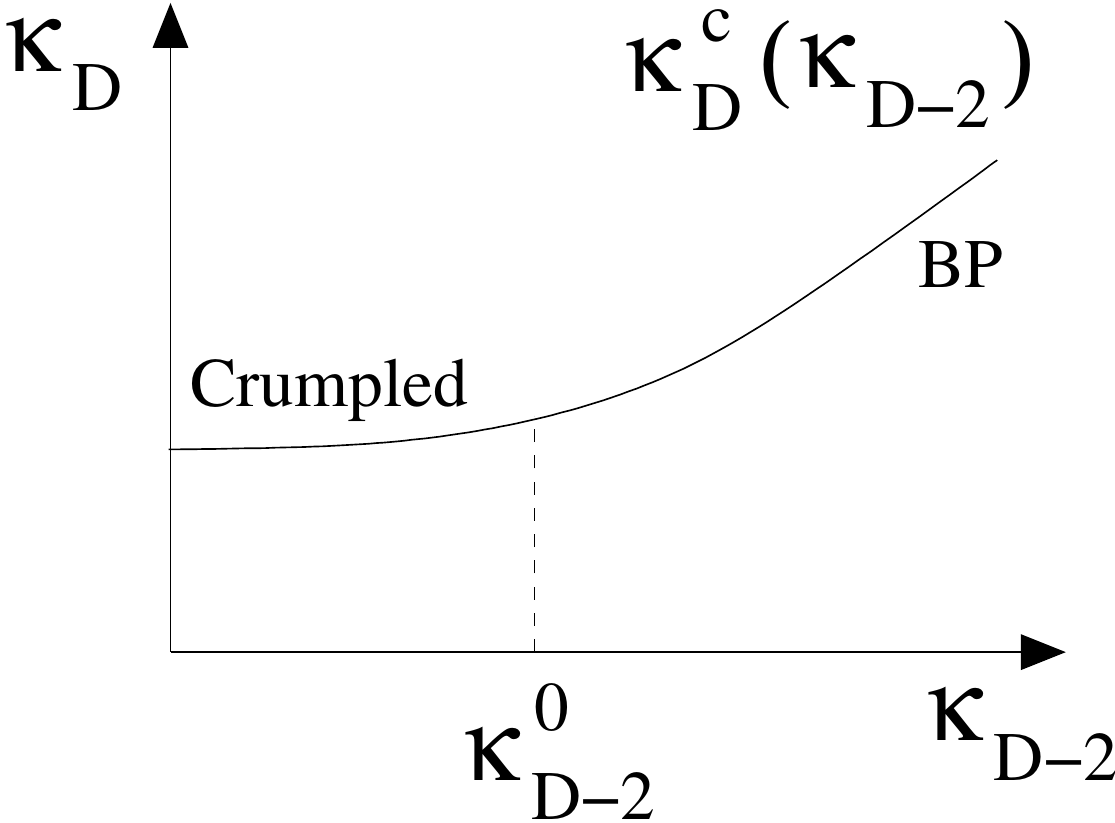}
\caption{The phase portrait of dynamical triangulations.}\label{fig:BPphaseportrait}
\end{figure}

The family of melonic triangulations shares a  number of similarities with the branched polymers (BP) phase  of dynamical triangulations
\cite{ambjorn-revueDT,ambjorn-BP, ambjorn-d>1, ambjorn-scaling4D, david-revueDT}.
Indeed, BP are known to dominate the regime of large positive curvature, $\kappa_D>\kappa_D^c(\kappa_{D-2})$
for suf\/f\/iciently large values of~$\kappa_{D-2}$. In Fig.~\ref{fig:BPphaseportrait} we present
the phase portrait of dynamical triangulations. The continuum phase is reached for $\kappa^c_D(\kappa_{D-2})$.
The BP phase corresponds to $\kappa_{D-2}> \kappa_{D-2}^0$. At  $\kappa_{D-2}^0$ dynamical triangulations undergo a (f\/irst order) phase transition
to the crumpled polymer phase.
Since $\kappa_{D-2} = \ln N$, one would expect that the large $N$ limit of tensor models yields the BP phase
of dynamical triangulations.

The case $D=2$ is special. Although melonic graphs appear in the large $N$ limit,
the critical behavior is driven by the whole planar sector~\cite{bessis} (and not only by the melonic graphs). Thus
$\gamma_{\rm string} = -\frac12$. Only beyond the $c=1$ barrier the dominant
family is reduced to~BP, with a~susceptibility exponent $\gamma_{\rm BP} = \frac12$.
In higher dimensions, the large $N$ limit yields directly a~phase of large positive curvature
with critical exponent $\gamma_{\rm melons} = \gamma_{\rm BP} =\frac12$, hence
with the statistical properties of BP.

The last comparison between melonic triangulations and BP is geometric. Indeed, it has been observed
in DT that in the BP phase the number of vertices and the number of $D$-simplices grow proportionally
 $N_0 \simeq N_D/D$ (see for example \cite{david-revueDT}). The natural geometric interpretation
 is the following. Start with a triangulation of the $D$-sphere and
 iterate random sequences of $(1\rightarrow (D+1))$ moves. Such moves add a vertex to a
 $D$-simplex and split it into $(D+1)$ new $D$-simplices. This way, the triangulation gets a
new vertex together with $D$ additional $D$-simplices. Hence, asymptotically, $N_0 \simeq N_D/D$.
Notice that in the crumpled polymer phase of DT, one has in contrast $N_0\sim N_D^\delta \ll N_D$.
Melonic graphs respect:
\[
N_0 =  \frac{N_D}{2} + D ,
\]
hence behave like BP. Irrespective of the dimension, the coef\/f\/icient $1/2$ arises in the f\/irst term on the right hand side of the above equation.
 Its occurrence has the following explanation. Melons are not obtained  just through $(1\rightarrow (D+1))$ moves\footnote{This move alone does
not respect the colors.}. Instead,
one adds a $D$-bubble with only two vertices to some initial melon by f\/irst performing a $(1\rightarrow (D+1))$
move, which is immediately followed by a $(D\rightarrow 2)$ move\footnote{This second move restores the colors
 spoiled by the f\/irst one.}. Such a sequence adds one vertex to the triangulation together with only two $D$-simplices.
Clearly melons provide a better balance between~$N_0$ and~$N_D$ (maximize~$N_0$ for a f\/ixed~$N_D$)
 than the usual set of BP. It may be surprising that they have never been considered as
 such in DT. We think that this is because the creation of a $D$-bubble with two vertices needs two
successive moves at the same place, which is unlikely to happen in Monte-Carlo algorithms.

The main reason we refrain for now from stating that the melonic graphs are BP it the following.
The BP phase is characterized by a Hausdorf\/f dimension $d_H=2$. Obviously, one could get
this value for melonic graphs by using the natural distance on the colored trees. However, the vertices of
the trees correspond to melons, which are non-local from the point of view of the triangulation. One
should instead use a dif\/ferent notion of distance (graph distance either for the triangulation or for the
melonic graph), and not the tree distance in the abstract colored tree.

\subsection{Large volume limit}

The continuum, large volume limit is obtained, like in matrix models, by simultaneously \mbox{sending}~$a$ to zero and
$g=\lambda\bar{\lambda}$ to its critical value. When $g \ll g_c$, triangulations with a
large number of $D$-simplices get exponentially suppressed by $(\frac{g}{g_c})^{N_D/2}$ and triangulations with a small number
 of $D$-simplices dominate the free energy $E_{\rm melons}$. When $g\to g_c$, the summability of $E_{\rm melons}$ is lost due to the triangulations
with an inf\/inite (i.e.\ large) number of simplices, whose individual sizes scales to zero.
Notice however that in this  large volume, continuum limit the coupling~$\kappa_{D-2}$ is sent to
inf\/inity (and thus~$G$ to zero) instead of tuning it to
some critical value. This is exactly like the standard large~$N$ limit of matrix models for pure two-dimensional
gravity and here, it is simply due to the fact that $\kappa_{D-2}=\ln N$. To access analytically a sector of f\/inite~$\kappa_{D-2}$ would require some kind of double scaling limit \cite{double,double1,double2} which remains still to be
uncovered in tensor models.

The average volume is:
\[
\langle  \Vol  \rangle = a^D\,\langle N_D\rangle = 2 a^D  \lambda \partial_\lambda  \ln E_{\rm melons}
= 3 g_c \frac{a^D}{g_c - g} + \text{less singular terms} ,
\]
and must be kept f\/inite in the continuum limit\footnote{Like in the case of matrix models, one needs in fact to look at
geometries with at least a boundary component in order to truly get an inf\/inite volume.}. In the
 large $N$ limit, we are left with a single
dimensionful parameter, $\widetilde{\Lambda}$ in~\eqref{log g}, which should be renormalized
when reaching the critical point. The average volume is written in terms of $\widetilde{\Lambda} $ as:
\[
\langle  \Vol  \rangle \sim -\frac{3 \ln g}{2 \widetilde{\Lambda}} \left(\frac{g_c - g}{g_c}\right)^{-1}
= \frac{3}{2 \widetilde{\Lambda}_{\rm R}}   , \qquad
\widetilde{\Lambda}_{\rm R} = \frac{\widetilde{\Lambda}}{-\ln g} \left(\frac{g_c - g}{g_c}\right) .
\]
In order to obtain a continuum phase with f\/inite physical volume, we need to take the continuum limit:
\[ a \rightarrow 0,\qquad g \rightarrow g_c,\qquad
\text{with} \quad \widetilde{\Lambda}_{\rm R} \sim a^{-D}(g_c-g) \ \text{f\/ixed}.
\]
Notice that this renormalization of $\widetilde{\Lambda}$ has already appeared in the DT literature,
for example~\cite{ambjorn-revueDT}, where it was given the equivalent form:
\[
a^D \widetilde{\Lambda}_{\rm R} = \kappa_D - \kappa_D^c(\kappa_{D-2}) .
\]

An important issue concerning the critical behavior we have observed is to understand its
universality, i.e.\ the relevance of the details of the microscopic model. In matrix models
for $2D$ gravity, one can describe the surfaces using either triangulations or quadrangulations.
Actually, any interaction with a positive coupling falls in that same (pure gravity) universality class~\mbox{\cite{Di_Francesco:1993nw, legall-miermont}}. We will
perform the f\/irst steps in the study of universality of colored models in arbitrary dimensions
in the sequel. Establishing the full (multi-)critical behavior of colored tensor models is
one of the active research directions in the f\/ield.

An interesting feature revealed in our analysis so far is a universality with respect
to the dimension $D$. Although the value of the critical coupling depends on~$D$ (and goes to zero
when~$D$ goes to inf\/inity), the susceptibility exponent does not. This is in agreement with the
almost systematic appearance of a branched polymer phase in dynamical triangulations.
Although we have restricted our analysis to the i.i.d.~model, the same melonic graphs dominate more
involved models like the colored Boulatov--Ooguri model~\cite{GurRiv}. In fact, as melonic graphs have
the maximal number of faces at f\/ixed number of vertices, they are likely to generically dominate
in tensor models, irrespective of the details of the covariance (propagator).
Since we have mapped melons to trees, it is appealing to think that they inherit all the
statistical properties  of trees. However, the mapping does not preserve the locality of the
triangulation, so that more investigations are needed in order to conclude whether the
physical correlations def\/ined on the triangulation agree with those of branched polymers. In
particular, it would be interesting to test Fisher's scaling relation directly from melons\footnote{Fisher's scaling relation occurs in the analysis of critical phenomena of f\/ilms of f\/inite thickness \cite{fisher-scaling}.  In its original context, it is a relation among the spin susceptibility exponent, the critical exponent of the spin-spin correlation length and the anomalous scaling dimension of the spin-spin correlation function.}.

\section{Bubble equations}\label{sec:bubbleq}

The critical behavior of matrix models can be systematically studied by means of the loop equations. In the sequel,
we shall generalize the loop equations to bubble equations in arbitrary dimensions.  Although general bubble equations can
be written, we shall present in detail only those equations relevant for the leading order in the $1/N$ expansion.

The bubble equations are a set of constraints satisf\/ied by the partition function of the tensor models, essentially, they are
the quantum equations of motion. They are derived starting from the Schwinger--Dyson equations. The constraints satisfy
 a Lie algebra. As this Lie algebra is somewhat involved, we shall start by detailing it.  Only subsequently, shall we relate it to the tensor models.

{\it $D$-ary tree Lie algebra:}
While at f\/irst sight, this subsection might seem rather technical, the ultimate aim is rather simple.  Our goal is to construct a
composition operator for rooted colored $(D+1)$-ary trees, that is, given two such trees, we can coherently identify another tree as
their composition.  This operation of composition must satisfy certain properties, the most important of which is that one can def\/ine a
 set of operators indexed by these trees, which form a Lie algebra such that the Lie bracket is compatible with this composition rule.

{\it From one to an infinity of coupling constants:}
In this part, we explicitly integrate out $D$ of the $D+1$ tensors in the path integral to get an ef\/fective action and path integral
 for the single remaining color.  This ef\/fective action contains many ef\/fective interaction terms, indexed by the possible $D$-bubbles
of that remaining color.  Each such interaction has a $(\lambda\bar\lambda)$-dependent coupling constant.  We relax this dependence,
rendering the couplings independent.

{\it Schwinger--Dyson equations:}
We explicitly describe how the large N limit of the i.i.d.~probability measure provides a representation of this algebra.

\subsection{A Lie algebra indexed by colored, rooted, $D$-ary trees}\label{sec:algebra}

We have already seen that the leading order graphs are in close correspondence with colored rooted $D$-ary trees. We shall see
that the bubble equations we derive are indexed by such trees. We f\/irst present a set of def\/initions and properties of the trees.
Although technically involved, this section is required in order to derive the algebra of constraints satisf\/ied by the partition
function at leading order (in the large $N$ limit).

\begin{definition}
 A {\it colored rooted $D$-ary tree} $\cT$ with $|\cT|$ vertices is a tree with the following properties:
\begin{itemize}\itemsep=0pt
 \item it has a $D$-valent root vertex, denoted $(\;)$;
 \item it has $|\cT|-1$ vertices that are $(D+1)$-valent (i.e.\ each of them has $D$ descendants);
 \item it has $(D-1)|\cT|+1 $ vertices that are $1$-valent (i.e.\ with no descendants). These are the leaves we mentioned earlier;
 \item all lines have a color index, $0,1,\dots ,D-1$, such that the $D$ direct descendants (leaves or vertices)
  of a given vertex (or of the root) are connected by lines with dif\/ferent colors.
\end{itemize}
\end{definition}

We shall subsequently ignore the leaves of the tree, as they can automatically be added once the vertices and lines of the colored tree are known.

\begin{figure}[htb]
\centering
 \includegraphics[scale = 1.3]{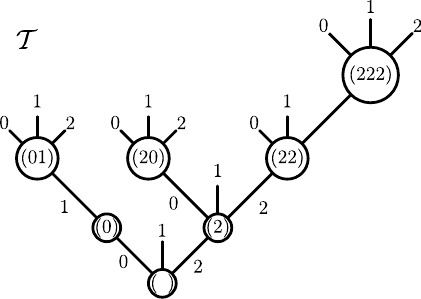}
\caption{A colored rooted $D$-ary tree.  We have included the leaves in this case.}\label{fig:Dary}
\end{figure}

{\it Canonical labeling:}
A colored rooted $D$-ary tree admits a canonical labeling of its vertices. Namely, every vertex can be labeled by the list of colors  $V=(i_1 \dots i_n)$ of
 the lines in the unique path connecting
$V$ to the root $(\;)$. The f\/irst color, $i_1$, is the color of the line in the path ending on the root
(and~$i_n$ is the color of the line ending on $V$).
For instance the vertex $(0)$ is the descendant connected to the root $(\; )$ by the line
of color $0$, and the vertex $(01)$ is the descendant of the vertex $(0)$ connected to it by a line
of color $1$ (see Fig.~\ref{fig:Dary} for an example of a canonically labeled $3$-ary tree with
$|\cT|=7$ vertices).

{\it Some notation:}
In the sequel, `tree' will always mean a colored rooted $D$-ary tree. A tree is completely identif\/ied by its canonically
labeled vertices, hence it is a set $\cT=\{ (\;), \dots\}$. We shall use the shorthand notation~$V$ for the list of labels identifying a vertex.

Moreover, we denote by $\underbrace{i,\dots,i}_{n}$ a list of~$n$ identical labels~$i$.

 {\it Successor functions:}
Consider a tree $\cT$ and one of its vertices:
\[
V = l, \dots, k , \underbrace{i, \dots, i}_{n}   ,
\]
with $k \neq i$. The colors $l,\dots, k$ might be absent.
The {\it successor of color $j$} of $V$, denoted $s^j_{\cT}[V]$, is the vertex:
\[
 s^j_{\cT}[ ( l, \dots, k , \underbrace{i, \dots, i}_{n }) ]
                = \begin{cases}
                ( l, \dots, k , \underbrace{i, \dots, i}_{n}, j)
                  \text{ if it exists }, \\
                \text{ if not }   \begin{cases}
                                   ( l, \dots, k , \underbrace{i, \dots, i}_{n })
                                              & \text{ if } j \neq i, \\
                                   ( l, \dots, k ) & \text{ if } j=i.
                                  \end{cases}
                 \end{cases}
\]
The colored successor functions are cyclical, namely if a vertex does not have a descendant in the tree
of color $j$, then its `successor of color $j$' is the f\/irst vertex one encounters, when going form $V$ to the root,
whose label does {\it not} end by $j$. For the example of Fig.~\ref{fig:Dary} we have
\begin{alignat*}{4}
& s^0[(\; )]  = (0), \qquad &&  s^{1}[ (\;) ] =   (\;), \qquad && s^{2} [(\;)]  =  (2), & \\
& s^0[(0)]  =  (\;), \qquad && s^1[(0)] =  (01), \qquad & & s^2[(0)]  =  (0), & \\
& s^0[(01)]  =  (01), \qquad && s^1[(01)] =  (0), \qquad && s^2[(01)]  =  (01),& \\
& s^0[(2)]  =  (20),\qquad && s^1[(2)] =  (2),\qquad && s^2[(2)]  =  (22),& \\
& s^0[(20)]  =  (2),\qquad && s^1[(20)] =  (20),\qquad && s^2[(20)]  =  (20), & \\
& s^0[(22)]  =  (22), \qquad && s^1[(22)] =  (22), \qquad && s^2[(22)]  =  (222), & \\
& s^0[(222)]  =  (222), \qquad && s^1[(222)] =  (222),\qquad && s^2[(222)]  =  (\;)   .&
\end{alignat*}

 {\it Maximal vertices:}
We call $V$ the {\it maximal vertex of color $i$} in a tree $\cT$ if
\[
V = \underbrace{i\dots i}_{n} \qquad \text{such that}
\qquad s^i_{\cT}[ ( \underbrace{i\dots i}_{n} ) ] = (\;)  .
\]
In Fig.~\ref{fig:Dary} the maximal vertex of color~$2$ is~$(222)$, the maximal
vertex of color~$0$ is~$(0)$ and the maximal vertex of color $1$ is the root $(\;)$ itself.

 {\it Branches and branch complements:}
We def\/ine the {\it branch of color $i$ of $\cT$}, denoted $\cT^i$, the tree
\[
 \cT^i = \big \{  (X) \, \vert \, (i\,X) \in \cT \big \}  .
\]
The branch $\cT^i$ can be empty. The root of the branch $\cT^i$, $(\;)\in \cT^i$
corresponds to the vertex $(i) \in \cT$.
The rest of the vertices of~$\cT$ (that is, the root $(\;)$ and all vertices of the form
$(k,U)\in \cT$, $k\neq i$) also form a canonically labeled tree.  Denoted $\tilde \cT^i$,
it is the {\it complement in~$\cT$} of the branch~$\cT^i$.

In Fig.~\ref{fig:Dary}, the branch of color $2$ is the tree
$\cT^2=\{ (\;), (2), (22), (0)\}$, as all the vertices $(2)$, $(22)$, $(222)$
and $(20)$ belong to~$\cT$. Its complement is $\tilde \cT^2=\{ (\;), (0), (01)\}$.

\begin{figure}[htb]
\centering
 \includegraphics[scale = 1.2]{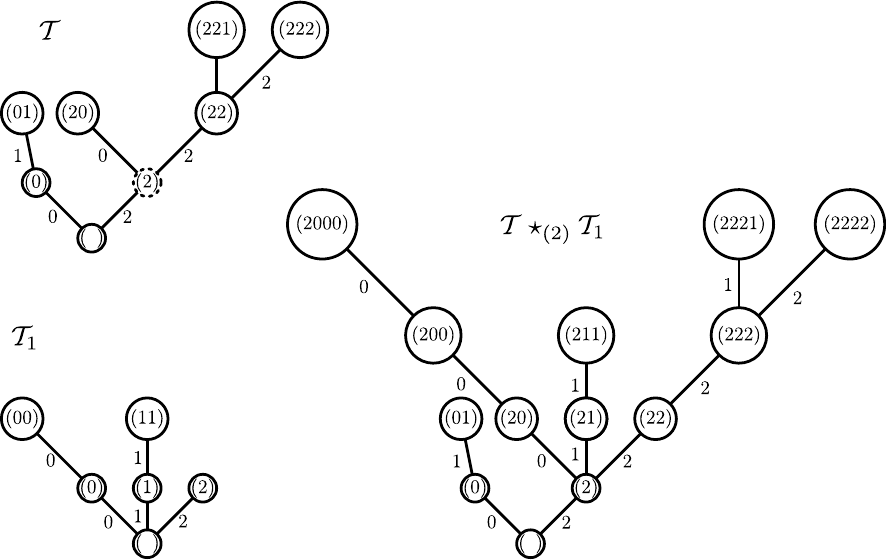}
\caption{Gluing of two trees at a vertex $\cT \star_{(2)} \cT_1$.}\label{fig:glue}
\end{figure}

 {\it Joining trees:}
Two colored rooted $D$-ary trees $\cT$ and $\cT_1$ can be {\it joined} (or glued) at a vertex $V\in \cT$.
For all colors $i$, denote the maximal vertices $\cT_1$ of color $i$:
\[
 (\underbrace{i,\dots,i}_{n_i} )   ,
 \qquad s^i_{\cT_1}[ ( \underbrace{ i,\dots,i }_{n_i } ) ] = (\;)   .
\]
The glued tree, $\cT \star_{V} \cT_1$, is the tree canonically labeled by:
\[
 \cT \star_{V}\cT_1 = \begin{cases}
                       (X)     &\text{ for all } (X) \in \cT  , \  (X) \neq (V, \dots), \\
                       (V,Y)   &\text{ for all } (Y) \in \cT_1, \\
                       (V, \underbrace{i,\dots,i}_{n_i + 1 } , Z )
                         &\text{ for all } (V,i,Z) \in \cT.
                      \end{cases}
\]
This operation can be seen as cutting all the branches starting at $V$ in $\cT$, gluing the tree~$\cT_1$ at~$V$, and then gluing back the branches
at the maximal vertices of the appropriate color in~$\cT_1$. The vertices of $\cT\setminus (V)$ and $\cT_1\setminus (\;)$ map one to one onto the
 vertices of $ (\cT \star_V \cT_1) \setminus (V)$, and both
$(V)\in \cT, (\;)\in \cT_1$ map to $(V) \in  \cT \star_V \cT_1 $, thus
$ | \cT \star_V \cT_1 | = |\cT| + |\cT_1| -1$.
An example is given in Fig.~\ref{fig:glue}, where the leaves
are not drawn.

\begin{figure}[htb]
\centering
 \includegraphics[scale = 1.2]{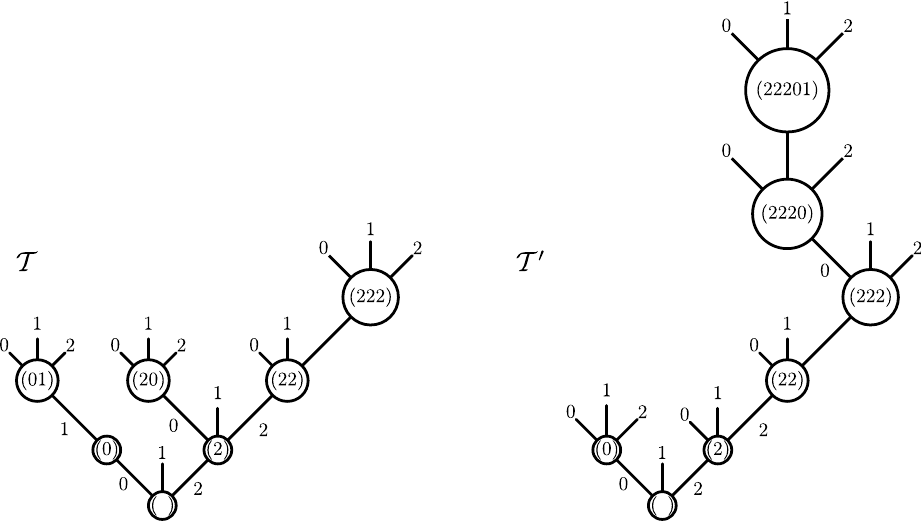}
\caption{Two equivalent trees $\cT \sim \cT'$, with
$\cT' = \Bigl( \cT^2 \star_{(22)} \{(\;),(2)\} \Bigr) \star_{(222)} \tilde \cT^2$.}
\label{fig:equiDary}
\end{figure}

 {\it Equivalence:}
For any tree $\cT$, with maximal vertex of color $i$, $(\underbrace{i,\dots,i}_{n_i})$,
the maximal vertex of color $i$ in the branch $\cT^i$ will have one less label $(\underbrace{i,\dots,i}_{n_i-1})$.
One can glue the tree $\{ (\;), (i)\}$ on $(\underbrace{i,\dots,i}_{n_i-1} ) \in \cT^i$. All the vertices of $\cT^i$ are unchanged by this gluing,
 its only ef\/fect being to introduce a new vertex, $(\underbrace{i,\dots,i}_{n_i})$ which becomes the new maximal vertex of color $i$. Subsequently,
one can glue the complement of the branch $i$, $\tilde \cT^i$, on this new maximal vertex:
\[
 \cT' = \Bigl( \cT^i \star_{ (\underbrace{i,\dots, i}_{n_i-1} ) } \{(\;),(i)\} \Bigr)
\star_{ (\underbrace{i,\dots, i}_{n_i} ) } \tilde \cT^i   .
\]
The two trees $\cT$ and $\cT'$ have the same number of vertices $|\cT'| = |\tilde \cT^i|+| \cT^i |=|\cT|$, and the vertices of the initial
 tree $\cT$ map one to one on the vertices of the f\/inal $\cT'$.  As none of the vertices of $\tilde \cT^i$ starts by a label $i$, it follows
that the maximal vertex of color $i$ in $\cT'$ is $( \underbrace{i,\dots,i}_{n_i} ) $. Thus:
\begin{gather*}
 (i,V) \in \cT  \leftrightarrow  (V) \in \cT'   , \  (V)\neq ( \underbrace{i,\dots, i}_{n_i} , U )  , \\
 (W)\in \cT , \  (W)\neq (i,V)  \leftrightarrow  ( \underbrace{i,\dots, i}_{n_i} , W ) \in \cT'  .
\end{gather*}
Most importantly, it is straightforward to check that the mapping is consistent with the successor functions:
\[
 V,W \in \cT \leftrightarrow V',W' \in \cT' \qquad \text{such that} \qquad s^i_{\cT}[V]=W \Leftrightarrow s^i_{\cT'}[V']=W'   .
\]
We will say that the two trees $\cT$ and $\cT'$ are {\it equivalent}, $\cT \sim \cT'$, and extend it by the transitivity  of $\sim$ to an equivalence
relation among rooted trees. An example is presented in Fig.~\ref{fig:equiDary}.

The equivalence class of a tree $[\cT]$ has at most $|\cT|$ members,  all of which are obtained by choosing a vertex $V=i,j,k, \dots, l$
in $\cT$ and performing the elementary operation $\sim$ on the colors~$i$ followed by~$j$ followed by $k$ and so on up to $l$.

 {\it Properties of colored rooted $D$-ary trees:}
We list a number of lemmas concerning the gluing of trees,~$\star$. All these properties can be readily understood in terms of the
 graphical representation of the trees. In the sequel, we shall deal with three trees~$\cT$,~$\cT_1$ and~$\cT_2$. We denote
by $( \underbrace{i,\dots, i}_{n_i} )$ the maximal vertex of color~$i$ in~$\cT_1$
and by $( \underbrace{i,\dots, i}_{q_i} )$ the maximal vertex of color~$i$ in~$\cT_2$.
The proofs of these lemmas are straightforward and all rely on carefully tracking the
canonically labeled trees under various joinings. They can be found in~\cite{Gurau:2011tj}.

\begin{lemma} \label{lem:lema1}
 If $ (V) =(k,U) \in \cT $ then  $ (\cT\star_V \cT_1)^k =\cT^k \star_U \cT_1$,
  and, for $i\neq k$, $ (\cT\star_{V} \cT_1)^i  = \cT^i$.
\end{lemma}

\begin{lemma}\label{lem:lema2}
  For any three trees, $(\cT \star_V \cT_1) \star_{V} \cT_2= \cT\star_{V} (\cT_1 \star_{(\;)} \cT_2) $.
\end{lemma}

\begin{lemma}\label{lem:lema3}
 For any two distinct vertices $V \neq W \in \cT$,
we denote $ W' $ the image of $W$ in the tree $ \cT \star_V \cT_1 $
and $V' $ the image of $V$ in $ \cT \star_ W \cT_2  $. Then
$(\cT \star_V \cT_1) \star_{W'} \cT_2= (\cT \star_ W \cT_2) \star_{V'} \cT_1$.
\end{lemma}

\begin{lemma}\label{lem:lema4}
  Let $V\in \cT$ and $W \in \cT_1$ then
$(\cT \star_V \cT_1) \star_{W'} \cT_2 = \cT \star_{V} (\cT_1 \star_{W} \cT_2)$, where
again $W'$ denotes the image of $W$ in the tree $ \cT \star_V \cT_1 $.
\end{lemma}

\begin{lemma}\label{lem:lema5}
  We have $(\cT_2 \star_{(\;)} \cT_1)^i \sim (\cT_1 \star_{(\;)} \cT_2)^i $.
\end{lemma}

{\it The Lie algebra indexed by colored rooted $D$-ary trees:}
We now def\/ine a Lie algebra of operators indexed by the trees.  We associate to every tree a variable $t_{\cT}$,
and we denote $|R_1|$ the valency of the root of $\cT_1$. Consider the operators $\cL_{\cT_1}$ def\/ined as:
\begin{gather}
 \cL_{\cT_1}  =   (-)^{|R_1|}N^{D- D |R_1|} \;
  \frac{\partial^{|R_1|}} {
\prod_{i, \cT_1^i \neq \varnothing }
\partial t_{\cT_1^i}}
+ \sum_{\cT} t_{\cT} \sum_{V \in \cT }  \frac{\partial}{\partial t_{\cT \star_V \cT_1}}   ,\nonumber\\
\cL_{ \{(\;) \} } =  \dsty N^D +  \sum_{\cT} t_{\cT} \sum_{V \in \cT }  \frac{\partial}{\partial t_{\cT} }   ,\label{eq:defcl}
\end{gather}
where $N$ is some parameter (destined to become the large $N$ parameter of the tensor model).
We will not consider the most general domain of these operators. Instead of def\/ining them for arbitrary functions
$f(t_{\cT})$, we restrict their domain to class functions $f(t_{[\cT]})$, with $t_{[\cT]} = \sum_{\cT'\sim \cT} t_{\cT'}$.

\begin{theorem}\label{th:comutator}
 When restricted to class functions, the operators $\cL_{\cT}$ form a Lie algebra with commutator:
\[
 \bigl[ \cL_{\cT_2} , \cL_{\cT_1} \bigr] f(t_{[\cT]})    = \sum_{V \in \cT_2 } \cL_{ \cT_2 \star_V \cT_1 }f(t_{[\cT]}) -
  \sum_{V \in \cT_1} \cL_{ \cT_1 \star_V \cT_2 }f(t_{[\cT]})   .
\]
\end{theorem}
The proof of this statement is given in \cite{Gurau:2011tj}. As it is a straightforward, albeit long computation
(using all the Lemmas \ref{lem:lema1}, \ref{lem:lema2}, \ref{lem:lema3}, \ref{lem:lema4}, \ref{lem:lema5}), we do not reproduce it here.

\subsection{From one to an inf\/inity of coupling constants}\label{ssec:infty}

In this section, we generalize the i.i.d.~colored tensor model with one coupling to an i.i.d.~model with an inf\/inity of
couplings and derive the SDEs of the general model.  Our modus operandi here is to integrate all colors save one and then
to `free' the couplings of the operators in the ef\/fective action for the last color.  To get down to business, on integrating
all colors save one, the partition function becomes:
\begin{gather*}
 Z  =  \int d\phi^D d\bar \phi^D \; e^{-S^D(\phi^D, \bar \phi^D)}   ,\\
 S^D(\phi^D, \bar \phi^D)  =   \sum \bar \phi^D_{ \vec n_D } \phi^D_{\vec n_D }
 + \sum_{ \cB^{\widehat{D}} } \frac{(\lambda\bar\lambda)^{p} }{ \text{Sym}(\cB^{\widehat{D}}) }
  \tr_{\cB^{\widehat{D}}} [\bar\phi^D, \phi^D ]
   N^{-\frac{D(D-1)}{2}p + \cF_{\cB^{\widehat{D}} } } ,
\end{gather*}
where the sum over $\cB^{\widehat{D}}$ runs over all connected vacuum graphs with colors $0,\dots, D-1$
(i.e.\ over all the possible $D$-bubbles with colors $0,\dots, D-1$) and $p$ vertices.
The operators $\text{Tr}_{\cB^{\widehat{D}} } [\bar\phi^D, \phi^D ] $ in the ef\/fective action for the last color
are the tensor network operators corresponding to the boundary graphs of bubbles, hence provide a natural set
of observables of the model.

Every vertex of $\cB^{\widehat{D}} $ is decorated by a tensor $\phi^D_{\vec n_D}$ or $\bar \phi^D_{ \bar {\vec  n}_D }$, and the tensor indices
are contracted as dictated by the graph $\cB^{\widehat{D}} $.  We denote by $v$, $\bar v$ the positive (resp. negative) vertices
of~$\cB^{\widehat{D}} $  and by~$l^i_{v\bar v}$ the edges (of color~$i$)
connecting the positive vertex $v$ with the negative vertex~$\bar v$.   The interaction operators are written as:
\begin{gather} \label{eq:tensnet}
 \text{Tr}_{\cB^{\widehat{D}}} [\bar\phi^D, \phi^D ] = \sum_{n}
 \left( \prod_{v,\bar v \in \cB^{\widehat{D}}} \bar \phi^D_{\bar {\vec   n}^{\bar v}_D } \phi^D _{\vec n^v_D } \right)
 \left( \prod_{i=0}^{D-1} \prod_{ l^i_{v\bar v} \in \cB^{\widehat{D}}  }
\delta_{n^{v }_{Di} {\bar n}^{\bar v }_{Di} } \right)   ,
\end{gather}
where all indices $n$ are summed. Note that, as all vertices in the bubble belong to an unique line of a given color, all
the indices of the tensors are paired. The scaling in $N$ of an operator is determined by the degree of the associated $D$-bubble graph:
\begin{gather*}
 N^{-\frac{D(D-1)}{2}p + \cF_{\cB^{\widehat{D}} }} = N^{-\frac{D(D-1)}{2}p + \frac{(D-1)(D-2)}{2} p + D-1-\frac{2}{(D-2)!}\omega(\cB^{\widehat D}) }
 = N^{ - (D-1) p + D-1 -\frac{2}{(D-2)!}\omega(\cB^{\widehat D}) }  ,
\end{gather*}
where we used equation \eqref{eq:faces2} in the f\/irst equality.
Thus, the ef\/fective action for the last color (dropping the index $D$) can be written as:
\begin{gather}\label{eq:actint}
 S^D ( \phi , \bar \phi) =  \sum \bar \phi_{ \vec  n }   \phi_{  \vec n }
 + N^{D-1} \sum_{ \cB }  \frac{ (\lambda \bar\lambda)^{ p  } }{ \text{Sym}( \cB  )}
  N^{-(D-1)p -\frac{2}{(D-2)!}\omega(\cB) } \,
\text{Tr}_{ \cB  }  [ \bar\phi  , \phi ].
\end{gather}
Attributing to each operator its own coupling constant and rescaling the f\/ield to $T =  N^{-\frac{D-1}{2}}  \phi$, we
obtain the partition function of a colored tensor model with generic potential:
\begin{gather}
 Z  =   e^{N^D F(g_{\cB } ) }= \int d\bar T dT \, e^{-N^{D-1} S(\bar T, T)}   ,\nonumber\\
 S (\bar T, T )  =    \sum \bar T_{\vec n} T_{\vec n} +   \sum_{ \cB  }  t_{ \cB  }
N^{ -\frac{2}{(D-2)!}\omega(\cB) } \, \text{Tr}_{ \cB } [ \bar T ,  T ]    .\label{eq:genmod}
\end{gather}

It is worth noting that, although in the end we deal with an unique tensor $T$, the colors are crucial to the
def\/inition of the tensor network operators in the ef\/fective action. The initial interaction term of the tensor model
described a $D$-simplex. These later tensor network operators describe (colored) polytopes in $D$ dimensions obtained
by gluing simplices along all save one of their faces around a point (dual to the bubble $\cB $).  This is in strict
parallel with matrix models, where higher degree interactions represent polygons obtained by gluing triangles around a vertex.

When evaluating amplitudes of graphs, obtained by integrating the last tensor $T$, the tensor network operators act
as ef\/fective vertices (for instance each comes with its own coupling constant). It is still more convenient, however,
to represent the Feynman graph of the path integral~\eqref{eq:genmod} using graphs with $D+1$ colors. The ef\/fective
 vertices are the subgraphs with colors $0,\dots,D-1$, and encode the connectivity of the tensor network operators.

The partition function  \eqref{eq:genmod} provides a natural set of observables for the model, the multi-bubble correlations, def\/ined as:
\begin{gather*}
  \Big\langle  \text{Tr}_{\cB_{(1)} } [\bar T, T] \, \text{Tr}_{\cB_{(2)}  } [\bar T, T]
\cdots \text{Tr}_{\cB_{(\rho )} } [\bar T, T] \Big\rangle
  =  \prod_{i=1}^{\rho} \left( -N^{- \bigl[D-1- \frac{2}{(D-2)!}\omega(\cB_{(i) }) \bigr]}
\frac{\partial} {\partial t_{ \cB_{(i)} } } \right) Z   .
\end{gather*}

When introducing an inf\/inity of coupling constants, we did not change the scaling with $N$ of the
operators. The graphs $\cG$ contributing to the connected multi-bubble correlations
are connected vacuum graphs with $D+1$ colors and with~$\rho$ marked subgraphs corresponding to the insertions
$\tr_{\cB_{(\nu)} } [\bar T, T]$. Taking into account the scaling of the insertions, the global scaling of such graphs is:
\begin{gather*} %\label{eq:scaling}
\left\langle  \text{Tr}_{\cB_{(1)}  } [\bar T, T] \, \text{Tr}_{\cB_{(2)}   } [\bar T, T]
\cdots \text{Tr}_{\cB_{(\rho )}  } [\bar T, T] \right\rangle_c \\
\qquad{}  \le  N^{D - \frac{2}{(D-1)!} \omega(\cG)} N^{- \rho (D-1) + \sum_{\rho} \frac{2}{(D-2)!} \omega(\cB_{(\rho)})}
\le N^{ D - \rho (D-1)  - \frac{2}{D!} \omega(\cG) }
  ,
\end{gather*}
where we use Lemma~\ref{lem:ddegg}.

In the large $N$ limit, the connected correlations receiving contributions from graphs of degree~$0$ (melonic graphs) dominate
the multi-bubble correlations. All their bubbles are necessarily melonic, in particular, the insertions $\text{Tr}_{\cB_{(i)}} [\bar T, T] $.
  As we have seen in Section~\ref{sec:melons}, the melonic $D$-bubbles (i.e.\ melonic graphs with $D$ colors) are in one-to-one correspondence
with colored rooted $D$-ary trees~$\cT$. The tensor network operators~\eqref{eq:tensnet} of melonic bubbles can be written directly in terms
of~$\cT$. When building a $D$-bubble
starting from~$\cT$, each time we insert a melon corresponding to a vertex $V\in \cT$ we bring a~$T$ and a~$\bar T$ tensor for the two external
 points of the melon. We denote the indices of  $T$ by $\vec n_V$ and the ones of~$\bar T$ by~$\bar {\vec  n}_V$.   We get:
\begin{gather*}
\text{Tr}_{ \cB } [ \bar T ,  T ] \equiv
\text{Tr}_{ \cT} [\bar T, T] = \prod_{V \in \cT } \left( T_{ \vec n_V }
\bar T_{ \bar {\vec  n}_{V} }
\prod_{i=0}^{D-1} \delta_{n^i_{V} \bar n^i_{s^i_{ \cT }[V] }  } \right)  ,
\end{gather*}
where $s^i_{\cT}$ is exactly the colored successor function def\/ined in Section~\ref{sec:algebra}. As this operator depends exclusively of
the successor functions, it is an invariant for an equivalence class of trees: $\cT\sim\cT' \Rightarrow \tr_{ \cT} [\bar T, T] = \tr_{ \cT'} [\bar T, T]$.
Hence, the action and the partition function depend only on the class  variables $t_{[\cT]} = \sum_{\cT'\sim \cT} t_{\cT}$. Taking into account that
the melonic bubbles have degree~$0$ (and redef\/ining the coupling of the tree $\cT=\{(\;)\}$), the action may be written as:
\begin{gather*}
 S (\bar T, T ) = \sum_{\cT}   t_{\cT }
\tr_{ \cT } [ \bar T ,  T ] + S^{r} (\bar T, T)   ,
\end{gather*}
where $S^{r}$ corresponds to non-melonic bubbles.

\subsection{Schwinger--Dyson equations}

Consider a melonic bubble corresponding to the tree $\cT_1$ with root $(\;)_1$. We denote:
\[
\delta^{\cT_1}_{n,\bar n} = \prod_{V\in \cT_1} \prod_{i=0}^{D-1} \delta_{n^i_{ V} \bar n^i_{s^i_{ \cT_1 }[V] } }   .
\]
The SDEs are deduced starting from the trivial equality:
\[
\sum_{\vec p ,n} \int  \frac{\delta}{\delta  T_{\vec p } }  \left[
 T_{  \vec n_{ (\;)_1 } } \delta_{ \bar {\vec  n}_{ (\;)_1} \vec p}
\left(  \prod_{  V_1 \in \cT_1 \setminus (\;)_1  } T_{ {\vec n}_{V_1} }
\bar T_{ \bar {\vec  n}_{V_1} } \right)
\delta^{\cT_1}_{n,\bar n}
e^{-N^{D-1} S } \right] =0  ,
\]
which expands as:
\begin{gather}
 \sum_n \int \left\{ \delta_{\vec n_{(\;)_1} \bar {\vec  n}_{(\;)_1}  }
 \left( \prod_{ V_1 \in \cT_1 \setminus (\;)_1   }  T_{ {\vec n}_{V_1} }
\bar T_{ \bar {\vec  n}_{V_1} } \right)
 \delta^{\cT_1}_{n,\bar n} \right.
\nonumber\\
\qquad{}
 + \sum_{V_2\neq (\;)_1}
T_{  \vec n_{(\;)_1} } \bar T_{ \bar {\vec  n}_{V_2} }  \delta_{\bar {\vec  n}_{(\;)_1} \vec n_{V_2}}
\left( \prod_{ V_1 \in \cT_1\setminus \{ (\;)_1,V_2 \}   } T_{ {\vec n}_{V_1} }
\bar T_{ \bar {\vec  n}_{V_1} } \right) \; \delta^{\cT_1}_{n,\bar n}\nonumber\\
\qquad{}
 - N^{D-1}   T_{  \vec n_{ (\;)_1} }  \delta_{ \bar {\vec  n}_{ (\;)_1} \vec p}
\left(  \prod_{ V_1 \in \cT_1\setminus (\;)_1   }  T_{ {\vec n}_{V_1} }
\bar T_{ \bar {\vec  n}_{V_1} } \right) \;\delta^{\cT_1}_{n,\bar n} \nonumber\\
\left. \qquad{}
\times
\left[ \sum_{\cT} t_{\cT} \sum_{V \in \cT }
  \bar T_{ \bar {\vec  n}_{V} }
\delta_{ \vec n_V  \vec p  }
 \left(  \prod_{ V' \in \cT \setminus V  }
T_{ \vec n_{V'} }  \bar T_{ \bar {\vec  n}_{V'}  } \right)    \delta^{\cT}_{n,\bar n}
+ \frac{\delta S^r}{\delta T_{\vec p}} \right] \right\}   e^{-N^{D-1}S }  .\label{eq:SDEfull}
\end{gather}
The second line in  \eqref{eq:SDEfull} represents graphs in which a line of color $D$ on a melonic
bubble connects the $\bar T$ on the root melon $(\;)$ to a $\bar T$ on a distinct
melon~$V_2$. Hence, it can not be a melon (see the end of Section~\ref{sec:melons}).
The last term represents a melonic bubble connected via a line to a~non-melonic bubble (coming from $\frac{\delta S^r}{\delta T}$).
Thus, it cannot be a melon either. Taking into account that we have one line explicit in both graphs,
(hence a factor $N^{-(D-1)}$), and that the scaling of $\tr_{\cT} [T,\bar T]$ is $N^{D-1}$, we f\/ind that in both cases the correlations scale at most like
\begin{gather}\label{eq:nonmelonbound}
\frac{1}{Z} \big{\langle} \cdots \big{\rangle} \le N^{D-\frac{2}{(D-1)!} }   .
\end{gather}

The f\/irst term in~\eqref{eq:SDEfull} factorizes over the branches $\cT_1^i$ of $\cT_1$. We denote by $(\;)_{1,i}$ the root of the branch
$\cT_1^i$. For a non-empty branch $\cT^i$, recall that the vertex $  s^i_{\cT}[(\;)_1]= (i)\in \cT$ maps
to the root $ (\;)_{1,i} \in \cT^i $. For each branch, we evaluate:
\begin{gather*}
 \sum_{ n^i_{R_1}, {\bar n}^i_{ (\;)_1}  } \delta_{ n^i_{ (\;)_1} {\bar n}^i_{ (\;)_1}  }
\delta_{n^i_{ (\;)_1} \bar n^i_{s^i_{\cT_1}[ (\;)_1 ]} }
 \delta_{ n^i_{ [s^i_{\cT}]^{-1}[ (\;)_1]} \bar n^i_{ (\;)_1} } \\
\qquad{} =
\begin{cases}
   N & \text{if } s^i_{\cT}[ (\;)_1]= (\;)_1, \\
\displaystyle \delta_{ n^i_{ [s^i_{\cT}]^{-1}[ (\;)_1]} \bar n^i_{s^i_{\cT_1}[ (\;)_1 ]}   } =
\delta_{ n^i_{ [s^i_{\cT^i}]^{-1}[ (\;)_{1,i} ] }  \bar n^i_{ (\;)_{1,i} } } & \text{if not.}
\end{cases}
\end{gather*}
Thus, denoting the valency of the root $(\;)_1\in \cT_1$ by $|R_1|$, we get:
\[
 \sum_{\vec n_{R_1}, \bar {\vec } n_{R_1}} \delta_{\vec n_{R_1} \bar {\vec  n}_{R_1}  } \delta^{\cT_1}_{n,\bar n}
= N^{D-|R_1|} \prod_{ \stackrel{i=0}{\cT^i_1\neq\varnothing }}^{D-1} \delta^{\cT_1^i}_{n,\bar n}  .
\]
The third term in  \eqref{eq:SDEfull} simplif\/ies:
\begin{gather*}
\sum_{ \bar {\vec  n}_{ (\;)_1}  \vec n_V } \delta_{\bar {\vec  n}_{ (\;)_1}  \vec n_V }
\delta^{\cT_1}_{n, \bar n} \delta^{\cT}_{ n,\bar n}
 =\sum_{ \bar {\vec  n}_{ (\;)_1}  \vec n_V } \delta_{\bar {\vec  n}_{ (\;)_1}  \vec n_V }
\prod_{i=0}^{D-1}  \delta_{   n^i_{ [s^i_{\cT_1}]^{-1}[ (\;)_1]} \bar n^i_{(\;)_1} }
\prod_{i=0}^{D-1} \delta_{ n^i_V \bar n^i_{s^i_{\cT} [V] } }
\; \delta^{\cT_1\setminus (\;)_1 }_{n,\bar n}\delta^{\cT\setminus V}_{n,\bar n} \\
\hphantom{\sum_{ \bar {\vec  n}_{ (\;)_1}  \vec n_V } \delta_{\bar {\vec  n}_{ (\;)_1}  \vec n_V }
\delta^{\cT_1}_{n, \bar n} \delta^{\cT}_{ n,\bar n}}{}
= \prod_{i=0}^{D-1} \delta_{n^i_{ [s^i_{\cT_1}]^{-1}[ (\;)_1]}  \bar n^i_{s^i_{\cT} [V] } }
\delta^{\cT_1\setminus (\;)_1 }_{n,\bar n}\delta^{\cT\setminus V}_{n,\bar n}
=\delta^{\cT \star_V \cT_1}_{n,\bar n}   .
\end{gather*}
Hence, the SDEs,  for every rooted tree $\cT_1$ with $|R_1|$ non-empty branches, may be written starting from the root as:
\begin{gather*}
N^{D-|R_1|}\left\langle  \prod_{ \stackrel{i=0}{\cT_1^i \neq \varnothing}}^{D-1}\text{Tr}_{\cT_1^i}[\bar T, T] \right\rangle
-N^{D-1} \sum_{\cT}  t_{\cT}  \sum_{V \in \cT }
\Big{ \langle } \text{Tr}_{\cT \star_{V} \cT_1}[\bar T, T] \Big{ \rangle }  =  \big{\langle} \cdots \big{\rangle},
\\
(-)^{|R_1|}N^{D- D |R_1|}
\left(  \frac{\partial^{|R_1|}} {
\prod_{i, \cT_1^i \neq \varnothing }
\partial t_{\cT_1^i}}  \right) Z
+ \sum_{\cT} t_{\cT} \sum_{V \in \cT }  \frac{\partial}{\partial t_{\cT \star_V \cT_1}} Z  =
\big{\langle} \cdots
 \big{\rangle}   ,
\end{gather*}
where $\big{\langle} \cdots \big{\rangle} $ denotes the non-melonic terms of  \eqref{eq:nonmelonbound}.
Taking into account the def\/inition of $\cL_{\cT_1}$ in~\eqref{eq:defcl}, we spot that:
\[
\cL_{\cT_1}Z = \left\langle \cdots \right\rangle \ \Rightarrow \ \lim_{N\to \infty}
\left( N^{-D}   \frac{1}{Z}  \cL_{\cT_1} Z \right)= 0  ,  \qquad \forall \, \cT_1  \; .
\]
Recall that $Z = e^{N^DF(t_{\cB})}$ depends only on class variables $t_{[\cT]}$. At leading order in~$1/N$, only melonic
graphs contribute to the free energy $F(t_{\cB})$,  hence $\lim\limits_{N\to\infty} F(t_{\cB})= F_{\infty} ( [t_{[\cT]}])$.
The SDEs at leading order imply:
\begin{gather} \label{eq:SDfree}
\prod_{ \stackrel{i=0}{\cT_1^i \neq \varnothing}}^{D-1}
 \left(- \frac{\partial F_{\infty}(t_{[\cT]})}{\partial t_{\cT_1^i}} \right)  + \sum_{\cT} t_{\cT} \sum_{V \in \cT }
   \frac{\partial F_{\infty}(t_{[\cT]})}{\partial t_{\cT \star_V \cT_1}} =0  , \qquad \forall\, \cT_1     .
\end{gather}

The most useful way to employ the SDEs is the following. Consider a class function $\tilde Z =e^{-N^{D} \tilde F}$
satisfying the constraints at all orders in $N$, $\cL_{\cT_1} \tilde Z =0$. Its free energy in the large $N$ limit is
$\tilde F_{\infty}(t_{[\cT]}) = \lim\limits_{N\to \infty} \tilde F(t_{[\cT]})$ and respects  \eqref{eq:SDfree}.
As a result,  $ \tilde F_{\infty}(t_{[\cT]}) = F_{\infty}(t_{[\cT]})$, that is, the $N\to \infty$ limit of $\tilde Z$ and $Z$ coincide.

Note that an SDE at all orders can be derived for the trivial insertion:
\[
  \sum_{\vec p} \int  \frac{\delta}{\delta  T_{\vec p } }
\Big{[} T_{  \vec p }
e^{-N^{D-1} S } \Big{]} =0 \ \rightarrow \
\left( N^D + \sum_{\cB} p    t_{\cB }   \frac{\partial}{\partial t_{\cB}}  \right) Z=0  ,
\]
where $p$ denotes, as usual, half the number of vertices of the bubble $\cB$. The above operator, which at leading order is~$\cL_{(\;)}$,
should be identif\/ied with the generator of dilations~\cite{Dijkgraaf:1990rs}.

\section{Non-trivial classical solutions}\label{sec:solutions}

It seems unlikely that we shall ever have at our disposal an exhaustive catalog of classical solutions to these models,  since most quantum
f\/ield theories are in a similar position. Quite the opposite situation is more prevalent; it is non-trivial, especially for scalar theories, to
 have any solutions at all in closed form (other than the trivial solution).

Here, we shall formulate an ansatz to extract classical solutions to colored tensor models. On the one hand, the ansatz is rather general in the
 sense that it appears to be insensitive to the specif\/ic details of the covariance and vertex kernels, although we shall only apply it explicitly
 to the i.i.d.~model.  On the other hand, it is just an ansatz and so for a specif\/ic model it only captures a small fraction of the solution space.
It is based upon a splitting of the tensor model structure into a background and a foreground.  The PJ-factorizations of Section~\ref{ssec:factor} enable this mechanism.

In fact, we are not just interested in acquiring non-trivial classical solutions. In the sequel, once we have found explicit classical solutions
for the background, we shall be interested in analyzing the quantum theory resulting from f\/luctuations around this classical point.   It emerges
 that a fruitful area of research lies not just in analyzing the most generic f\/luctuations but also more restricted scenarios.

\subsection{Ansatz set-up}

The ansatz has three distinct steps: splitting the degrees of freedom, solution f\/inding and quantum analysis.

 {\it Splitting degrees of freedom:}
Sticking with the i.i.d.~probability measure, as usual, the f\/irst step we take is to separate the degrees of freedom of the tensor model
 into two types, {\it background} and {\it foreground}.  We recall from Section~\ref{sec:embed} that  a neat way to do this splitting is
 to utilize a PJ-factorization.  This packages the degrees of freedom of the tensor model into sets pertaining to jackets and patches.
  From this, one can split the jackets (and patch) into background jackets and foreground jackets. Let us take $D=3$ as an example. Every
PJ-decomposition contains one jacket and one patch. Thus, one has two choices, once a given factorization is chosen: either the jacket
forms the background and the patch the foreground or vice versa.

 {\it Classical solution:}
As a next step, one attempts to f\/ind solutions to the theory in the sector where one has trivialized the foreground.  In essence, one looks
for solutions to a lower rank tensor theory albeit still with $D+1$ colors.    In realistic terms, of course, we are not ready to examine
all such sectors.  Our algebraic techniques are suitably advanced only in the areas of vector and matrix models, that is, backgrounds consisting
of a patch or a jacket.  More involved backgrounds are still out of our reach.  For $D=3$ or $4$, however, these simpler backgrounds are suf\/f\/icient
 to exhaust the ansatz given above.

 {\it Effective quantum theory:}
The power of the splitting becomes amplif\/ied once one returns to the quantum theory.  It allows one to parse systematically through various
 types of quantum f\/luctuation.

The f\/irst and most general f\/luctuation corresponds simply to expanding the original tensor about our new solution.  In this way,   we do
 not depart from our original theory, we rather view it in a new and hopefully revealing light.

It is also interesting to consider less general quantum f\/luctuations. Perhaps the least general corresponds to restricting them to consist
of only foreground degrees of freedom.  For the i.i.d.~model, as we shall see shortly, these quantum f\/luctuations decouple from the background.

There exist also some intermediate scenarios.  We can choose to allow a mixture of background and foreground f\/luctuations.  This allows for
 some back-reaction onto the background.  In fact, once one has a specif\/ic background, one can analyze f\/luctuations corresponding to an
arbitrary collection of jackets.  These jackets need not even be in the original PJ-factorization.  Such a~jacket contains degrees of
freedom which are partially foreground and partially background with respect to the original slicing.

\begin{remark}
We refer the reader to \cite{dev2} and~\cite{livorirya}, where a systematic analysis of the quantum f\/luctuations around classical solutions
 of the Boulatov--Ooguri model in 3 dimensions was conducted.  In both cases, the background is a patch, while the foreground is a jacket.  Due to
the non-trivial covariance of the model, one f\/inds that even purely foreground f\/luctuations do not decouple from the background. Interestingly,
these jacket f\/luctuations can be re-interpreted as the quantum degrees of freedom of a non-commutative massive scalar f\/ield theory~\cite{prriii}.
Moreover, expanding the action to quadratic order in the f\/luctuating f\/ields, these jacket f\/luctuations along with pure patch f\/luctuations provide
 a basis for the full quantum f\/luctuations. In other words, at quadratic order,  the quantum f\/luctuations are diagonalized in terms of
 a PJ-factorization.  In the full action of course, interference occurs due to higher order interaction terms.
\end{remark}

\begin{remark}
These classical solutions are often referred to as instantonic solutions in the group f\/ield theory literature, although this is a slight misnomer.
Genuine instantons in f\/ield theory lie at minima of the potential, meaning that the value of the action at these solutions is {\it less}
than that for the null conf\/iguration.  Thus, they contribute heavily to the path integral, although this would be missed at low orders in
the na\"ive perturbative expansion.  It makes sense then to treat the quantum theory as a perturbation about these instantonic solutions,
which provide the dominant contribution to the semi-classical physics.  We are not in this scenario here. Our classical solutions are
 not minima of the potential.  To see this, one can appeal to the derivation of the ef\/fective action in Section~\ref{ssec:infty}.  One
is left with a single color, $0$~say, once all the other colors $\{1,\dots, D\}$ have been integrated out.  As we saw in~\eqref{eq:actint},
 all the coupling constants $(\lambda\bar\lambda)^j/j!$ are positive.  Thus, the true minimum of the ef\/fective potential is $\phi^{0} = 0$.
 This is also the case for the simplest matrix models, but one may alter the coupling constants in the ef\/fective action to get to a dif\/ferent theory with true instantonic
 contributions. In other words, some of the coef\/f\/icients need to become negative.  To see that we are not hiding anything by integrating out
 these f\/ields, note that the positivity of the coef\/f\/icients emerges even after integrating out just one color, $D$ say.  The potential is once
again positive def\/inite and the minimum is at: $\phi^i = 0$ for $i\in\{0, \dots, D-1\}$.
\end{remark}

\subsection{Patch background}

Let us consider the classical equations of motion of the $(D+1)$-colored i.i.d.~model.  There are $2D+2$ such equations:
\[
\frac{\delta S}{\delta \phi^i} = 0 = \frac{\delta S}{\delta\bar \phi^i}  ,
\]
which expand as:
\[
\sum_{\vec{n}_i} \phi^i_{\vec{n}_i}\; \delta_{\vec{n}_i,\bar {\vec {n}}_i} + \frac{\bar\lambda}{N^{\frac{D(D-1)}{4}}}
 \prod_{j\neq i}\sum_{\bar {\vec {n}}_j} \bar\phi_{\bar {\vec {n}}_j}^j \; K_{\bar {\vec {n}}_0\dots \bar {\vec {n}}_D} = 0
\]
and their conjugates.  In its present incarnation, we have not lost the manifest division of degrees of freedom among the jackets of
 a PJ-factorization as in~\eqref{eq:pjfactorize}.  Thus, having decided on a background, we can pick solutions which trivialize the
foreground sector.  The simplest case is where the background corresponds to the patch in a PJ-factorization.  To possess a patch, $D$ must
be odd. We choose the specif\/ic PJ-factorization constructed earlier in Proposition \ref{lem:pjfactor} and consider solutions of the form:
\[
\Phi^i_{\bar {\vec {n}}_i} = f^i_{n^{ii+\frac{D+1}{2}}}  \delta_{n_\cJ} ,
\]
where $\delta_{n_\cJ}$ is shorthand for a product of identity matrices  associated to the background jackets  in our PJ-factorization. This is our ansatz.
 These functions must satisfy conditions imposed by the equations of motion:
\begin{gather} \label{eq:eompatch}
f^i = \frac{\bar\lambda}{N^{\frac{D(D-1)}{4}}}\left[\prod_{\substack{j = 0\\j\notin\{ i, i+\frac{D+1}{2}\}}}^{(D-1)/2} \bar f^j
\cdot \bar f^{j+ \frac{D+1}{2}}  \right] \bar f^{i+\frac{D+1}{2}},
\end{gather}
where we have introduced vectors $f^i = f^i_{n^{i i + \frac{D+1}{2}}}$.
One can explicitly realize functions satisfying these equations but we shall leave our treatment  in a  rather generic state here.
   Now, one can expand the f\/ield theory about this non-trivial classical solution as  $\phi^i = \Phi^i + \varphi^i$, leading to:
\begin{gather*}
S_\Phi[\varphi^i,\bar\varphi^i]  \equiv  S[\phi^i,\bar\phi^i] - S[\Phi^i,\bar\Phi^i]\\
\hphantom{S_\Phi[\varphi^i,\bar\varphi^i]}{}
=  \left[\sum_{m = 1}^{D}
\prod_{l = 0}^m\sum_{j_l = 0}^D \varphi^{j_l} \frac{\delta }{\delta \phi^{j_l}} \right]
\left[ \sum_{\bar m = 1}^{D} \prod_{\bar l = 0}^{\bar m}\sum_{\bar j_{\bar l} = 0}^D
\bar\varphi^{\bar j_{\bar l}} \frac{\delta }{\delta \bar\phi^{\bar j_{\bar l} }}  \right]
S[\phi^i, \bar\phi^i]\Bigg|_{ \substack{\phi = \Phi,\\ \bar\phi = \bar\Phi}}\\
\hphantom{S_\Phi[\varphi^i,\bar\varphi^i]}{}
 =  \sum_i \varphi^i  \bar\varphi^i +\sum_i \varphi^i  C_{\Phi}^i  \varphi^{i+\frac{D+1}2}
+ \sum_i \bar\varphi^i  \bar C_{\Phi}^i  \bar \varphi^{i+\frac{D+1}2}  +  \textrm{higher order},
\end{gather*}
is the action with respect to this background $\Phi$ and we have used the equations of motion~\eqref{eq:eompatch} to remove the terms linear in $\varphi^i$ and $\bar\varphi^i$. Most clearly, we see that we now have a f\/ield theory with a non-trivial
covariance $C_{\Phi}$ dependent upon the background~$\Phi$.  This could be of use when one wishes to devise a renormalization prescription
for the theory~\cite{livorirya, GFTren}. Of course, one can restrict  to purely foreground f\/luctuations. However, in this general form one
 does not see anything particularly striking. We refer the reader once again to~\cite{dev2, livorirya} for specif\/ic results relating such
quantum f\/luctuations to matter f\/ield theories and the Hamiltonian constraint operator of quantum gravity in three dimensions.

\begin{remark}
As already mentioned, one need not take the background to be a patch.  The next simplest case, corresponds to a background consisting of a single
 jacket, that is a single Riemann surface, and is possible in any
 dimension $D$.  One needs to solve a set of coupled algebraic matrix equations in that case.
\end{remark}

\section{Extended discussion and conclusion}\label{sec:conclu}

We have by no means attempted to compile a historical or complete review of higher-dimensional tensor models.  On the contrary, we wished to commence from what we believe
are the core mathematical precepts upon which the whole theory is based.  Thence, we felt it was perhaps more benef\/icial that the review developed in a rather linear deductive style, obviously guided by particular prescribed motivations.  While these of course are bound to be somewhat idiosyncratic but they are by no
means uncommon; they are some of the basic questions one should expect a~quantum f\/ield theory to answer.  With these provisos, we have pushed right up to the cutting edge.
It does mean, however,  that we arrive at a rather singular focus, perhaps at the expense of other topics.  With this in mind, we present here a short summary of results that lie at a slight tangent to, although by no means disconnected from, the main text.
We divide this section in three parts. The f\/irst deals with further developments on critical behavior(s) in colored tensor models. The second
details the connection between colored tensor models and spin foams. The third and f\/inal part draws the conclusions of our work.

\subsection{Critical behavior in more complex models}

Now that the main ideas behind the large $N$ behavior of colored tensor models are in place, it is only natural to apply
them to particular models describing specif\/ic physical situations. In particular a very active research direction is
to identify more complex critical behavior in various models.

 {\it The Ising model on a random lattice in higher dimensions.}
One of the most impressive early successes of random matrix models was the analytic solution by Kazakov et al.~\cite{Kazakov:1986hu},
followed by~\cite{Boulatov:1986sb, Brezin:1989db},
of the Ising model on a random planar surface.
In particular, the solution provides analytic proof that the system undergoes a third order phase transition between the low tempe\-ra\-tu\-re magnetized phase
and the high temperature non-magnetized phase. A similar study~\cite{bonzom:matter} can now be undertaken in arbitrary dimension for the Ising model
on a random melonic lattice. As we explain below, in the continuum limit, the spin system {\it does not} exhibit a phase transition at f\/inite temperature,
 which is in agreement with numerical investigations.

The two-matrix model \cite{Kazakov:1986hu} used to describe an Ising spin system generalizes to a two tensor model, both colored,
$(X^i,Y^i)$ (and $(\bar X^i,\bar Y^i)$) def\/ined by the partition function:
\begin{gather}
  e^{  N^D F(x,\bar x, y, \bar y, c)}  =
Z(x,\bar x, y, \bar y, c) = \int   d\bar X   d X  d\bar Y   d Y
\, e^{- S (X, \bar X , Y  \bar Y )}  , \nonumber  \\
S (X, \bar X , Y,  \bar Y )  =  \sum_{i=0}^{D} \sum_{\vec k_i}   \Big [ \bar X^i_{\vec k_i} X^i_{\vec k_i}
+ \bar Y^i_{\vec k_i} Y^i_{\vec k_i}  -
c  \bar X^i_{\vec k_i} Y^i_{\vec k_i}  - c \bar Y^i_{\vec k_i} X^i_{\vec k_i}   \Big]
\nonumber  \\
\hphantom{S (X, \bar X , Y,  \bar Y )  =}{} +  \frac{x}{ N^{D(D-1)/4} } \sum_{\vec k} \prod_{i=0}^D X^i_{ \vec k_i } +
\frac{\bar x}{ N^{D(D-1)/4} } \sum_{\vec k}
\prod_{i=0}^D \bar X^i_{ \vec k_i }   \nonumber  \\
\hphantom{S (X, \bar X , Y,  \bar Y )  =}{}
 +  \frac{y}{ N^{D(D-1)/4} } \sum_{\vec k} \prod_{i=0}^D Y^i_{ \vec k_i } +
\frac{\bar y}{ N^{D(D-1)/4} } \sum_{\vec k}
\prod_{i=0}^D \bar Y^i_{ \vec k_i }   .
 \label{eq:ising-tensor}
\end{gather}

The Feynman graphs of the action \eqref{eq:ising-tensor} are identical to the ones of
the i.i.d.~model, up to the fact that a vertex now involves either $X$ or $Y$ tensors (both coming with
positive and negative orientations). Tensor indices $k_{ij}$ and colors are preserved along the strands, and the propagator between the $X$ and $Y$ sectors is
\[
C = \frac1{1-c^2} \begin{pmatrix} 1&c\\c&1\end{pmatrix}  .
\]
Lines in a graph can join a f\/ield $\bar X$ (resp. $\bar Y$) to a f\/ield $X$ (resp.~$Y$), with weight $1/(1-c^2)$,
or a f\/ield $\bar X$ (resp.~$\bar Y$) to $Y$ (resp. $X$), with weight $c/(1-c^2)$. This is the only coupling
between $X$-tensors and $Y$-tensors. As in the i.i.d.~model, the free energy organizes
in powers of $1/N$, with melonic graphs dominating the large $N$ limit, so that
\[
F (x,\bar x, y, \bar y, c)
=\sum_{\text{melons}\,\cG} s(\cG)\ x^{p_x} \bar x ^{p_{\bar x}} y^{p_y} \bar y^{p_{\bar y} }
  \frac{ c^{L_{\bar XY} + L_{X \bar Y}} }{(1-c^2 )^{L} }  + O\big(N^{-1}\big)   ,
\]
with $p_x$ (resp.~$p_{\bar x}$, $p_y$ and $p_{\bar y}$) the number of~$X$ (resp.~$\bar X$ $Y$ and $\bar Y$) vertices,
$L$ is the total number of lines, $L_{\bar X Y}$ (resp.~$L_{X\bar Y}$) is the number of lines from a $\bar X$ to a $Y$
(resp. from a~$X$ to a~$\bar Y$) and~$s(\cG)$ is a symmetry factor. We have $L=(D+1)(p_x+p_y)$, $p_x+p_y = p_{\bar x} + p_{\bar y}$.

The $X$-vertices and $Y$-vertices represent the Ising spins on random lattices. One, say $X$, mimics the
random world of the up spins, while the other the random world of the down spins. On a graph~$\cG$, the total number
of vertices $n$ is the number of Ising spins. Through the propagator, two parallel spins are coupled with the
same weight $1/(1-c^2)$ and two anti parallel spins with $c/(1-c^2)$, from which one can extract the Ising coupling $\beta J$
where $\beta$ is the inverse temperature. Taking $c\in[0,1]$ corresponds to the ferromagnetic model $J>0$.
Setting $x=\bar x$ and $y=\bar y$, the free energy of the tensor model equates the grand canonical partition function
of the Ising model on random lattices:
\begin{gather*}
  Z(\mu, \beta J,\beta h)  =  \sum_{n} e^{-n\mu}  Z_n(\beta J, \beta h) ,\\
Z_n(\beta J, \beta h) = \sum_{\cG, 2(p_x+p_y) = n} s(\cG)  \sum_{s_i=\pm 1} e^{\beta J \sum_{\langle i,j\rangle } s_i s_j +\beta h\sum_{i} s_i } .
\end{gather*}
$Z_n$ is the canonical partition function for the Ising system on random lattices with a f\/ixed number of
spins~$n$, $h$ is the magnetic f\/ield and~$\mu$ the chemical potential. The relationship between the Ising parameters
and those of the tensor model is:
\[
e^{ - 2 \beta J} =c ,\qquad e^{-\mu} = \left( \frac{ 1 } { c^{D+1}  (1-c^2)^{ D+1 } }
\sqrt{x y}\right)^{1/2}   , \qquad e^{\beta h} = \left(\frac{ x}{ y  } \right)^{1/2}
 .
\]
As the model depends on two complex tensors, there are four types of 2-point functions,
denoted $\big\langle\bar X X \big\rangle$, $\big\langle\bar Y Y\big\rangle$,
$\big\langle\bar X Y\big\rangle$ and $\big\langle\bar Y X \big\rangle$, with which  one must deal.
In the melonic sector, the self-energy factorizes into convolutions of 2-point functions. Some algebra
leads one to a system of equations for the two point functions (at zero magnetic f\/ield),
$ \big\langle\bar X X \big\rangle = \big\langle\bar Y Y\big\rangle =U$ and
$\big\langle\bar X Y\big\rangle = \big\langle\bar Y X \big\rangle= V$:
\begin{gather}
 U - cV - g \bigl( U^{D+1} + V^{D+1} \bigr) =1, \qquad
  V - cU  - g UV \bigl( U^{D-1} + V^{D-1}\bigr) =0   ,\label{eq:syst}
\end{gather}
The solution for the noninteracting model (g=0) at arbitrary temperature is:
\[ %\label{eq:condinit}
   U(c,0) = \frac{1}{1-c^2}, \qquad V(c,0) = \frac{c}{1-c^2}   ,
\]
which is a boundary condition for the system of equations~\eqref{eq:syst}. For arbitrary dimensions, one can obtain a parametrized
solution of~\eqref{eq:syst}. To see this, we introduce the following pair of variables:
\[
 z = g^{1/D} U   ,\qquad w = \sqrt{1- \frac{1}{U(1-gU^D)}}   .
\]
and rewrite equations \eqref{eq:syst} as:
\[ %\label{maineq}
 c = (1-z^D) w - z^D w^D   , \qquad g^{1/D} = z(1-z^D) (1-w^2)  .
\]
For suf\/f\/iciently small values $g$, the free energy (related to $z$ in the new parameterization) is analytic. We now
send the lattices to criticality. We obtain a critical curve~$g_c(c)$, on which the free energy
$F(c)\equiv F(c,g_c(c))$ describes the Ising system in the large volume limit as a function of the temperature.

The critical curve is def\/ined by the vanishing of $ \big(\partial g/\partial z\big)_c$, so that the expansion of~$g$ in
terms of~$z$ only begins at the second order. This equation has four solutions $z_{1,2,3,4}(w)$, and one can show that the physical
solution~$z_1(w)$ is isolated form the others at f\/inite temperature, hence the free energy is analytic for
all $0\le c <1$. In the  {\it infinite} temperature limit~$c\to 1$, the physical root~$z_1(w)$
collapses to~$0$ and the free energy acquires a non-analytic behavior.
The Ising spin system on random melonic lattices undergoes a phase transition at
inf\/inite temperature.

 {\it Dually weighted models.}
A second model, which we shall discuss here, consists of a class of colored tensor models with a modif\/ied propagator
which allows us to associate weight factors to the faces of the graphs~\cite{EDT}.
Such models correspond to dynamical triangulations in three and higher dimensions
with generalized amplitudes. We present a model which undergoes a~third-order phase transition in the
continuum characterized by a jump in the susceptibility exponent.
This model is inspired by the dually weighted matrix models introduced by Kazakov et al.~\cite{Benedetti:2008hc,
Kazakov:1995gm,Kazakov:1995ae,Kazakov:1996zm,Szabo:1996fj}.
 The dually weighted colored tensor model in dimension $D$ is def\/ined by the partition
function:
\begin{gather*}
  e^{ N^D E }  =  Z_N(\lambda, \bar{\lambda}) = \int   d\bar \psi   d \psi
\, e^{- S (\psi,\bar\psi)}   , \\
  S (\psi,\bar\psi)  =  \sum_{i=0}^{D} \sum_{ {\vec p}_i, \bar {\vec n}_i } \psi^i_{ {\vec p}_i} \left(
\prod_{j} (C^{-1})_{ p_{ij}  \bar n_{ij}  } \right)  \bar \psi^i_{\bar {\vec  n}_i}
\\
\hphantom{S (\psi,\bar\psi)  =}{} +
\frac{\lambda}{ N^{D(D-1)/4} } \sum_{ n} \prod_{i=0}^D \psi^i_{ \vec n_i } +
\frac{\bar \lambda}{ N^{D(D-1)/4} } \sum_{ \bar n}
\prod_{i=0}^D \bar \psi^i_{ \bar {\vec n}_i }   . %\label{eq:EDT}
\end{gather*}
Using the counting of faces of a colored graph established for the i.i.d.~model, the amplitude of a graph can be written:
\[
  A(\cG) = (\lambda \bar \lambda)^p \left(  \prod_{(ij), \rho}
\frac{\text{Tr}[ (CC^T)^{p^{ij}_{(\rho)} }]}{N} \right)
N^{ D - \frac{2}{(D-1)!} \omega(\cG) }   ,
\]
where $2 p^{ij}_{(\rho)}$ is the number of vertices of the $\rho$'th face with colors $(ij)$ and  $C^T$ is the matrix transpose $C$.
The full connected 2-point function of the dually weighted model:
\[
 \langle  \bar \psi^i_{ \bar {\vec  n}_i} \psi^i_{ {\vec p}_i}   \rangle
 =  \prod_{j} P_{  \bar n_{ij}  p_{ij}  }(g,C)   ,
\]
is independent of the color $i$ and factorized along strands. Surprising as it may seem, the self-consistency equations in
the melonic sector can be solved analytically for~$P$~\cite{EDT}. A particular model one can consider is def\/ined by a covariance $C$ such that:
\[ %\label{betaC}
 \text{Tr}\big[ (C^TC)^q\big] = N q^{-\beta}  .
\]
Such a choice corresponds to the DT amplitude:
\begin{gather} \label{eq:ampliC}
A(\cG) = e^{\kappa_{D-2}N_{D-2}  -  \kappa_D N_D} \prod_{i} q_i^{-\beta}   ,
\end{gather}
where the product is over all $(D-2)$-dimensional simplices (bones) of the triangulation, with~$q_i$ being the number of
 $D$-simplices to which the bone $i$ belongs. The DT amplitude~\eqref{eq:ampliC} was studied via numerical simulations in~\cite{Bilke:1998vj},
and more recently in~\cite{Laiho:2011ya}.

The self-consistency equation (and its physical initial condition) appear in the melonic sector
(denoting $g\partial_g  E  \equiv  U $) as:
\begin{gather}
  U  =    S\bigl(\beta, z(g,U) \bigr)   ,  \qquad
 S(\beta,z)  =  \sum_{q=1}^\infty \frac{1}{q^{\beta} (q+1) } \binom{2q}{q} z^q   , \nonumber\\
z(g,U)  =  \frac{ U^{1-\frac{2}{D(D+1) } }   }
{   ( 1+ U  )^{2 \frac{D-1}{D}}  }   g^{\frac{2}{D(D+1)}}   , \qquad
g( \beta, 0  )  = 0   .\label{eq:sys}
\end{gather}
We shall denote the solution of the equation \eqref{eq:sys} by  $g(\beta, U)$. The series $S(\beta,z)$ has radius of convergence:
$z_b=\frac{1}{4}$ for all values of~$\beta$.
\begin{figure}[htb]\centering
 \includegraphics[width=5cm]{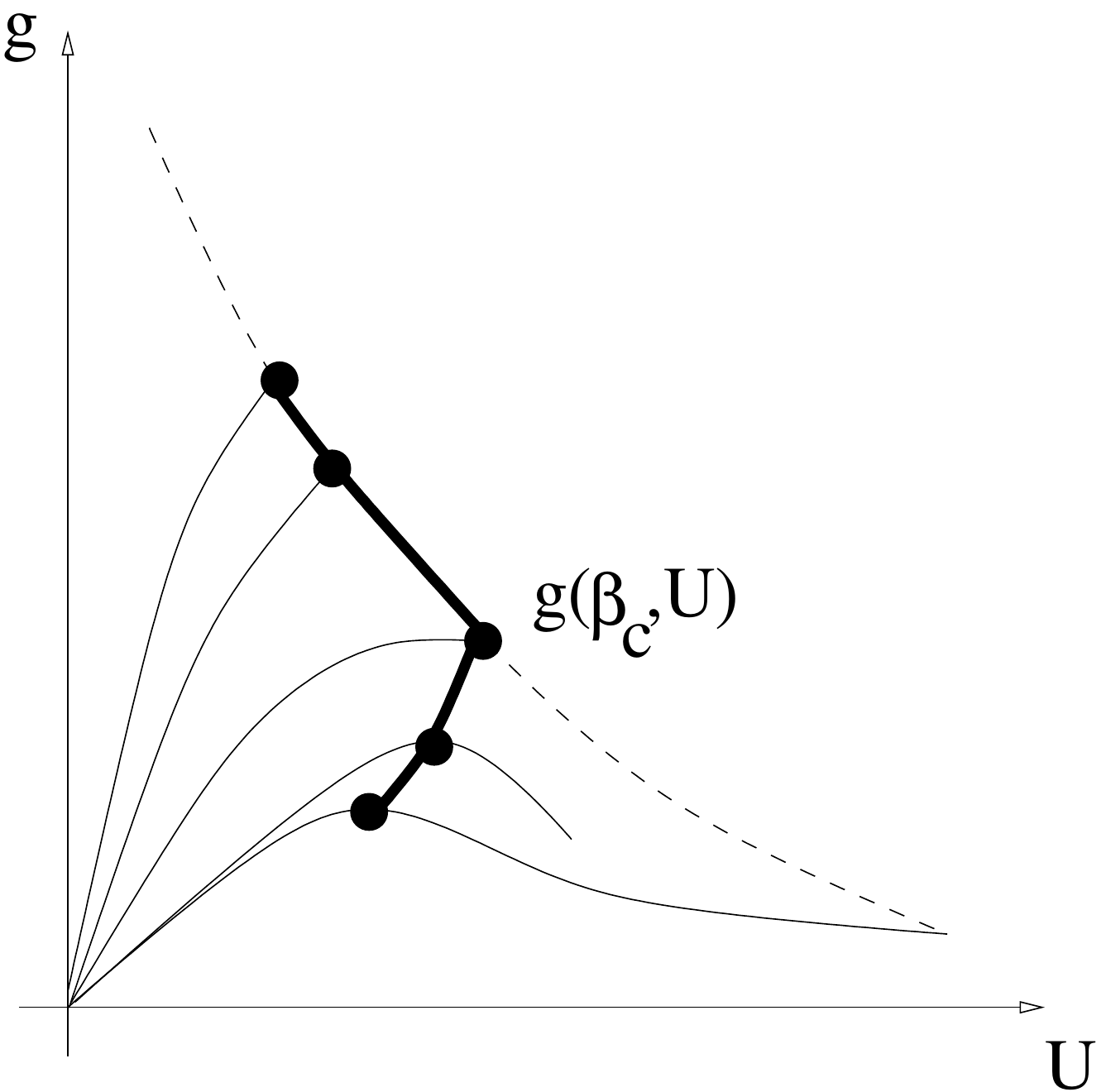}
  \caption{A schematic representation of the phase transition and the low $\beta$ and high $\beta$ regimes.}
\label{fig:gbg0mare}
\end{figure}

The self-consistency equation \eqref{eq:sys} admits two regimes corresponding to two distinct phases in the continuum limit.
Either, for small~$\beta$, the curve $g(\beta, U)$ achieves its maximum before exiting the analyticity domain of $S$, in which case, one recovers
the usual critical behavior:
\[ %\label{smallbeta}
  E_{\text{non-analytic}} \sim (g-g_c)^{\frac{3}{2}}   .
\]
Alternatively, for large enough $\beta$, the curve $g(\beta, U)$ exits the analyticity domain of $S$ before reaching its maximum, in which case,
 we recover a non-analytic behavior due to $S(\beta, z)$ itself:
\[ %\label{largebeta}
 E_{\text{non-analytic}}
\sim (g-g_b)^{\beta + \frac{3}{2}}   .
\]
The critical value $\beta_c$, where the phase transition occurs can be determined numerically:
$\beta_c\approx 1.162$ for $D=3$, $\beta_c\approx 1.216$ for $D=4$, and  $\beta_c\approx 1.134$ for $D=\infty$.
A schematic portrait of the transition is presented in Fig.~\ref{fig:gbg0mare}.
The critical behavior is dif\/ferent at $\beta=\beta_c$, namely~\cite{EDT}:
\[ %\label{critbeta}
 E_{\text{non-analytic}} \sim (g-g_c)^{\frac{\beta_c +\f32}{\beta_c +\f12}}   .
\]
In conclusion the susceptibility exponent, def\/ined by $E_{\text{non-analytic}} \sim (g-g_c)^{2-\gamma}$, is:
\[ %\label{gamma}
\gamma = \begin{cases}
\f12 &   \text{for} \  \beta <\beta_c   ,\vspace{1mm}\\
\dfrac{\beta_c-\f12}{\beta_c+\f12} &   \text{for} \  \beta=\beta_c   ,\vspace{1mm}\\
 \f12-\beta &    \text{for} \  \beta>\beta_c   .
\end{cases}
\]
We observe that for a large range of values of $\beta$,  that is,  for $\beta<\beta_c$, universality holds: the critical exponent
$\gamma$ is independent of $\beta$.
On the contrary, for $\beta\geq \beta_c$, one has a one-parameter family of dif\/ferent critical behaviors. A possible interpretation
of such unusual behavior is that
for $\beta$ suf\/f\/iciently large the new term starts behaving as a non-local or long-range interaction, for which universality
is not expected to hold.
This $\beta$-dependent behavior, which we encounter here, is reminiscent of certain models of branched polymers~\cite{ambjorn-BP,Bialas:1996ya}.

One can further show \cite{EDT} that the phase transition is third order. This result represents the f\/irst time that a phase
transition in the continuum limit is accessed by analytical means in dimensions higher than two. Furthermore, it is the f\/irst time that a
\emph{third-order} phase transition is observed in DT, which opens up the possibility of obtaining a continuum
limit with $2<d_H<\infty$, without the causality condition employed in causal DT
 \cite{Ambjorn:2010ce,Ambjorn:2011ph,Ambjorn:2011cg,Ambjorn:2005qt}.
The link to the numerical results of \cite{Laiho:2011ya}  deserves further exploration, either by pushing the
simulations to larger $\kappa_{D-2}$ and larger volumes, or by trying to extend the analytical tools to f\/inite $\kappa_{D-2}$.

   There are of course many interesting open questions regarding the critical behavior of colored tensor models. One would like
for instance to f\/ind multi-critical points generalizing the classical
 multi-critical points of matrix models \cite{Di_Francesco:1993nw}. Also a better understanding of
the continuum limit can be gained via some sort of double scaling limit including contributions from sub-leading terms in the $1/N$ expansion
(corresponding to a f\/inite $\kappa_{D-2}$ in the DT interpretation). Third, in two-dimensional matrix models the KPZ
correspondence \cite{david2, DK, Knizhnik:1988ak} relates statistical systems on random and f\/ixed lattices and led to
early computations of novel critical exponents~\cite{Dup}. One would like to f\/ind the appropriate
generalization of this correspondence in higher dimensions.

 {\it Renormalization and colored tensor models.}
    A direction of research in which the $1/N$ expansion of the colored tensor models is very consequential is the
renormalization of non-identically distributed tensor models. We shall brief\/ly draw the lessons one learns from the
results presented in this review. When studying non-i.i.d.~models, the divergences regularized by the re-scaling with $N$ of the coupling constant
become UV divergences in the amplitudes. Such divergences must be dealt with using renormalization
group tools. A similar situation appears in matrix models where the renormalization of non-i.i.d.~models
\cite{GW1,GW2,GW4,GW5,GW3} is intimately related to the $1/N$ expansion. As some non-i.i.d.~tensor models
are UV complete \cite{beta3,beta2, beta1}, one hopes that a non-i.i.d.~tensor model can provide a non-trivial
complete quantum f\/ield theory.

Renormalization is a recipe relying on three intertwined ingredients
\cite{GFTren}: a scale decomposition, a power counting, and a locality principle. In the case of matrix models,
both the locality principle and the power counting can be read of\/f already for the i.i.d.~models. In the $1/N$
expansion:
\begin{itemize}\itemsep=0pt
 \item {\it planar graphs with one external face dominate}. This is a key property, since in non-i.i.d.~models, only planar graphs with
one external face can be subtracted by adding local counter-terms in the action (the original trace interaction of the action
are planar with only one external face). The `locality' of usual quantum f\/ield theories becomes, in matrix models, `planarity
with one external face'.
  \item under the appropriate scaling of the coupling constant, the {\it planar graphs with one external face are uniformly
   divergent with $N$} yielding an inf\/inite family of uniformly divergent graphs. In non-i.i.d.~models, this leads to a non-trivial
    subtraction and running of the parameters.
\end{itemize}

   In non-i.i.d.~models the propagator {\it must scale} for large matrix indices exactly like the appropriate power of $1/N$ in order
to conserve these two features. Planar graphs will dominate any model, but any other scaling of the propagator leads either
to non-renormalizable or super-renormalizable models.

   The melonic graphs act in higher dimensions as the planar graphs of matrix models. They dominate the $1/N$ series having
the largest number of faces for a f\/ixed number of vertices. As such they will dominate any model. In order to
write a sensible non-identically distributed model one must propose a propagator with the appropriate scaling
for large indices. which render the melonic family uniformly divergent with respect to the cutof\/f. Such divergences can then
be subtracted by adding melonic counter-terms in the action.

One can ask if it is possible in any way to subtract the melonic family. That is, is it possible
to write some tensor models dominated by non-melonic  conf\/igurations? It is premature to give a def\/inite answer to this question,
but we doubt that this is possible. Non-trivial topologies are generated by (core) graphs with very few vertices.
For instance in $D=3$, already at 6 and 8 vertices one encounters not only the $\mathbb{R}P^3$ manifold but also pseudo-manifolds.
If one subtracts the melons it is very likely that non-trivial (non-manifold) topologies will dominate the
ef\/fective behavior of the series.

\subsection{Tensor models and spin foams}

Each graph of a colored tensor model represents a triangulation. If the base group is $SU(2)$, for instance, the faces of the
graphs are then labeled by spins (which one sums to recover the amplitudes of the graph). Such a triangulation with
spins associated to its $(D-2)$-simplices is called a `spin foam'. The spin foam models are then particular realizations of tensor
models on f\/ixed graphs.

 {\it A brief prelude -- the colorless world:}
When concentrating on particular graphs, one can of course chose a preferred graph on which to conduct a study. As one need not deal with the full series
 generated by a path integral, in this setting, one can consider also non-colored graphs.
In this case, the connection to cellular topological spaces is by no means direct, since the cellular structure is no longer intrinsically
encoded in the graph. Actually, these graphs are only 2-complexes, possessing just  0-, 1- and 2-bubbles. In other words,  the higher simplices alone have
a matching prescription.  This leaves a lot of implicit freedom in how one identif\/ies lower-dimensional simplices as one moves to ever higher dimensions.
Essentially, there exists a~class of diagrams (dominant both by weight and entropy) which cannot be immediately related to simplicial pseudo-manifolds,
 rather only after suf\/f\/icient barycentric subdivision \cite{DePietri:2000ii}. This barycentric subdivision can not be encoded in the Feynman rules of a tensor model.

As a result, one is not in a position to perform a $1/N$ expansion, extract the dominant order, examine
quantum equations of motion; it appears that the colors are indispensable for this purpose. Despite that, however, uncolored models have a place in the
 scheme of higher tensor models since in a certain sense they allow one to test whether or not certain properties depend intrinsically on the stricter
topological control one exerts in colored models.  For example, the Borel summability of some models does not necessarily require that the model is
colored \cite{Borel1,Borel2}.  Unfortunately, even this result requires a modif\/ication of the theory by adding an extra term (the pillow term) which
 changes the perturbative power-counting.

 {\it The `projector' models:}
Undoubtedly, a motivation for the study of tensor models comes from a discrete approach to quantum gravity known as spin foam quantization
\cite{Lrev,Drev1,Drev2}.  This particular brand
of theory relies on a simplicial (pseudo-)manifold, upon which the gravitational degrees of freedom are f\/irst placed and then quantized.

Three-dimensional gravity is ideally suited to this procedure, since it may be recast a topological gauge theory.  In this case, the absence of local degrees
 of freedom facilitates the passage to a discrete theory without any serious anomalies, in the sense that the local gauge algebra is preserved at the simplicial
 level.  Quantization in a lattice gauge theory sense is then immediate.

This case cannot be made for gravity theories in higher dimensions.  There, the simplicial manifold implies a severe truncation of the local degrees of freedom.
Also, there appears to be no way to preserve dif\/feomorphisms. One might hope to still possess some seed which would lead to dif\/feomorphism symmetry in an
inf\/inite ref\/inement limit, or one could hope to regain these degrees of freedom by summing over inf\/initely many triangulations. An initial step for the f\/irst
method is to f\/ind a coherent ref\/inement strategy while for the second, it is to f\/ind a summing prescription.  Colored tensor models of\/fer a
canonical summing prescription (alternative, less canonical, summing prescriptions have also been proposed \cite{Rovelli:2010qx}).

All current spin foam models place weights just on the 0-, 1- and 2-bubbles.  In particular, we are talking about
the Boulatov--Ooguri \cite{Boul,Ooguri:1992eb} and EPRL/FK models \cite{dev4,EPR2,EPR1,FreKra,LivineSpeziale}. To give a f\/lavour,
let us give the general f\/lavour of these non-i.i.d.~models with specif\/ic projectors provided for the Boulatov--Ooguri (BO) class of models:
\begin{gather}
 d\nu(\phi,\bar\phi )  =  \prod_{i=0}^D d\mu_{ \mathcal{P}^{\rm BO}}(   \phi^i,  \bar \phi^i )   e^{  -
\frac{\lambda}{ N^{\alpha(D)} } \sum_{\{\vec n\}} \prod_{i=0}^D \phi^i_{ \{\vec n_i\} } -
\frac{\bar \lambda}{ N^{\alpha(D)} } \sum_{\{\vec {\bar n} \}}
\prod_{i=0}^D \bar \phi^i_{ \vec {\bar n}_i  } }   ,\nonumber\\
\mathcal{P}^{\rm BO}_{ \vec {\bar n}_i , \vec n_i }  =  \sum_{m_i}\delta_{ \vec {\bar n}_i , \vec n_i + m_i} ,  \qquad \alpha^{\rm BO}(D)= \frac{(D-1)(D-2)}{4} .\label{eq:boulatov-ooguri}
\end{gather}
One can see that the Boulatov--Ooguri projects out a diagonal component, while the value of~$\alpha^{\rm BO}$ is chosen to make a $1/N$-expansion meaningful.
From the tensor model point of view, they represent some particular choices of non-i.i.d.~covariance, while the
vertex kernel remains unchanged. Note that in the literature such models are sometimes seen as a modif\/ication of the vertex while the propagator remains unchanged. Of course, by dressing the f\/ield itself with the square root of the propagator one can shift the non-triviality from the
covariance to the vertex kernel and vice versa. The two formulations are perfectly
equivalent, but we favor the former. The rationale behind this is that for the usual $\phi^4$ model, such a rescaling leads to an action with
$\delta(p_1-p_2)$ as a propagator and $ \delta(p_1+p_2+p_3+p_4) \prod\limits_{a=1}^4\frac{ 1 }{\sqrt{ p_a^2  +m^2}} $ interaction kernel.  One can
of course establish renormalizability of the $\phi^4$ model in this setting, but it is more dif\/f\/icult to prove that the vertex kernel is in
fact reproduced under the renormalization group f\/low. In addition and more importantly, since the quadratic part is just the identity,
one might be tempted to search for the semiclassical behavior in the large (semi-classical) $p$ limit.  In the case of the $\phi^4$ theory, this is
certainly a misguided endeavor, since the renormalization group f\/lows in the opposite direction.  For spin foams, this argument at the very least
cautions against prematurely drawing conclusions from results obtained in the large spin limit.

The Boulatov--Ooguri tensor model assigns a $BF$ theory weight to each graph \cite{blau}\footnote{$BF$ theory is a topological gauge f\/ield theory dependent on two f\/ields: a~$\mathfrak{g}$-valued $(D-2)$-form $B$ and a~$\mathfrak{g}$-valued 1-form connection $w$, which takes part via its curvature $F[w]$. $\mathfrak{g}$ is the particular algebra of interest. Its action is of the form:
\[
 S[B,w] = \frac{1}{k}\int_{\mathcal{M}}\textrm{tr}_{\mathfrak{g}}(B\wedge F[w]) .
 \]
In the case of 3d gravity $\mathcal{M}$ is a smooth manifold, $\mathfrak{g} = \mathfrak{su}(2)$ and $k = -16\pi G$. For such a theory, one replaces $\mathcal{M}$ by a simplicial presentation $\Delta$ of the same topology and the smooth f\/ields by counterparts which are distributional on relevant sub-simplices.  Thereafter, one arrives at an analogous discrete theory, which nonetheless encodes the relevant information of the continuum theory, since both capture just topological features of the manifold in question.  In the example given in equation~\eqref{eq:boulatov-ooguri}, $\mathfrak{g}= \mathbb{Z}_n$, but one may formulate the tensor model for gravity inspired gauge groups.}. Therefore, in the 3-dimensional case, it is appropriate for describing
gravity on a simplicial pseudo-manifold.  One may wonder why one would care about summing over graphs in this case, given that one has a well-def\/ined
quantum theory on  a graph-by-graph basis. One may appeal to the 2-dimensional case for a certain amount of motivation.  The Einstein--Hilbert action
in that case is a topological action; it is sensitive only to  the Euler characteristic of the underlying manifold.  Upon quantization, however, the theory
acquires the non-trivial property that quantum f\/luc\-tuations may induce tunneling between topologies.  Thus, a sum over topologies is appropriate to capture
the true quantum nature of the theory. Moreover, in the presence of a cosmological term the theory becomes highly non-trivial~\cite{DK}.
 While such a direct motivation is less persuasive in 3 dimensions, the sum over topologies can induce non-trivial ef\/fects.
For the colored version of the Boulatov--Ooguri models, one can again construct a topolo\-gi\-cal~$1/N$ expansion and again melo\-nic
graphs are the sole contributors to the leading order \cite{Gur3,Gur4,GurRiv}. One can hope for a critical behavior reminiscent of
i.i.d.~models, but one should keep in mind that cutof\/f ef\/fects, although not modifying the power-counting of graphs, might inf\/luence the
(previously combinatorial) f\/inite coef\/f\/icients of the graphs in the melonic series. Various bounds and amplitude evaluations
have been established for Boulatov--Ooguri like tensor models
\cite{BOpcont6,BOpcont3,BOpcont2,BOpcont4,BOpcont5,BOpcont7,BOpcont8,BOpcont1,Borel2}.

The EPRL-FK model assigns a constrained BF amplitude to each graph, intended to mimic the gravitational theory for $D\ge 4$.   Here, of course,
one hopes to use the full power of tensor models to regain a classical gravity on a dif\/ferential manifold (symmetries and all) in the con\-tinuum
semi-classical limit. A $1/N$ expansion has not yet been constructed for these models and constitutes an open research question.
Some intriguing results on graphs amplitudes have been established in the literature \cite{PcontEPRL2,PcontEPRL1} for these models.

{\it Baby spin foams:}
A recent development in the spin foam literature is the advent of so-called `baby spin foams' \cite{bahr-dittrich-ryan:baby}.
These are spin foam models utilizing f\/inite groups (for example, $\Z_N$) rather than the Lie groups (as used in Boulatov--Ooguri and
related models).   The hope is that the simplif\/ied group structure might allow one to perform more complicated tasks within the spin
foam formalism: to perform calculations with a large number of simplices, at least numerically and to utilize methods from statistical
lattice theories to coarse-grain spin foams~\cite{dittrich-eckert-martin:coarse}.  Moreover, these models provide a nice
 middle ground between tensor models and spin foams, since both the i.i.d.~and non-i.i.d.~potentials detailed earlier generate spin foam
amplitudes based on the group $\Z_N$ or a compact Lie group with a cut-of\/f.

{\it Symmetries and Ward-identities:}
In analyzing various colored tensor models one needs to address the question of the symmetries of the action. Indeed, they are
hugely consequential for the renormalizability of the theories. One can adopt a more algebraic approach and relate the
symmetries of the tensor to an underlying $n$-ary algebra~\cite{Sasa4, Sasa2,Sasa3} of functions. In this setting, every given tensor
conf\/iguration corresponds to an algebra whose structure constants are given by the tensor. In a more quantum f\/ield theoretic perspective,
the issue of symmetries is neatly clarif\/ied in the colored models. Indeed, as the tensors have no symmetry properties one can perform local
shifts in various arguments without af\/fecting the others. This allows the analysis of such symmetries at the
classical~\cite{Geloun:2011cz} and quantum~\cite{BenGeloun:2011xu} level. Moreover, the colors have proved instrumental in the identif\/ication
of a peculiar symmetry of projector models under shifts of the arguments corresponding to 0-simplices (vertices) of the triangulation~\cite{di2, di1}. Understanding
in more depth the symmetries of tensor models is a very active research direction.

There are many other topics in tensor models to which this review does not do justice. From alternative models~\cite{furt1} to the intimate relation
 with non-commutative geometry via a~Fourier transform~\cite{furt2} (which naturally sheds light on the geometry of the quantum space time~\cite{furt3});
 from the emergence of matter~\cite{furt5, dev2} on a non-commutative background to collective be\-haviors~\cite{furt6} or topological polynomials for tensor
 graphs~\cite{furt7}; these and may other topics are left for the dedicated reader.

\subsection{Recapitulation}

Alas, we have arrived at the moment to close of\/f this review of colored tensor models.  The preceding sections have been largely focused on adapting
 and applying methods well-known in statistical physics and matrix models.  We paid less attention to the physical and mathematical applications of
the colored tensor model framework. For sure, there is a multitude of toy-model scenarios of
physical theories, which have been known for quite some time to be captured by tensor models, but as for real-world examples \dots.
Of course, we hope that rapid progress in this area will rectify this situation sooner rather than later.  All the same,  we expect that the core
mathematical techniques, which we presented here, will continue to act as a linchpin for subsequent developments, no matter what direction they
might take.  With this in mind, we set down our roots in the theory of probability measures, stripping many of the implicit physical connotations
one inherits from a specif\/ic application.  This allowed us to concentrate on the core principles without the necessity to detail the physical
interpretation at every step.

Thereafter, we provided a brisk review of some familiar techniques employed in multi-matrix models with a view to extending them to higher dimensions.
This set the scene for the subset of quantum f\/ield theories to which this review is devoted, the $(D+1)$-colored tensor models.  From a~perturbative point
 of view, the central feature is that the Feynman graphs encode, as succinctly as possible, $D$-dimensional simplicial pseudo-manifolds.  The form of the
tensor models permits a sharp division of labor: the colors are solely responsible for the cellular topology, while the index structure is responsible for
 how the sub-cells are weighted in each cell complex.  Since this has so much potential, we embarked on a detailed study of a subset of structures embedded
in the $D$-complex.  The bubbles obviously play an important role no matter what the choice of tensor, covariance or vertex kernel. One needs them to grasp
the processes of dipole creation and deletion, which constitute graph manipulations, including, as a subset, the homeomorphisms. Convergence degree,
 combinatorial and topological equivalence are introduced along with the ubiquitous Riemann surfaces known as jackets.  Depending on our preference, we can
 utilize all or just a subset of the jackets to capture the degree, although in the most minimal case, one must sometimes introduce other 2-dimensional
objects known as patches for completeness.    Despite all this, we have just touched on the wealth of insights to be garnered from higher-dimensional cellular topology.

The model discussed exclusively in the main text is the i.i.d.~model for which the combinatorial and topological apparatus developed earlier was apt to
characterize the amplitudes.
Their scaling in the parameter $N$ (the size of the tensor) is controlled by the degree of the underlying graph.  Using the combinatorial and topological
 equivalence moves, we establish the concept of core graph at a given order.  It becomes apparent that one can order two distinct $1/N$ expansions, combinatorial
 or topological,  according to the core graphs of the appropriate type.  The important result is the at leading order, only the spherical topology contributes.

\looseness=-1
We made a brief excursion into the land of matrix models where we used the concept of factorization introduced earlier to partition the degrees of freedom of
 the tensor model into subsets attached to specif\/ic jackets (and possibly patches).  These subsets take the form of matrix models embedded inside the tensor
 structure.  This identif\/ication provides a staging post to subsequently apply matrix model methods directly to certain sub-sectors of a given tensor model.

Afterwards, we returned to analyze the leading order of our i.i.d.~model.   The leading order graphs all exhibited a characteristic `melonic' structure
 and could be mapped to a class of trees.  This re-indexing of the series allowed us to resum the dominant contributions.  This was already enough to reveal
the presence of a critical behavior leading to a continuum limit.   With this result, who could not venture to speculate on its physical implications.
Some insight comes from dynamical triangulations, which can give a more geometrical interpretation to graph conf\/igurations.   The melonic sector displays
 similar behavior to the branched polymer phase.

In addition, the leading order correlation functions also satisfy some Schwinger--Dyson equations.  These in turn form a Lie algebra another representation
of which is carried by the space of rooted colored $D$-ary trees.

Before concluding, we developed a rather abstract ansatz to obtain classical solutions to the tensor models.  It is based on the various possible factorizations
 of the $(D+1)$-colored graphs.  Ultimately, the main aim is to reveal properties of the tensor models by expanding around non-trivial classical solutions.

\subsection{Towards future endeavours}

To bring us in a full circle, colored tensor models enumerate random walks where the walks are themselves simplicial pseudo-manifolds. The feature that
instigated the breakthrough for tensor models was the coloring.  It allows one to identify the topology of the graphs, enables a~$1/N$ expansion ordered,
at least roughly, by topological features.  Moreover, it is the coloring once again that facilitates an in-depth analysis of the dominant behavior and a
 method to index the algebra associated to the quantum equations of motion.  But where do we see the theory progressing from this point on.  Clearly,
developments will be divided into two broad categories:  generic methods to analyze tensor theories and applications to pertinent physical and mathematical problems.

For the f\/irst category, their are a large number avenues to pursue. Let us list a few.
\begin{itemize}\itemsep=0pt
\item  We should ref\/ine our knowledge of the topological features encoded by the various graphs, e.g.\ via the def\/inition of (colored) homotopy invariants.  This has an number of side benef\/its. For the various extant tensor models, it would allow us  to investigate the nature of their sub-dominant contributions.  Moreover, it may allow us to develop models which weight graphs according to these topological properties.

\item At the moment, we possess a rather nice map between melonic graphs and a class of trees.   Of course, sub-leading graphs do not possess this melonic property. However, we should examine whether they possess a def\/ining characteristic that we might aid in our task of resumming these orders.  In any case, upon resummation, we should like to check the position of the f\/irst critical point for these higher order contributions.  A nice property of certain matrix models is that the sub-dominant contributions all diverge at the same critical point as the spherical sector~\cite{Di_Francesco:1993nw}.  This allows for a certain tuning of the parameters so that on passing to the continuum limit, one incorporates contributions from all 2d topologies. This is known as the {\it double scaling limit}.

 \item The double scaling limit that we mentioned a moment ago is a remarkable achievement of the matrix model program.  Since higher dimensional topology is so much richer in structure, generic tensor models are unlikely to possess just a single double scaling limit but rather double and multiple scaling limits. One possibility has already been investigated in \cite{gurau:double-scaling}.

 \item Another point of paramount importance is the analysis of the Schwinger--Dyson equations and Ward Identities.  These not only encapsulate the quantum dynamics but succinctly describe the quantum symmetries.  At leading order in the $1/N$-expansion of the i.i.d.~model, these symmetry generators have been shown to form a Lie algebra.  A pertinent question is whether this algebra forms a representation of the dif\/feomorphism algebra of the $D$-sphere, or perhaps even a sub-algebra.

 \item The wholescale incorporation of matrix model techniques into colored tensor models is certainly possible in the following sense.  They may be utilized as-is for certain sub-classes of tensor models which weight only one particular Riemann surface embedded in each $(D+1)$-colored graph.  One has enough control at the level of the action to ensure such a weighting.  The task is then is to repeat the study done throughout this article for such models.  They certainly have dif\/ferent contributions at leading order and therefore, dif\/ferent critical behaviors.  While there are several technical issues pertaining to this analysis, one must also attempt to give these models some physical interpretation.

On the other hand, various matrix model techniques are not obviously generalizable. For example, while there is certainly enough freedom in the choice of basis to diagonalize an Hermitian matrix, there is seemingly not enough freedom to diagonalize a tensor without f\/irst drastically reducing the number of degrees of freedom (essentially from $N^D$ down to $N^2$).  However, recent work in relating these tensors to $D$-ary algebras \cite{Sasa4, Sasa2, Sasa3} might help to motivate the study of both issues.

 \end{itemize}
 Of course, this is just a non-exhaustive list of active topics, but somehow the most immediate.

The problems of the second category may be summarized in the manner of a recent (in)famous press statement: there are known knowns, known unknowns,
unknown knowns and unknown unknowns.  The later subcategories are truly the most populated.    However, let us wax lyrical for a moment on some possibilities.

\begin{itemize}\itemsep=0pt
\item Given the obvious link between colored tensor models and simplicial (pseudo-)manifolds, a~natural application centres on (simplicial) quantum gravity, or rather, simplicial quantum physics in general.

With regards to quantum gravity, we described earlier recent endeavours to describe quantum space-time.  The natural space-time interpretation of i.i.d.~like models is via a map to dynamical triangulations, while the projector group f\/ield theory models are more directly interpreted via spin foams.  The general thrust of colored tensor models dif\/fers slightly from that of group f\/ield theories although there is a high level of cross pollination.  On the one hand, colored tensor models are well adapted to analysis via analogues of matrix model techniques, upon whose development this review has remained largely concentrated.  On the other hand, group f\/ield theories, as models with genuinely inf\/initely many degrees of freedom from the outset, set their sights on incorporating techniques from standard quantum f\/ield theory, e.g.\ renormalization and so forth.

Moreover, the incorporation of quantum matter is done slightly dif\/ferently in colored tensor models and group f\/ield theories.  For the former, the most readily feasible matter coupling is of Ising/Potts or hard dimer type \cite{bonzom:matter1,bonzom:matter}.  Once again, much inspiration can be taken from previous matrix models studies.  For the latter, there is a tentative yet ambitious proposal that matter might be incorporated via topological singularities charged with the relevant quantum numbers of the matter in question \cite{dev2, livorirya}.  These topological singularities are naturally present in the Feynman diagrams of the theory~-- they put the pseudo into pseudo-manifold.  This has been concretely studied in 3 dimensions with favorable results.  We should investigate this idea in more physical regimes.

\item Matrix models have become a rather convenient tool with which to study certain problems in mathematical biology, in particular, the secondary and tertiary structures of ribonucleic acid (RNA) \cite{david-biology, orland-zee}. RNA is a sequence of molecules, called nucleotides, which encodes genetic information. The sequence itself is known as the primary structure.  However, the molecules in the sequence are attracted to each other, which causes the sequence to fold.  This folding is the secondary structure.  Moreover, these folds arrange themselves not just on the plane but in 3-dimensional space, generating the tertiary structure.  Matrix models take advantage of their built-in topological expansion to model the tertiary structure.  Colored tensor models provide a natural 3-dimensional setting within in which to embed the various folded sequences and may give a more realistic picture for the structures in question.

\item As we motivated at the beginning, in our line of work it is often important to be able to count, and moreover, to count precisely.  We saw in Section~\ref{sec:critiid} that the leading order of a~class of tensor models could be mapped to the space of colored rooted $(D+1)$-ary trees, which in turn aided the leading order resummation.  No doubt as one investigates higher and higher orders, one shall encounter maps between dif\/ferent spaces of graphs.  Thus, one can expect a f\/low of information among tensor models, graph theory and combinatorics.
 \end{itemize}

As one can see, the general formalism of colored tensor models is broad enough to be applicable in principle to a wide variety of challenges.  The task remains, however, to pinpoint more exactly such situations.

%\pdfbookmark[1]{References}{ref}
\LastPageEnding

\end{document}